\documentclass{extarticle}

\newcommand*{\acknowledgesection}{}

\newcommand*{\usebiblatex}{}

\newcommand{\moffatbeta}{4.76}
\newcommand{\convdemoa}{1}
\newcommand{\convdemob}{3}
\newcommand{\convdemoc}{5}
\newcommand{\convdemod}{10}
\newcommand{\convdemoe}{20}
\newcommand{\mockfwhm}{3}
\newcommand{\senna}{0.5}
\newcommand{\sennb}{1}
\newcommand{\sennc}{4}
\newcommand{\sennd}{10}
\newcommand{\senq}{0.5}
\newcommand{\senre}{10}
\newcommand{\truncr}{5}
\newcommand{\purityn}{0.5}
\newcommand{\sentheta}{45}
\newcommand{\puritymag}{-10.5}
\newcommand{\senmagdiff}{-0.06}
\newcommand{\senfaintmag}{-9.2}
\newcommand{\senbrightmag}{-10.64}
\newcommand{\scccdncha}{4}
\newcommand{\scccdlmesh}{256}
\newcommand{\puritysethresh}{0.6}
\newcommand{\figtexdir}{/mnt/Work/NoiseChiselPaper/tex/}
\newcommand{\scccdname}{CORR01269690}
\newcommand{\backmag}{-10.00}
\newcommand{\backstd}{100}
\newcommand{\backcount}{10000}
\newcommand{\zeropoint}{0.00}
\newcommand{\backthresh}{10100}
\newcommand{\onelargefracabove}{19.83}
\newcommand{\severallargemaxpix}{39.67}
\newcommand{\datanoiselow}{-250}
\newcommand{\datanoisehigh}{1000}
\newcommand{\convsampbnumhigher}{44}
\newcommand{\ncpuritya}{27}
\newcommand{\ncpurityb}{26}
\newcommand{\ncpurityc}{32}
\newcommand{\ncpuritydemonfalse}{7}
\newcommand{\ncpurityd}{27}
\newcommand{\ncpurity}{0.89}
\newcommand{\sepuritya}{79}
\newcommand{\sepurityb}{90}
\newcommand{\sepurityc}{94}
\newcommand{\sepuritydemonfalse}{69}
\newcommand{\sepurityd}{87}
\newcommand{\sepurity}{0.29}
\newcommand{\crinnoisemax}{150}
\newcommand{\crinnoisenumhigher}{84}
\newcommand{\MEMORYOBJSTACK}{3000}
\newcommand{\MEMORYPIXSTACK}{300000}
\newcommand{\MEMORYBUFSIZE}{1024}
\newcommand{\DETECTTYPE}{CCD}
\newcommand{\GAINKEY}{GAIN}
\newcommand{\PIXELSCALE}{1.0}
\newcommand{\FITSUNSIGNED}{N}
\newcommand{\BACKTYPE}{AUTO}
\newcommand{\BACKSIZE}{64}
\newcommand{\BACKFILTERSIZE}{3}
\newcommand{\BACKPHOTOTYPE}{LOCAL}
\newcommand{\BACKPHOTOTHICK}{24}
\newcommand{\WEIGHTTYPE}{BACKGROUND}
\newcommand{\WEIGHTGAIN}{N}
\newcommand{\FILTER}{Y}
\newcommand{\FILTERNAME}{PSF.conv}
\newcommand{\MASKTYPE}{CORRECT}
\newcommand{\INTERPMAXXLAG}{16}
\newcommand{\INTERPMAXYLAG}{16}
\newcommand{\INTERPTYPE}{ALL}
\newcommand{\DETECTTHRESH}{1}
\newcommand{\ANALYSISTHRESH}{1}
\newcommand{\THRESHTYPE}{RELATIVE}
\newcommand{\DETECTMINAREA}{5}
\newcommand{\DEBLENDNTHRESH}{64}
\newcommand{\DEBLENDMINCONT}{0.001}
\newcommand{\CLEAN}{Y}
\newcommand{\CLEANPARAM}{1.5}
\newcommand{\FLAGIMAGE}{flag.fits}
\newcommand{\FLAGTYPE}{OR}
\newcommand{\SEEINGFWHM}{1.2}
\newcommand{\STARNNWNAME}{default.nnw}
\newcommand{\MAGZEROPOINT}{0.0}
\newcommand{\PHOTAPERTURES}{5}
\newcommand{\PHOTAUTOPARAMS}{3,3.5}
\newcommand{\PHOTAUTOAPERS}{0,0}
\newcommand{\PHOTFLUXFRAC}{0.5}
\newcommand{\PHOTPETROPARAMS}{2.0,3.5}
\newcommand{\SATURLEVEL}{50000.0}
\newcommand{\SATURKEY}{SATURATE}
\newcommand{\CHECKIMAGETYPE}{SEGMENTATION,APERTURES,BACKGROUND}
\newcommand{\CHECKIMAGENAME}{seg.fits,aper.fits,back.fits}
\newcommand{\CATALOGNAME}{SEresults.txt}
\newcommand{\CATALOGTYPE}{ASCII\_HEAD}
\newcommand{\PARAMETERSNAME}{./parameters/sextractor/sextractor.param}
\newcommand{\VERBOSETYPE}{NORMAL}

\newcommand{\hdu}{0}
\newcommand{\khdu}{0}
\newcommand{\skysubtracted}{0}
\newcommand{\minbfrac}{0.5}
\newcommand{\minnumfalse}{100}
\newcommand{\smeshsize}{32}
\newcommand{\lmeshsize}{200}
\newcommand{\lastmeshfrac}{0.51}
\newcommand{\numnearest}{10}
\newcommand{\smoothwidth}{3}
\newcommand{\fullconvolution}{0}
\newcommand{\fullinterpolation}{0}
\newcommand{\fullsmooth}{0}
\newcommand{\mirrordist}{1.5}
\newcommand{\minmodeq}{0.49}
\newcommand{\qthresh}{0.3}
\newcommand{\erode}{2}
\newcommand{\erodengb}{4}
\newcommand{\opening}{1}
\newcommand{\openingngb}{8}
\newcommand{\sigclipmultip}{3}
\newcommand{\sigcliptolerance}{0.2}
\newcommand{\dthresh}{-0.1}
\newcommand{\detsnminarea}{15}
\newcommand{\detsnhistnbins}{0}
\newcommand{\detquant}{0.99}
\newcommand{\dilate}{3}
\newcommand{\segsnminarea}{25}
\newcommand{\segquant}{0.99}
\newcommand{\segsnhistnbins}{0}
\newcommand{\gthresh}{0.5}
\newcommand{\minriverlength}{15}
\newcommand{\objbordersn}{1}

\newcommand{\invqthresh}{70}
\newcommand{\cpughz}{3.06}
\newcommand{\nconscccdtime}{4.43}
\newcommand{\cpunumcore}{4}
\newcommand{\sextractorversion}{2.19.5}
\newcommand{\imgstatmirrordist}{1.5}
\newcommand{\modeonex}{100}
\newcommand{\modeoney}{140}
\newcommand{\modeonemag}{-20}
\newcommand{\modeonen}{2.5}
\newcommand{\modeonere}{50}
\newcommand{\modeonet}{2}
\newcommand{\modetwox}{110}
\newcommand{\modetwoy}{285}
\newcommand{\modetwomag}{-28}
\newcommand{\modetwon}{0.5}
\newcommand{\modetwore}{100}
\newcommand{\modetwot}{5}
\newcommand{\moderealonesym}{0.43}
\newcommand{\moderealonemodeq}{43.37}
\newcommand{\moderealonesympoint}{13.032}
\newcommand{\moderealsixsym}{0.25}
\newcommand{\moderealsixmodeq}{34.35}
\newcommand{\moderealsixsympoint}{7.4477}
\newcommand{\modemodeonesym}{0.84}
\newcommand{\modemodeonemodeq}{15.05}
\newcommand{\modemodeonesympoint}{202.4}
\newcommand{\modemodetwosym}{0.74}
\newcommand{\modemodetwomodeq}{4.16}
\newcommand{\modemodetwosympoint}{1023}
\newcommand{\setenththresh}{11296}
\newcommand{\setenthkronff}{87.33}
\newcommand{\sehalfthresh}{319}
\newcommand{\sehalfkronff}{80.22}
\newcommand{\seonethresh}{0}
\newcommand{\seonekronff}{78.21}
\newcommand{\setwothresh}{0}
\newcommand{\setwokronff}{74.30}
\newcommand{\setenminarea}{8989}
\newcommand{\setwofiveminarea}{3880}
\newcommand{\sefiftyminarea}{966}
\newcommand{\sesevenfiveminarea}{350}
\newcommand{\sehundredminarea}{260}
\newcommand{\txtrandomsmallmode}{9.90}
\newcommand{\txtrandomsmallmodesym}{1.26}
\newcommand{\txtonelargemode}{9.50}
\newcommand{\txtonelargemodesym}{0.48}
\newcommand{\txtseverallargemode}{156.10}
\newcommand{\txtseverallargemodesym}{0.44}
\newcommand{\onelargekronstd}{16.02}
\newcommand{\onelargekronmean}{4.69}
\newcommand{\issigclipnum}{5}
\newcommand{\issigclipmultip}{4}

\newcommand{\naxisa}{1000}
\newcommand{\naxisb}{1000}
\newcommand{\rkron}{3,3.}

\newcommand{\datanoisersstarmag}{-12}
\newcommand{\datanoisersngals}{95}
\newcommand{\datanoisersgminmag}{-11}
\newcommand{\datanoisersgmaxmag}{-6}
\newcommand{\datanoiseslngals}{6}
\newcommand{\datanoiseslgminmag}{-18}
\newcommand{\datanoiseslgmaxmag}{-15}
\newcommand{\onelargemag}{-16}
\newcommand{\onelargen}{4.00}
\newcommand{\onelargere}{30}
\newcommand{\onelargeq}{0.25}
\newcommand{\randomsmallsca}{11.09}
\newcommand{\randomsmallscb}{10.58}
\newcommand{\randomsmallscc}{10.37}
\newcommand{\randomsmallscd}{10.37}
\newcommand{\randomsmallsce}{10.37}
\newcommand{\onelargesca}{20.16}
\newcommand{\onelargescb}{19.41}
\newcommand{\onelargescc}{17.93}
\newcommand{\onelargescd}{17.22}
\newcommand{\onelargesce}{17.09}
\newcommand{\severallargesca}{199.33}
\newcommand{\severallargescb}{196.68}
\newcommand{\severallargescc}{194.16}
\newcommand{\severallargescd}{192.60}
\newcommand{\severallargesce}{191.86}
\newcommand{\fourdettfnum}{132}
\newcommand{\fourdettfmax}{4.77}
\newcommand{\fourdettfqnt}{8.50}
\newcommand{\onelargedettfnum}{113}
\newcommand{\onelargedettfmax}{1.89}
\newcommand{\onelargedettfqnt}{3.82}
\newcommand{\sensitivitycdettfnum}{111}
\newcommand{\sensitivitycdettfmax}{2.37}
\newcommand{\sensitivitycdettfqnt}{3.61}
\newcommand{\dettfsmallestsnqnt}{3.61}
\newcommand{\scccdnaxisa}{2048}
\newcommand{\scccdnaxisb}{4177}
\newcommand{\onedgepa}{45}
\newcommand{\onedgedist}{35}
\newcommand{\senamed}{5.6}
\newcommand{\senamode}{5.7^{LS}}
\newcommand{\senascconv}{5.4\pm99.9}
\newcommand{\senascfixed}{5.6\pm99.9}
\newcommand{\senasemean}{5.1\pm99.9}
\newcommand{\senancmean}{2.2\pm99.2}
\newcommand{\senanfalse}{3}
\newcommand{\senasencfrac}{2.3}
\newcommand{\senbmed}{6.6}
\newcommand{\senbmode}{6.6}
\newcommand{\senbscconv}{6.4\pm100.7}
\newcommand{\senbscfixed}{6.6\pm100.7}
\newcommand{\senbsemean}{5.9\pm100.7}
\newcommand{\senbncmean}{3.5\pm100.1}
\newcommand{\senbnfalse}{4}
\newcommand{\sencmed}{4.8}
\newcommand{\sencmode}{2.2}
\newcommand{\sencscconv}{5.6\pm101.2}
\newcommand{\sencscfixed}{4.8\pm101.2}
\newcommand{\sencsemean}{3.9\pm100.5}
\newcommand{\sencncmean}{3.1\pm100.4}
\newcommand{\sencnfalse}{2}
\newcommand{\sendmed}{5.9}
\newcommand{\sendmode}{5.9}
\newcommand{\sendscconv}{6.3\pm102.2}
\newcommand{\sendscfixed}{5.9\pm102.2}
\newcommand{\sendsemean}{3.1\pm100.4}
\newcommand{\sendncmean}{2.2\pm100.1}
\newcommand{\sendnfalse}{2}
\newcommand{\realamed}{3.4}
\newcommand{\realamode}{1.3}
\newcommand{\realascconv}{5.1\pm29.0}
\newcommand{\realascfixed}{3.1\pm28.6}
\newcommand{\realasemean}{-0.9\pm25.0}
\newcommand{\realancmean}{-1.2\pm25.0}
\newcommand{\realbmed}{2.0}
\newcommand{\realbmode}{0.7}
\newcommand{\realbscconv}{3.1\pm27.9}
\newcommand{\realbscfixed}{1.9\pm27.8}
\newcommand{\realbsemean}{-0.7\pm25.8}
\newcommand{\realbncmean}{-0.8\pm25.9}
\newcommand{\realcmed}{2.1}
\newcommand{\realcmode}{0.7^{LS}}
\newcommand{\realcscconv}{2.5\pm22.0}
\newcommand{\realcscfixed}{1.6\pm21.7}
\newcommand{\realcsemean}{-0.9\pm19.7}
\newcommand{\realcncmean}{-1.1\pm19.8}
\newcommand{\realdmed}{4.1}
\newcommand{\realdmode}{3.0}
\newcommand{\realdscconv}{5.2\pm28.6}
\newcommand{\realdscfixed}{4.1\pm28.5}
\newcommand{\realdsemean}{-1.1\pm25.5}
\newcommand{\realdncmean}{-1.2\pm25.6}
\newcommand{\onedgemed}{38.9}
\newcommand{\onedgemode}{32.3}
\newcommand{\onedgescconv}{41.1\pm115.8}
\newcommand{\onedgescfixed}{36.1\pm114.9}
\newcommand{\onedgesemean}{23.7\pm102.8}
\newcommand{\onedgencmean}{8.3\pm100.2}
\newcommand{\onedgenfalse}{3}
\newcommand{\onedgesencfrac}{2.9}
\newcommand{\onelargemed}{20.2}
\newcommand{\onelargemode}{9.5}
\newcommand{\onelargescconv}{23.1\pm114.2}
\newcommand{\onelargescfixed}{17.1\pm113.1}
\newcommand{\onelargesemean}{8.3\pm100.9}
\newcommand{\onelargencmean}{1.8\pm99.9}
\newcommand{\onelargenfalse}{3}
\newcommand{\onelargettsencfrac}{4.6}
\newcommand{\onelargetonesemean}{6.9\pm100.7}
\newcommand{\onelargetosencfrac}{3.8}
\newcommand{\onelargethalfsemean}{5.3\pm100.3}
\newcommand{\onelargettenthsemean}{-5.0\pm99.9}
\newcommand{\dnintxtrandomsmallmed}{11.1}
\newcommand{\dnintxtrandomsmallscconv}{11.1\pm103.3}
\newcommand{\dnintxtrandomsmallscfstd}{167.30}
\newcommand{\dnintxtonelargemed}{20.2}
\newcommand{\dnintxtonelargescconv}{23.1\pm114.2}
\newcommand{\dnintxtonelargescfstd}{455.90}
\newcommand{\dnintxtseverallargemed}{199.3}
\newcommand{\dnintxtseverallargescconv}{236.9\pm215.6}
\newcommand{\dnintxtseverallargescfstd}{341.05}

\usepackage[letterpaper, includeheadfoot, body={18.7cm, 24.5cm}]{geometry}

\usepackage[T1]{fontenc}
\usepackage{txfonts}

\usepackage{./tex/aas_macros}

\usepackage{array}
\usepackage{tabularx}
\newcolumntype{L}[1]{>{\raggedright\let\newline\\\arraybackslash\hspace{0pt}}m{#1}}
\newcolumntype{C}[1]{>{\centering\let\newline\\\arraybackslash\hspace{0pt}}m{#1}}
\newcommand\Tstrut{\rule{0pt}{0.35cm}}         

\usepackage{xcolor}
\usepackage{pst-all}

\definecolor{mymg}{cmyk}{0,1,0,0}
\definecolor{myblack}{cmyk}{1,1,1,1}
\definecolor{mywhite}{cmyk}{0,0,0,0}
\definecolor{myblue}{cmyk}{1,0.6,0,0}
\definecolor{mypurp}{cmyk}{0.75,1,0,0}
\definecolor{trueblue}{cmyk}{1,0.7,0,0}
\definecolor{myred}{cmyk}{0,0.99,1.0,0}
\definecolor{myyellow}{cmyk}{0,0,1.0,0}
\definecolor{mygreen}{cmyk}{0.78,0.05,0.71,0}
\definecolor{myredorange}{cmyk}{0,0.82,0.70,0}
\definecolor{myyelloworange}{cmyk}{0,0.37,0.83,0}

\color{myblack}

\usepackage{fixltx2e}

\usepackage{alltt}

\usepackage[hang]{footmisc}
\usepackage{datetime}

\usepackage{courier}

\usepackage{enumerate}

\usepackage[utf8]{inputenc}

\usepackage{setspace}
\renewcommand{\baselinestretch}{0.95}

\usepackage{pslatex}

\usepackage[dvips]{graphicx}

\usepackage[english]{babel}

\usepackage{textcomp}

\usepackage{picture}
\usepackage{eso-pic}

\usepackage{caption}
\ifdefined\makeeps
\usepackage{tikz}
\usepackage{pgffor}
\usetikzlibrary{spy}
\usetikzlibrary{graphs}
\usetikzlibrary{arrows}
\usetikzlibrary{shadows}
\usetikzlibrary{shapes.misc}
\usetikzlibrary{backgrounds}
\usepgflibrary{shapes.symbols}
\usetikzlibrary{shapes.geometric}
\tikzset{ mybox/.style = { rectangle, rounded corners=1.5mm,
                           minimum width=2cm, minimum height=0.5cm,
                           anchor=north, align=center, draw=black },
                           mydiamond/.style = {diamond, aspect=3,
                           draw=black}, hv path/.style = {rounded
                           corners=0.8mm, to path={-|
                           (\tikztotarget)}}, vh path/.style =
                           {rounded corners=0.8mm, to path={|-
                           (\tikztotarget)}} }
\usetikzlibrary{external}
\tikzset{external/system call={rm -f "\image".eps "\image".ps "\image".dvi; latex \tikzexternalcheckshellescape -halt-on-error
-interaction=batchmode -jobname "\image" "\texsource";
dvips -o "\image".ps "\image".dvi;
ps2eps "\image.ps"}}
\usepackage{pgfplots}
\pgfplotsset{compat=1.10}
\usepgfplotslibrary{groupplots}
\tikzsetexternalprefix{./tikz/}
\pgfplotsset{axis line style={thick}, tick style={semithick}}
\fi

\usepackage{setspace}

\usepackage{authblk}
\renewcommand\Authfont{\normalsize\scshape}
\renewcommand\Affilfont{\footnotesize\normalfont}
\usepackage[
    style=authoryear,
    hyperref=true,
    doi=false,
    url=false,
    eprint=false,
    uniquelist=false,
    maxbibnames=4,
    minbibnames=1,
    maxcitenames=2,
    mincitenames=1,
    backend=biber,natbib]{biblatex}
\AtEveryCitekey{\clearfield{title}}
\AtEveryBibitem{\clearfield{title}}
\DeclareFieldFormat[article]{pages}{#1}
\DeclareFieldFormat{pages}{\mkfirstpage[{\mkpageprefix[bookpagination]}]{#1}}
\renewbibmacro{in:}{}
\renewcommand*{\bibfont}{\footnotesize}
\DefineBibliographyStrings{english}{references = {REFERENCES}}

\usepackage[colorlinks,urlcolor=blue,
            citecolor=blue,linkcolor=blue]{hyperref}
\hypersetup
{
    pdfauthor={Mohammad Akhlaghi},
    pdfsubject={The Astrophysical Journal Supplement Series, 220(2015) 1. doi:10.1088/0067-0049/220/1/1. arXiv: 1505.01664},
    pdftitle={Noise Based Detection and Segmentation of Nebulous Objects},
    pdfkeywords={galaxies: irregular -- galaxies: photometry
    -- galaxies: structure -- methods: data analysis
    -- techniques: image processing -- techniques: photometric}
}

\DeclareSourcemap{
  \maps[datatype=bibtex]{
    \map{
      \step[fieldsource=adsurl,fieldtarget=iswc]
    }
  }
}

\newcommand{\doihref}[2]{\href{#1}{\color{mymg}{#2}}}
\newcommand{\adshref}[2]{\href{#1}{\color{mypurp}{#2}}}
\newcommand{\blackhref}[2]{\href{#1}{\color{black}{#2}}}

\DeclareFieldFormat{doilink}{
    \iffieldundef{doi}{#1}{\doihref{http://dx.doi.org/\thefield{doi}}{#1}}}
\DeclareFieldFormat{adslink}{
    \iffieldundef{iswc}{#1}{\adshref{\thefield{iswc}}{#1}}}
\DeclareFieldFormat{arxivlink}{
    \iffieldundef{eprint}{#1}{\href{http://arxiv.org/abs/\thefield{eprint}}{#1}}}

\DeclareBibliographyDriver{article}{%
  \usebibmacro{bibindex}%
  \usebibmacro{begentry}%
  \usebibmacro{author/translator+others}%
  \newunit%
  \printtext[doilink]{\usebibmacro{journal}}%
  \addcomma%
  \printtext[adslink]{\printfield{volume}}%
  \addcomma%
  \printtext[arxivlink]{\printfield{pages}}%
}

\usepackage{fancyhdr}
\renewcommand{\headrulewidth}{0pt}
\fancypagestyle{firststyle}
{ \lhead{\footnotesize{\scshape The Astrophysical Journal Supplement
  Series}, 220:1 (33pp), 2015 September \\\scriptsize \textcopyright
  2015, The American Astronomical Society. All rights reserved.}
\rhead{\footnotesize\href{http://dx.doi.org/doi:10.1088/0067-0049/220/1/1}{doi:10.1088/0067-0049/220/1/1}\\} }

\usepackage{titlesec}
\titlelabel{\thetitle}
\titleformat{\section}
  {\centering\normalfont\uppercase} {\thesection.}  {0em} { }
\titlespacing\section{0pt}{20pt plus 2pt minus 2pt}{5pt plus 2pt minus 2pt}
\titleformat{\subsection}
  {\centering\normalsize\slshape}
  {\thesubsection.}
  {0em}
  { }
\titlespacing\subsection{0pt}{20pt plus 2pt minus 2pt}{5pt plus 2pt minus 2pt}
\titleformat{\subsubsection}
  {\centering\small\slshape}
  {\thesubsubsection.}
  {0em}
  { }
\titlespacing\subsubsection{0pt}{20pt plus 2pt minus 2pt}{5pt plus 2pt minus 2pt}

\title{\large \uppercase{Noise-based detection and segmentation of nebulous objects} \vspace{-0.2cm}}
\author{{Mohammad Akhlaghi,}}
\author{Takashi Ichikawa}
\affil{Astronomical Institute, Tohoku University,
  Aoba, Sendai 980-8578, Japan; \url{akhlaghi@astr.tohoku.ac.jp}\\
  \emph{Received 2014 November 17; accepted 2015 April 6; published 2015 August 26}}
\date{}

\begin{document}
\twocolumn[
  \begin{@twocolumnfalse}
    {\setstretch{1.0}
      \maketitle
    }
    \renewcommand{\baselinestretch}{0.8}
    \vspace{-1.2cm}
    \begin{spacing}{0.9}
      \begin{center}
        \setlength{\leftskip}{1.25cm}
        \setlength{\rightskip}{1.25cm}
        {
          \begin{center}ABSTRACT\end{center}

          \vspace{0.2cm}

          \noindent A noise-based non-parametric technique for
          detecting nebulous objects, for example, irregular or clumpy
          galaxies, and their structure in noise is
          introduced. ``Noise-based'' and ``non-parametric'' imply
          that this technique imposes negligible constraints on the
          properties of the targets and that it employs no regression
          analysis or fittings. The sub-sky detection threshold is
          defined and initial detections are found, independently of
          the sky value. False detections are then estimated and
          removed using the ambient noise as a reference. This results
          in a purity level of $\ncpurity$ for the final detections as
          compared to $\sepurity$ for \textsf{SExtractor} when a
          completeness of 1 is desired for a sample of extremely faint
          and diffuse mock galaxy profiles. The difference in the mean
          of the undetected pixels with the known background of mock
          images is decreased by $\onelargettsencfrac$ times depending
          on the diffuseness of the test profiles, quantifying the
          success in their detection. A non-parametric approach to
          defining substructure over a detected region is also
          introduced. \textsf{NoiseChisel} is our software
          implementation of this new technique. Contrary to the
          existing signal-based approach to detection, in its various
          implementations, signal-related parameters such as the image
          point spread function or known object shapes and models are
          irrelevant here. Such features make this technique very
          useful in astrophysical applications such as detection,
          photometry, or morphological analysis of nebulous objects
          buried in noise, for example, galaxies that do not
          generically have a known shape when imaged.\hfill{}

          \vspace{0.2cm} \noindent \emph{Key words:} galaxies:
          irregular -- galaxies: photometry -- galaxies: structure --
          methods: data analysis -- techniques: image processing --
          techniques: photometric \hfill{}

          \vspace{0.2cm} \noindent \emph{Free reproduction:} All the
          data-generated numbers and figures in this paper are exactly
          reproducible/configurable with a \texttt{make}
          command. See \texttt{\small reproduce/README} in the arXiv
          source files. All results generated
          by \blackhref{http://www.gnu.org/philosophy/free-sw.en.html}{free
          software}. \hfill{}

          \vspace{-0.3cm}
      }
      \end{center}
    \end{spacing}
    \vspace{0.75cm}
  \end{@twocolumnfalse}
]
\setlength{\leftskip}{0pt}
\setlength{\rightskip}{0pt}

\thispagestyle{firststyle}

\section{Introduction}\label{intro}

  Galaxies are one of the most prominent forms of nebulous or
  amorphous objects in astrophysics and astronomical image
  processing. They appear to have a very diverse kinematic history and
  thus display a very large variety of shapes and forms. They also
  have complex three-dimensional structures which we can observe in
  two dimensions. Galaxy images can host an arbitrary number of clumps
  positioned anywhere over their light profiles. These clumps might be
  minor or major mergers, supernovae, other galaxies, or stars on the
  same line of sight, or highly localized unobscured star-forming
  regions. With such hurdles, astronomical image processing, and in
  particular galaxy morphological analysis, can be considered one of
  the most demanding of all disciplines in image processing.

  Observations are inevitably diluted with noise. Any systematic bias
  in measuring the noise properties will directly propagate to all
  higher level measurements on the detected signal or scientific
  targets. Therefore, accurately measuring the noise characteristics
  is the most fundamental and primary issue in data analysis and
  subsequent astrophysical interpretations. However, noise can only be
  accurately characterized when the signal or object(s) buried in the
  noise are detected and removed. Therefore any difficulties in
  detection can directly translate into a systematic bias in the
  measured noise properties. For example, objects that are very faint
  and diffuse overall are very hard to detect. For brighter objects
  that can be detected, identifying the possible boundaries is a major
  concern. The uniformity in shape, or morphology, between various
  objects can also play a significant role in the detection ability.

  Detection of nebulous astronomical targets significantly suffers
  from all the problems mentioned above. Because of the instrument and
  atmospheric point spread function ({\small PSF}) and also the
  intrinsic shapes of galaxies, their profiles sink deep into the
  noise very slowly with a hardly detectable clear
  cutoff. Astronomical targets also have an extremely wide range of
  apparent luminosity and surface brightness profiles, from very
  bright nearby stars and galaxies to faint objects far below the
  detection limits.

  The most commonly used detection methods in astronomy, regardless of
  their implementation, can be classified as \emph{signal-based}
  detection (see Appendix \ref{existingmethods} for a review). The
  signal's a priori known properties are the basis of this
  method. Thus the inherent shape (for example, being an ellipse or a
  functional radial profile) of the galaxies and the atmospheric and
  instrumental effects, or {\small PSF}, have to be known prior to
  running the detection algorithm. This approach to detection will
  therefore be most accurate for objects that satisfy the a priori
  known shapes and will become less reliable as the inherent shapes of
  the targets deviate from the expected shape. Failing to correctly
  find the {\small PSF} will also hamper the accuracy of the results.

  Systematic biases that are caused by cosmological surface brightness
  dimming along with {\small K} corrections and morphological
  evolution can hamper the accuracy of any study attempting to compare
  galaxy populations at different redshifts. Therefore it is very
  important that the detection and measurement tools used can provide
  less biased results. An example of such comparisons is the observed
  morphological evolution of blue galaxies where it has been observed
  that at higher redshifts ($z>1$), a larger fraction of galaxies
  display irregular or clumpy morphologies compared to their lower-$z$
  counterparts \citep{cowie95, abraham96, lotz06, elmegreen2007,
    Murata14}. This observed morphological evolution biases the
  photometric measurements of any detection method that is based on a
  fixed elliptical shape and surface brightness profile for all
  redshifts, for example, the \citet{petrosian} and \citet{kron}
  methods. It is currently thought that the the peak of cosmic star
  formation density was at $z\sim2$ \citep{hopkins06, kaj2010,
    lundgren14}. Hence, the spatial variation of star formation
  activity within these earlier galaxies or their star formation
  morphology is instrumental in understanding star formation in the
  galaxies.

  The concordance or {\small $\Lambda$CDM} cosmological model is
  another field that heavily depends on detection ability. It predicts
  much fainter, lower surface brightness dwarf galaxies than currently
  observed and may be lurking below the currently used detection
  thresholds \citep{Kauffmann93, klypin1999}. From the theoretical
  point of view, changing the astrophysical constraints of the
  semi-analytic models results in the simulations and observations
  becoming more consistent \citep[see][and references
    therein]{maccio2010}. However, an observational confirmation is
  still lacking. Since such low mass galaxies are very faint and have
  irregular/amorphous morphologies, existing detection methods that
  rely on an ellipse and a priori known radial profiles are
  insufficient and new detection methods that do not have this
  limitation might provide the solution from the observational point
  of view.

  In the study of individual galaxies, the faint outer wings are a
  treasure trove for expanding our knowledge of galaxy
  evolution. Being much less affected by internal processes, the
  fainter outskirts of galaxies preserve valuable information about
  the dynamic history of the galaxy.  For example, finding tidal tails
  present in these regions can be a very good indicator of possible
  previous merger events \citep[for example,
    see][]{ellison13}. Accurate detection of the galaxy light profiles
  at large radii is also very dependent on the detection technique and
  Sky subtraction (see Section \ref{sky}). It has been observed that
  the radial profiles of star-forming galaxies deviate from a purely
  exponential profile at large radii \citep[for example,
    see][]{degrijs2001, pohlen06}. Successfully detecting this
  behavior in the outskirts can help explain the dependency of the
  star formation on gas surface density \citep{kennicutt89,
    kregel2004}.

  Intra-cluster light is another field in galaxy evolution that relies
  heavily on the ability to accurately detect and subtract the
  faintest parts of cluster galaxies. These regions harbor stars that
  have been stripped from the galaxies but are gravitationally bound
  to the cluster \citep[see][and references there
    in]{intraclusterlight}. Using creative and simple hardware
  \citep{abrahamvandokkum}, \citet{vanDokkum14} recently found 47 very
  low surface brightness but large galaxies in the comma cluster. Some
  of them could be visually confirmed in archived images but remained
  undetectable using the existing detection methods due to their very
  diffuse structure.

  Increasing the detection ability by using larger and more accurate
  hardware is one the primary reasons that new telescopes and cameras
  are being commissioned. Over the past several decades the
  instruments used for astronomical observations have undergone
  significant advancements from analog photographic plates to digital
  {\small CCDs}. However, the most commonly used detection techniques
  still rely on the \citet{petrosian} or \citet{kron} approximations
  or \citet{deVaucClassification} or \citet{sersic} profile
  fittings. These functions are defined on a continuous space, so they
  can be considered analog techniques. The detection technique
  introduced in this paper exploits the digital nature of modern data
  sets. Most of the operation is done on the noise (undetected
  regions), which is also used as the basis for identifying false
  detection and segmentation results. Therefore statistically
  significant prior knowledge of signal-related properties becomes
  irrelevant. It also makes no use of any regression analysis or
  fitting which will bias the results when the targets deviate from
  the required model functional forms.

  In Section \ref{definitions} the properties of the images used and
  some definitions are provided. The new noise based solution is fully
  elaborated with images showing every step in Section
  \ref{noisechisel}. In Section \ref{largeimg} application of the
  concepts on a large image for increased accuracy and efficiency is
  explained. In Section \ref{analysis} the success in detection and
  the purity of the proposed algorithm on mock images with mock noise
  is analyzed. Results on mock profile and noise are only used for
  this ``proof of concept'' paper, a full completeness, purity, number
  count, and etc, analysis on real images with no mock profiles will
  be provided in a companion paper (M. Akhlaghi et al. 2015. in
  preparation). Finally, in Section \ref{discussion} the usefulness of
  this approach for existing and future astronomical research is
  discussed. In Appendices \ref{existingsky} and \ref{existingmethods}
  the existing approaches to calculating the Sky and detection and
  segmentation/deblending are critically reviewed. Appendix
  \ref{findingmode} explains a new algorithm for finding the mode of a
  distribution. This algorithm lies at the heart of our novel sub-Sky
  thresholding which is independent of the Sky value.  Appendices
  \ref{SEconfig} and \ref{NCconfig} show the parameter lists used for
  running \textsf{SExtractor} and \textsf{NoiseChisel}.

  \begin{figure*}[tbh]
    \centering
    \ifdefined\makeeps
    \input{./tex/dataandnoise}
    \else
    \includegraphics[width=\linewidth]{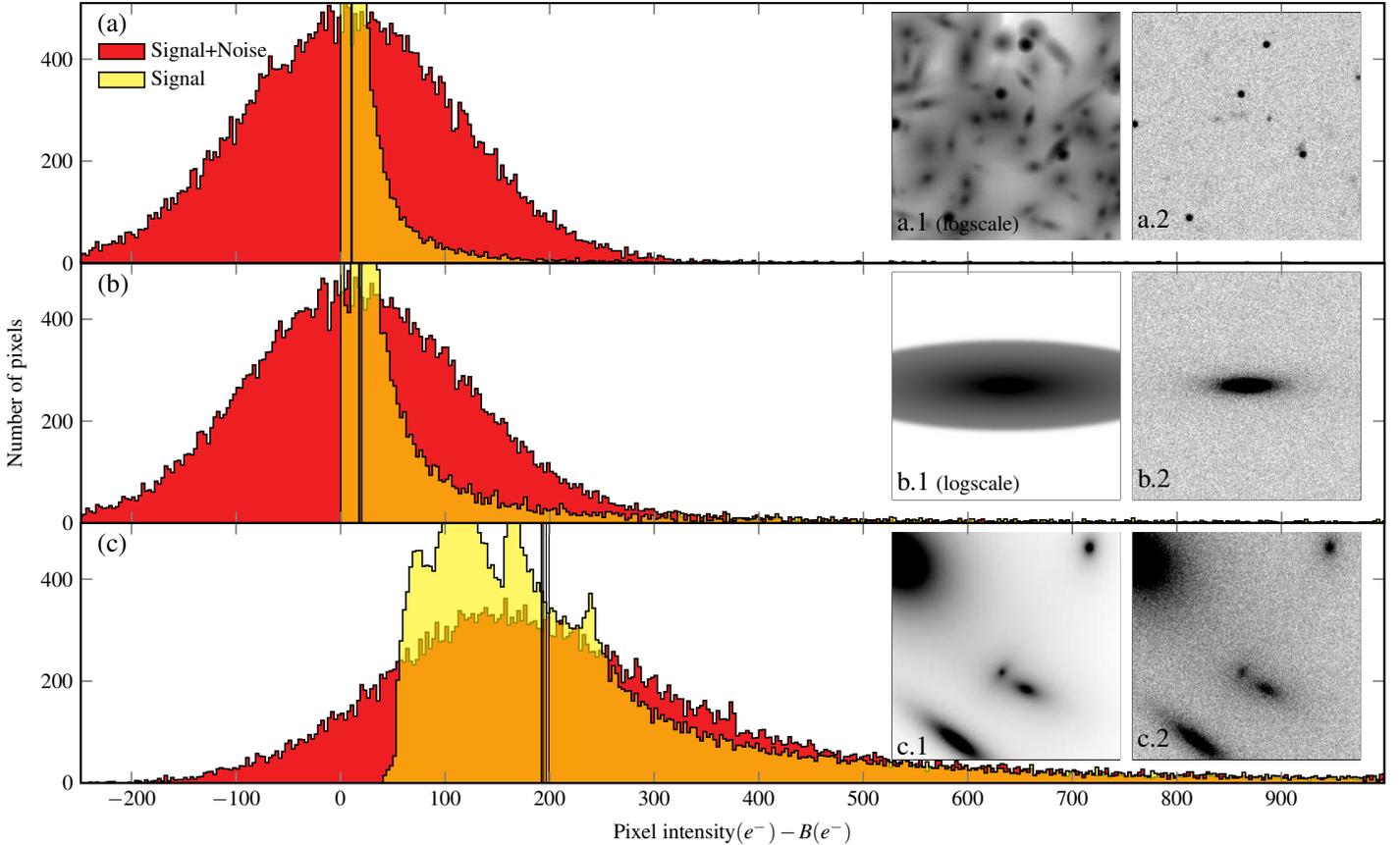}
    \fi
    \caption{\label{dataandnoise} Mock galaxies and their pixel
      distributions with and without noise. The plots have two inset
      images. The left insets show the image of mock galaxies after
      convolution with a PSF prior to adding noise. Logarithmic
      scaling has been used for (a.1) and (b.1) to show the full
      extent of the light profiles. The other insets all use a linear
      scale.  The right insets show the image of the mock galaxies
      with noise added. The histogram starting from or after zero
      (yellow) shows the pixel count distribution of the non-noised
      mock image (left inset). The histogram is vertically truncated
      for a clear comparison with the noised histogram.  The wider
      histogram (red) shows the histogram of the noised image (right
      inset).  The five vertical lines on each histogram show the
      median of the noised image after
      (\issigclipmultip,\issigclipnum) $\sigma$-clipping (see Section
      \ref{sigmaclipping}). Insets have been truncated to be in the
      range of the plot's horizontal axis: white is $\datanoiselow$
      counts and black is $\datanoisehigh$ counts. (a)
      $\datanoisersngals$ random S\'{e}rsic profiles with total
      magnitudes between $\datanoisersgminmag$ to
      $\datanoisersgmaxmag$ and five point sources (stars) of
      $\datanoisersstarmag$ magnitude. The zero-point magnitude for
      all images in this paper is set to $\zeropoint$.  (b) A
      $\onelargemag$ magnitude S\'{e}rsic profile with $n=\onelargen$,
      $r_e=\onelargere$ pixels, and $q=\onelargeq$, (c)
      $\datanoiseslngals$ large and bright S\'{e}rsic profiles with
      $\datanoiseslgmaxmag$ to $\datanoiseslgminmag$ magnitude.  The
      PSF is a circular 2D Moffat profile with FWHM$=\mockfwhm$ pixels
      and $\beta=\moffatbeta$ prior to adding noise. All the galaxies
      and the PSF are truncated at $\truncr$ times the effective
      radius and FWHM respectively. $B$ is the background count, see
      Section \ref{sky}. }
  \end{figure*}

  \textsf{NoiseChisel} is our software implementation of the proposed
  technique in this paper. Throughout this paper, \textsf{NoiseChisel}
  is used to refer to the proposed new approach for detection and
  segmentation. \textsf{NoiseChisel} is distributed as part of the
  {\small GNU} Astronomy
  Utilities\footnote{\url{https://www.gnu.org/software/gnuastro/}}
  which is a new collection of tools for astronomical data analysis
  and manipulation. It is being introduced here as the first
  astronomical software package that fully conforms to the {\small
    GNU} coding standards\footnote{The {\scriptsize GNU} coding
    standards define specific generally accepted requirements for
    software configuration, compilation, and installation as well as
    coding style, command line user interface, and a manual in various
    web-based, print, and command line formats.}  and has been
  evaluated and endorsed by {\small GNU}. The copyright was also
  assigned to the Free Software Foundation to guarantee its
  ``freedom'' and facilitate its use and future development by the
  worldwide astronomical community similar to all their major ``free''
  software projects. It integrates nicely with the {\small GNU}/Linux
  operating system.  It can also be natively compiled and installed on
  most other major operating systems. The existing tools in {\small
    GNU} Astronomy Utilities were used for all the data analysis and
  displaying steps in this paper (making mock profiles, convolving,
  adding simulated noise, producing a catalog, cropping parts of large
  survey tiles, {\small FITS} image conversion to document-friendly
  image formats, creating masks, and calculating image
  statistics). The {\small GNU} Astronomy Utilities is ``free''
  software, as defined in the {\small GNU} general public license
  ({\small GPL}),\footnote{\url{https://www.gnu.org/licenses/gpl.txt}}
  version 3.

\section{Definitions}\label{definitions}

  Prior to a complete explanation of the proposed new algorithm, some
  basic issues need to be addressed first. This new technique is
  non-parametric (with no functional profiles or regression-based
  fittings); therefore, demonstrating the concepts and steps on actual
  images plays a vital role in explaining the technique
  \citep{anscombe}. In Section \ref{aboutimages} a short explanation
  of the postage stamps used in this paper (including the appendices)
  is given.  Fundamentally, this new technique grew out of the
  definition of the Sky value, which is a very important concept in
  any measurement significantly affected by noise and is elaborated in
  Section \ref{sky}.

  \subsection{Displayed Images}\label{aboutimages}

    All the images in the figures are inverted: the darker the pixel,
    the larger its value. Unless otherwise stated, the displayed
    images are $200\times200$ pixels. They are all crops from original
    $\naxisa\times\naxisb$ pixel images. For mock images, the
    remaining area is filled with pure Gaussian noise and for real
    images, it is from the actual survey. When referring to an image
    by histogram, the histogram of the number of pixels at a given
    count level across an image or section of an image is implied,
    similar to those of Figure \ref{dataandnoise}. The histograms only
    use the pixels in the displayed region, not the larger image. The
    vertical axis values are not displayed in all histograms beyond
    Figure \ref{dataandnoise}, because their absolute values are
    irrelevant.

    Unless stated otherwise, the large images were used as input for
    \textsf{SExtractor} and \textsf{NoiseChisel}. The reason for
    setting these sizes was \textsf{SExtractor}'s modeling
    deficiencies in background interpolation and profile modeling on
    the corners and sides of the images (see Appendices
    \ref{SEbackinterp} and \ref{rkron}). Using larger images ensures
    that for both programs there is an abundant area of background
    pixels regardless of the central object shape and size. In the
    purity analysis of Section \ref{A_purity} the pure noise regions
    of the mock images are used to identify false detections. Unless
    otherwise stated, all the $200\times200$ pixel cutouts are from
    the center of the larger image.

    The real images used in this paper for real demonstrations are
    either cutouts from the {\small COSMOS} survey \citep{cosmos},
    \emph{Hubble Space Telescope} (\emph{\small HST}) {\small F814W}
    images \citep[see][]{cosmoshstiband, massy10}, or one full {\small
      CCD} image taken by Subaru Telescope's SuprimeCam
    \citep{suprimecam}.

  \subsection{Background, Sky, and Noise}\label{sky}
    Throughout this paper, pixel units are assumed to be in
    photon-counts or simply `count' ($e^-$). All the methods can be
    readily applied to count sec$^{-1}$ units. For example, all the
    displayed \emph{\small HST} images were in units of counts
    sec$^{-1}$.  Let us assume that all instrument defects---bias,
    dark and flat---have been corrected and the total count of a
    detected object, $O$, is desired. The sources of $e^-$ in a pixel
    $(i,j)$ of the image can be written as follows: (1) contribution
    from the target object, $O_{ij}$, (2) contribution from other
    detected objects, $D_{ij}$. (3) undetected objects or the fainter
    undetected regions of bright objects, $U_{ij}$, (4) cosmic rays,
    $C_{ij}$, and (5) the background count, which is defined to be the
    count if none of the others exists on that pixel, $B_{ij}$.  The
    total count in the pixel, $T_{ij}$, can thus be written as:

    $$T_{ij}=B_{ij}+D_{ij}+U_{ij}+C_{ij}+O_{ij}.$$

    Figure \ref{dataandnoise} shows the pixel count distribution before
    and after adding noise in three sample mock images. By definition,
    $D_{ij}$ is detected and it can be assumed that it is correctly
    subtracted, so that $D_{ij}$ can be set to zero. There are also
    methods to detect and remove cosmic rays \citep[for example, the
      method of][]{vDcosmicrays}, enabling us to set $C_{ij}=0$. Note
    that in practice, $D_{ij}$ and $U_{ij}$ are correlated because
    both directly depend on the detection algorithm and its input
    parameters. Also note that no detection or cosmic ray removal
    algorithm is perfect. With these limitations in mind, the observed
    Sky value for this pixel ($S_{ij}$) can be defined as

    \begin{equation}
      \label{skydefinition}
      S_{ij}=B_{ij}+U_{ij}.
    \end{equation}

    \noindent
    Therefore, as the detection process (algorithm and input
    parameters) becomes more accurate, or $U_{ij}\to0$, the Sky value
    will tend toward the background value or $S_{ij}\to B_{ij}$.
    Thus, while $B_{ij}$ is an inherent property of the data (pixel in
    an image), $S_{ij}$ depends on the detection process. Over a group
    of pixels, for example, in an image or part of an image, this
    equation translates to the average of undetected pixels.  With
    this definition of Sky, the object count in the data can be
    calculated with

    \begin{equation}
      T_{ij}=S_{ij}+O_{ij} \quad\Rightarrow\quad
      O_{ij}=T_{ij}-S_{ij}.
    \end{equation}

    Hence, the more accurately $S_{ij}$ is measured, the more
    accurately the count of the target object can be
    calculated. Similarly, any under- (over-)estimation in the Sky
    will directly translate into an over- (under-)estimation of the
    measured object's count.  In the fainter outskirts of an object a
    very small fraction of the photo electrons in the pixels actually
    belong to objects.  In Figure \ref{dataandnoise}(b.1) for example,
    only $\onelargefracabove\%$ of the pixels belonging to the central
    object have a value larger than $\backthresh$ counts while the
    background is $\backcount\pm\backstd$ counts.  Therefore even a
    small overestimation of the Sky value will result in the loss of a
    very large portion of this mock galaxy.

\begin{figure}
    \centering
    \ifdefined\makeeps

  \centering
  \begin{tikzpicture}

    \tikzstyle{every node}=[font=\small]

    \draw [rounded corners=5mm, line width=1mm, black!20]
    (-0.47\linewidth,0.1cm) rectangle (0.47\linewidth, -3.1cm);

    \path (-0.47\linewidth+0.05cm,-2.95cm)
          node [rotate=90, anchor=north west]
               {\textcolor{black!40}{\Large INITIAL}};
    \path (-0.47\linewidth+0.5cm,-3.2cm)
          node [rotate=90, anchor=north west]
               {\textcolor{black!40}{\Large DETECTIONS}};

    \draw [rounded corners=5mm, line width=1mm, black!20]
    (-0.47\linewidth,-3.8cm) rectangle (0.47\linewidth, -17.9cm);

    \path (-0.47\linewidth+0.05cm,-17.8cm)
          node [rotate=90, anchor=north west]
          {\textcolor{black!40}
            {\Large IDENTIFY AND REMOVE FALSE DETECTIONS}};

    \draw [rounded corners=2mm, black!40, fill=black!5]
    (0.2cm,-5.5cm) rectangle (3.5cm, -11.4cm);
    \path (0.1cm, -5.6cm) node[rotate=-90, anchor=south west, black!60]
          {SKY REGIONS ($R_s$)};

    \draw [rounded corners=2mm, black!40, fill=black!5]
    (-0.2cm,-5.5cm) rectangle (-3.5cm, -11.4cm);
    \path (-0.14cm, -5.6cm) node[rotate=-90, anchor=north west, black!60]
          {DETECTED REGIONS ($R_d$)};

    \graph [grow down=0.8cm, branch left=4cm, simple]
           {
             Convolve                                 [mybox]
             -> Find and apply quantile threshold on CI [mybox]
             -> Erode                                 [mybox]
             -> Open                                  [mybox]
             -- p0                                    [coordinate]
             -> "$I\leftarrow$ Number of CC.
                 $A\equiv\{A_0, ..., A_{I-1}\}$"       [mybox]
             -> "Average of undetected: \textbf{Sky} (initial)" [mybox]
             -- p1                                    [coordinate]
             -> [hv path]
             { [nodes={xshift=2cm, yshift=0.5cm}]
               {
                 / Sky threshold                      [mybox]
                 -> / Fill holes                      [mybox]
                 -> / Open                            [mybox]
                 -> / "$K\leftarrow$ Number of
                       CC"                            [mybox]
                 -> / "$C\equiv\{C_0, ..., C_{K-1}\}$" [mybox]
                 -> / "$S\equiv\{s_0, ..., s_{K-1}\}$" [mybox]
                 -> / "$s_q \leftarrow$ Quantile
                       of $S$"                        [mybox];
               },
               {
                 Sky threshold                        [mybox]
                 -> Fill holes                        [mybox]
                 -> / Open                            [mybox]
                 -> / "$J\leftarrow$ Number of CC"    [mybox]
                 -> / "$B\equiv\{B_0, ..., B_{J-1}\}$" [mybox]
                 -> / "$D\equiv\{s_0, ..., s_{J-1}\}$" [mybox]
               }
             }
             -- [vh path]
             p2                                      [coordinate]
             -> / $i\leftarrow0$                     [mybox]
             -> n1/ "$T\equiv\{j\in\{0, ..., J-1\}$ $|$
             $(D_j>s_q)$
             $\land$
             $(B_j\cap{}A_i\neq\emptyset)\}$"         [mybox]
           };

    \node [mydiamond, below of=n1] (d1){$T\neq\emptyset$};
    \path (d1.west) ++(3.7cm,-0.2cm) node [mybox, name=removeA]
          {$A_i$ is a false\\detection, remove it.};
    \path (d1.south) ++(0cm,-0.7cm) node [mybox, name=inci]
          {$i\leftarrow i+1$};
    \node [mydiamond, below of=inci] (d2){$i=I$};
    \path (d2.south) ++(0cm,-0.7cm) node [mybox, name=dilate]
          {Dilate};
    \node [mybox, below of=dilate, name=final]
          {Average of undetected: \textbf{Sky}};

    \draw [->] (n1) -- (d1);
    \draw [->, rounded corners=0.8mm] (d1.east) ++(-0.02cm,0cm) -| (removeA.north);
    \draw [->] (d1.south) ++(0cm,0.01cm) -- (inci);
    \draw [->, rounded corners=0.8mm] (removeA.south) |- (inci.east);
    \draw [->] (inci) -- (d2);
    \draw [->, rounded corners=0.8mm] (d2.east) ++(-0.02cm,0cm) -- ++(2.97cm,0cm)
                |- (n1.east) ++(0.2cm,0cm) -- (n1.east);
    \draw [->] (d2.south) ++(0cm,0.01cm) -- (dilate);
    \draw [->] (dilate) -- (final);

    \draw (d1.east) ++(-0.3cm,0cm) node[anchor=south west]
          {\footnotesize False};
    \draw (d1.south) ++(0cm,+0.12cm) node[anchor=north west]
          {\footnotesize True};
    \draw (d2.east) ++(-0.3cm,0cm) node[anchor=south west]
          {\footnotesize False};
    \draw (d2.south) ++(0cm,+0.12cm) node[anchor=north west]
          {\footnotesize True};

  \end{tikzpicture}
    \else
    \includegraphics{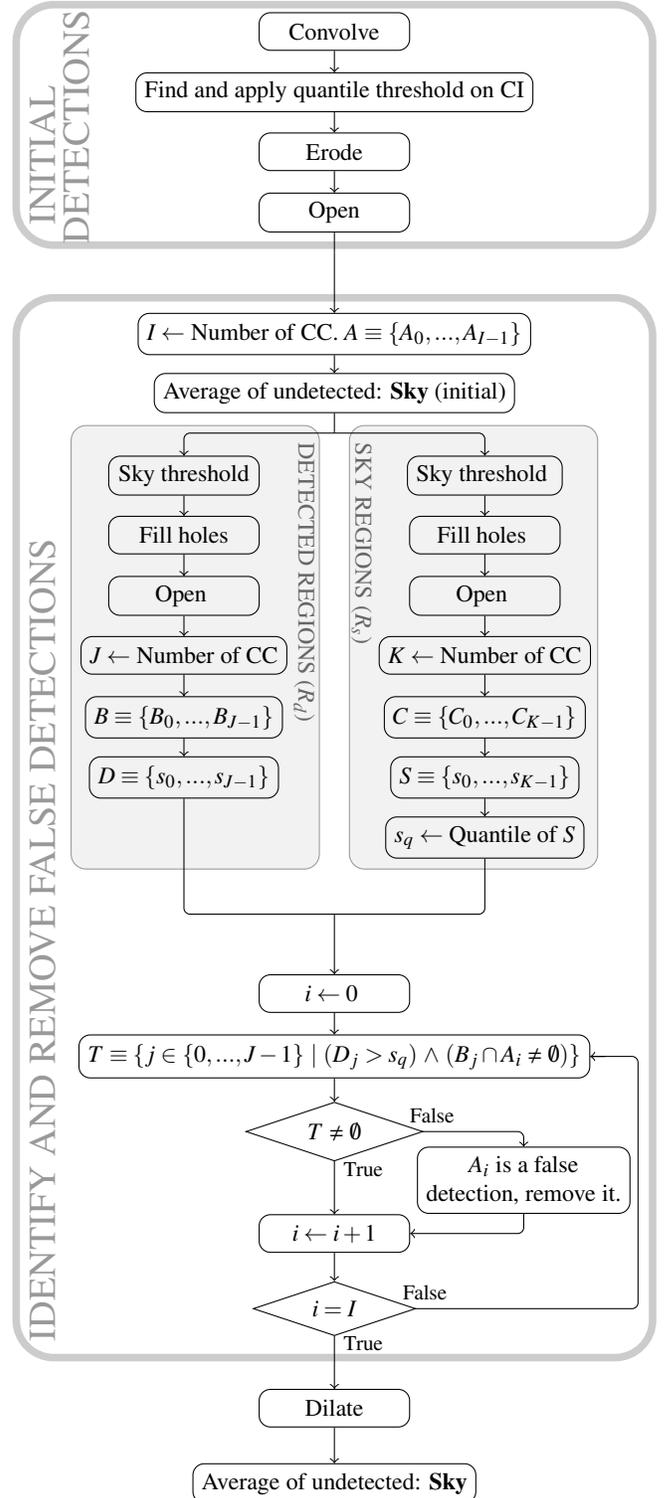}
    \fi
    \caption{\label{flowchart}Flowchart of the complete detection
      algorithm. The steps of the top and bottom boxes can be seen in
      the example images of figures \ref{det} and \ref{dettf}
      respectively. CI stands for convolved image, CC for connected
      components, $s$ for signal-to-noise ratio, $\land$ for logical
      `and', $\leftarrow$ for assignment, $=$ for comparison, and
      $\emptyset$ for the empty set. Each CC is defined as a set of
      image pixels. $I$ is the number of CCs in column 1 of Figure
      \ref{dettf}. $J$ and $K$ represent the number of CCs in the
      bottom and top of column 4 in Figure \ref{dettf},
      respectively. $A_i$, $B_i$, and $C_i$ are sets of pixels
      belonging to the $i$th connected component in their respective
      images. $A$, $B$, and $C$ are families of sets (connected
      components). $D_i$ and $S_i$ are the total signal-to-noise
      ratios of $B_i$, and $C_i$ respectively.}
  \end{figure}

    Based on the definition above, the Sky value is only correctly
    found when all the detected objects ($D_{ij}$ and $C_{ij}$) have
    been removed from the data. However, as shown in Section
    \ref{SEdetection}, existing methods define their detection
    threshold using the Sky value (see Section \ref{SExthresh} in
    particular). Therefore they cannot use this definition in finding
    the Sky. They have to apply the methods explained in Appendix
    \ref{existingsky} to approximate it. The foundation of the
    technique presented here is that the detection threshold and
    initial detection process is defined independent of the Sky value
    see Figure \ref{flowchart} and Section \ref{noisechisel}.

  \begin{figure*}[t]
    \ifdefined\makeeps
    \input{./tex/det}
    \else
    \includegraphics[width=\linewidth]{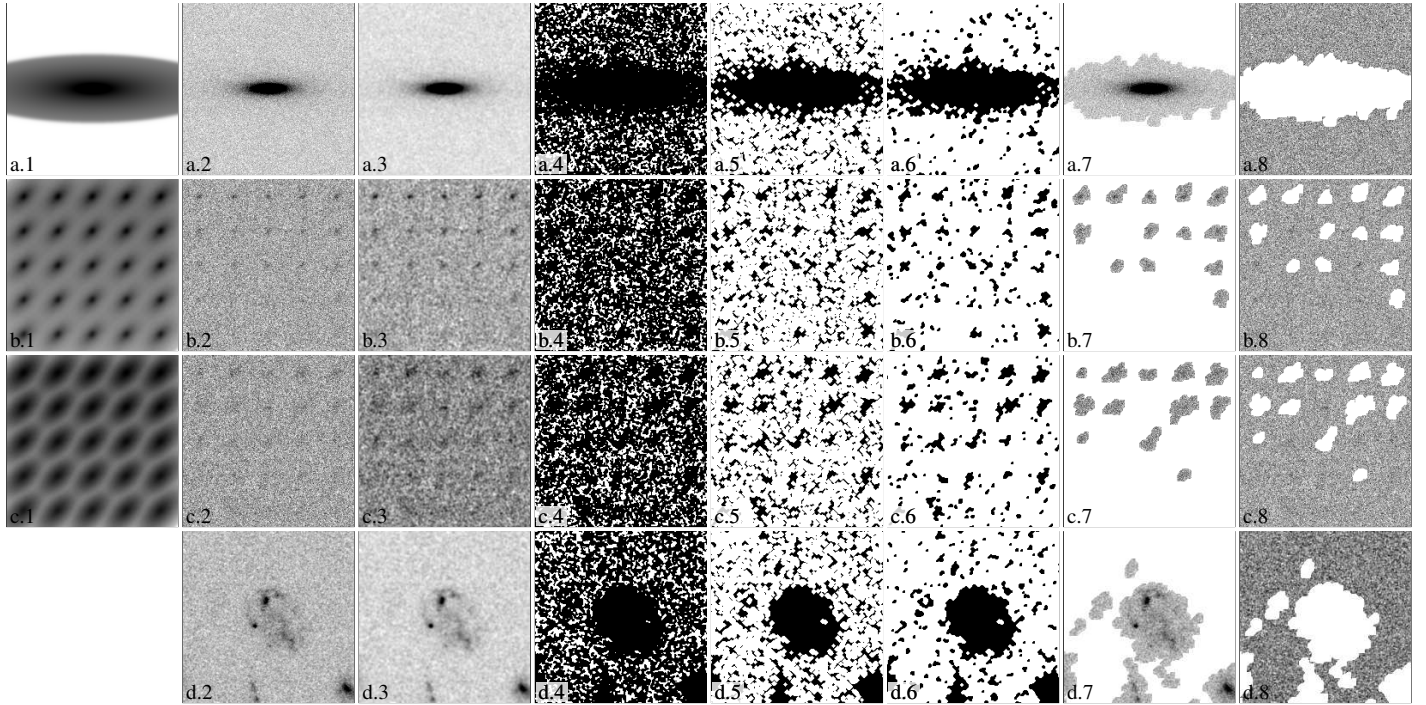}
    \fi
    \caption{\label{det} Demonstration of major detection steps in
      \textsf{NoiseChisel}. The rows are (a) the same mock profile as
      in Figure \ref{dataandnoise}(b). (b) 25 mock profiles with the
      same parameters with $n=\sennc$ (see Section \ref{mydetection}),
      (c) similar to (b), but with $n=\sennb$, (d) a real star-forming
      $z\sim1.5$ galaxy imaged in \emph{\small HST}/{\small ACS}
      {\small F814W}. The columns represent (1) mock profiles before
      the addition of noise and blank for the real image (it is blank
      since the signal without noise is not known in a real image; All
      mock images in this column are plotted in the logarithmic
      scale), (2) noisy image, (3) convolved with a 2D Gaussian kernel
      of FWHM$=2$ pixels, (4) threshold applied on the $\qthresh$
      quantile of the convolved image, (5) $\erodengb$ connected
      erosions applied $\erode$ times, (6) $\openingngb$ connected
      opening applied $\opening$ time, (7) estimated false detections
      removed (see Figure \ref{dettf}; true detections are dilated two
      times and undetected pixels masked; pixel values are taken from
      the noisy image in column 2), and (8) the inverse of column
      7. Columns 3 -- 6 (inclusive) correspond to the top four steps
      of the flowchart in Figure \ref{flowchart}. }
  \end{figure*}

    In the mock images in this paper each pixel value ($T_{ij}$) is
    composed of the object contribution, $O_{ij}$, and the background
    $B$: $T_{ij}=B+O_{ij}$; see Figure \ref{dataandnoise}. The noise
    is added to each pixel by replacing its value with a random value
    taken from a Gaussian distribution with a mean of $T_{ij}$ and
    $\sigma_{ij}=\sqrt{T_{ij}}$. In this paper, $B=\backcount{}e^-$.
    The magnitude of each object is defined as $-2.5\log{(F)}$, were
    $F$ (sum of pixel values) is in $e^-$. In other words, the
    zeropoint magnitude is \zeropoint{} since this paper is not
    limited to any particular instrument. So the background magnitude
    is $\backmag$. The total magnitude is calculated from the sum of
    all the object pixels in the image (not integrated to
    infinity). All profiles are truncated at $\truncr{}r_e$.

\section{NoiseChisel}\label{noisechisel}

  The general flowchart of the proposed technique can be seen in
  Figure \ref{flowchart}. In the proposed algorithm, a Sky-independent
  approach is used to arrive at an \emph{initial}
  detection. Individual initial detections are subsequently classified
  as true or false based on their signal-to-noise ratio ({\small S/N})
  using the surrounding undetected regions as a basis. As seen in
  Figure \ref{flowchart}, this second classification step requires an
  initial measurement of the Sky value as the average of the initially
  undetected pixels and is fully elaborated in Section
  \ref{mydetection}.  The substructure in each detection is then found
  and classified through defining ``clumps'' and ``objects'' which is
  explained in detail in Section \ref{mysegmentation}.

  \begin{figure*}[t]
    \centering
    \ifdefined\makeeps
    \input{./tex/convolution}
    \else
    \includegraphics[width=\linewidth]{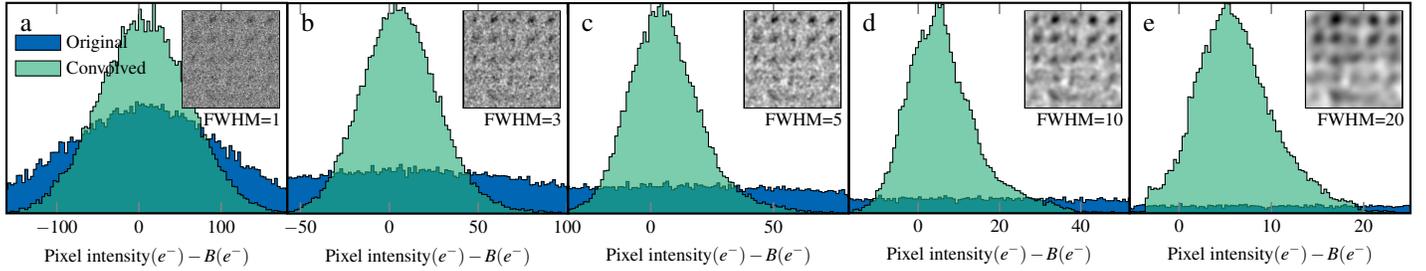}
    \fi
    \caption{\label{convolution} Effect of convolution with wider
      kernels on Figure \ref{det}(c.2).  The green histogram shows the
      count distribution of the convolved image.  The blue histogram
      shows the distribution of the original image pixel count values
      within the green histogram's range.  The convolution kernel used
      is a Moffat function with $\beta=\moffatbeta$ (same as the PSF
      used for the mock images), while the FWHM is shown for each in
      units of pixels. Each kernel is a square of side $\truncr\times$
      FWHM pixels.  Note how the convolved histogram becomes more
      skewed as a wider kernel is used for convolution.  The range of
      the horizontal axis significantly decreases with increasing
      kernel width. $B$ is the background count, see Section
      \ref{sky}.}
  \end{figure*}

  The input parameters\footnote{Not to be confused with the
    non-parametric basis of this algorithm. By non-parametric it is
    implied that there are no functions and fittings, not that there
    are no parameters.}  for \textsf{NoiseChisel} are specified in the
  text with \texttt{-{}-parametername=value}.\footnote{This is similar
    to the long format of how the options should be called on the
    command line in \textsf{NoiseChisel}.} The values used for the
  parameters of \textsf{NoiseChisel} reported here are chosen such
  that once applied to the mock and real image tests that are shown in
  this paper, accurate results are achieved. The full parameter list
  can be seen in Appendix \ref{NCconfig}. The criteria are to have
  fewer false detections in pure noise while detecting the faintest
  pixels of the mock profiles (see Section \ref{analysis}). We do not
  claim that these are ``the best'' set of parameters for any generic
  data set.

  \subsection{Detection and Sky}\label{mydetection}
    The basic idea behind this detection algorithm is that if there is
    a true signal buried in the noise over a certain \emph{contiguous}
    area, it will systematically (contiguously) augment that region
    (keeping the noise fluctuations). The steps explained here are
    specifically designed to best separate these connected, augmented
    regions deep in the noise. The initial detection steps in
    \textsf{NoiseChisel} are schematically shown in the top group of
    steps of the flowchart in Figure \ref{flowchart} and their
    application to sample images is demonstrated in Figure \ref{det}.
    First, a very small {\small FWHM} kernel is convolved with the
    image (Section \ref{conv}, column 3 in Figure \ref{det}). A very
    low threshold is then applied to the smoothed image (Section
    \ref{thresh}, column 4). The regions below the threshold are
    expanded by eroding those that are above it (Section
    \ref{erosion}, column 6). Since the objects are not completely
    separated yet, ``opening'' is applied (Section \ref{opening},
    column 7). False detections are estimated and removed (Section
    \ref{removenoise}) and the fainter regions of true detections,
    which were carved off during erosion are returned in a process of
    dilation (Section \ref{dilation}).

    Three of the examples in Figure \ref{det} are mock and one is a
    real galaxy.  Figure \ref{det}(a) is the same mock profile as in
    Figure \ref{dataandnoise}(b).  Figures \ref{det}(b) and (c)
    contain 25 mock 2D S\'{e}rsic profiles placed with equal spacing
    on a grid across the image. Other than their magnitude and
    position, all of the parameters are equal with a position angle of
    $\sentheta^\circ$, an axis ratio of $q=\senq$, and an effective
    radius $r_e=\senre$ pixels. The S\'{e}rsic index ($n$) of all the
    galaxies in each image is also the same. The bottom left profile
    is the faintest with $\senfaintmag$ magnitude. The next profiles
    are successively $\senmagdiff$ magnitudes brighter towards the top
    right. Such extremely faint profiles were intentionally chosen to
    push \textsf{NoiseChisel} (and \textsf{SExtractor} in Section
    \ref{SEminarea}) to the limits. Figure \ref{det}(d) shows the same
    steps with exactly the same parameters applied to a real
    $z\sim1.5$ star-forming galaxy in the {\small COSMOS} field imaged
    with the \emph{\small HST} Advanced Camera for Surveys ({\small
      ACS}) using the {\small F814W} filter.

    \subsubsection{Convolution}\label{conv}

      Convolution is used to maximize the ratio of an object's peak
      signal to the local noise level \citep{bijaoui70,
        irwin85}. Convolution smooths the image by removing high
      spatial frequencies. This means that after convolution,
      ``nearby'' pixels with large differences in value, which are
      commonly due to noise, will receive an intermediate
      value. Therefore, by smoothing the noise, the fraction of
      signal-to-noise is increased. Without convolution, the large
      spatial frequencies due to noise will significantly limit the
      ability to detect a contiguous region in noise.  Hence, in order
      to maximize the detection ability, convolution is necessary. The
      definition of a \emph{nearby} pixel and the functional form of
      how to define the intermediate value are defined with the kernel
      input into convolution.

      Figure \ref{convolution} shows the noised image of 25 very faint
      $n=\sennb$ profiles, convolved with increasingly wider (larger
      {\small FWHM}) kernels applied.  As the kernel used for
      convolution widens, the image becomes more and more blurred, its
      histogram becomes more skewed, and the range of pixel values
      (dynamic range) in the image decreases with more pixels having
      similar values. Comparing the blue and green histograms in
      Figure \ref{convolution} shows this effect. The skewness induced
      into the histogram is another manifestation of the increase in
      the {\small S/N} of connected objects. Due to the skewness,
      separation of the brightest sections of the objects from the
      noise is facilitated.  A wider convolution kernel has important
      consequences. (1) The shapes of the object tend to the shape of
      the kernel used to convolve the image.  (2) The dynamic range
      significantly decreases.  Therefore, separating the fainter
      parts of the brighter objects from noise becomes much harder.
      (3) Compact and faint objects, that is, those that are not wide
      relative to the convolution kernel, will be lost.

      The effects above are particularly harmful in ground-based
      images, where the {\small FWHM} is generally very large.  The
      decrease in the dynamic range is not a considerable issue in the
      signal-based technique because only the brightest pixels of the
      objects are sought and the fainter parts are modeled (see
      Section \ref{rkron}). As seen in the sequence of steps in Figure
      \ref{det}, the proposed detection technique is exactly the
      opposite and aims to impose negligible constraints and models on
      the detected signal. The dynamic range is hence extremely
      important to successfully find the faint but valuable object
      pixels. Therefore a sharp convolution kernel is used.

      Unlike contiguous (multi-pixel) signal, noise is ideally
      independent from pixel to pixel,\footnote{In processed images we
        have correlated noise which is addressed in Section
        \ref{removenoise}.} and thus is directly connected with
      sampling or pixel size. Inverting the Nyquist sampling theorem,
      the sharpest convolution kernel can be considered an {\small
        FWHM}$=2$ pixel kernel and a Gaussian is used for its
      functional form.  In all the examples in this paper, mock or
      real, ground-based or space-based, a Gaussian profile with
      {\small FWHM}$=2$ pixels truncated at five times the {\small
        FWHM} is used, so that the kernel is $11\times11$
      pixels\footnote{The convolution kernel is created with
        \textsf{MakeProfiles} which is part of the GNU Astronomy
        Utilities. The non-zero pixels in the square kernel have a
        circular shape.}. The fact that the same convolution kernel
      gives very accurate results on any type of image is one example
      of the objectivity of \textsf{NoiseChisel}'s output.

      Throughout \textsf{NoiseChisel}, the convolved image is only
      used for relative pixel values, for example in thresholding
      (Section \ref{thresh}) or oversegmentation (Section
      \ref{clumpseg}). The input image is used for absolute pixel
      values, for example, the Sky value, the magnitude of an object,
      or its {\small S/N}. This distinction is not commonly practiced
      in most existing techniques. For example, in
      \textsf{SExtractor}, the threshold value is found from the input
      image and applied to the convolved image; see Section
      \ref{SExthresh}.

      \begin{figure*}[t]
        \centering
        \ifdefined\makeeps
        \newcommand{\inputdir}{\figtexdir/NCthresh}

    \makeatletter \newcommand{\pgfplotsdrawaxis}{\pgfplots@draw@axis} \makeatother

    \pgfplotsset{axis line on top/.style={
      axis line style=transparent,
      ticklabel style=transparent,
      tick style=transparent,
      axis on top=false,
      after end axis/.append code={
        \pgfplotsset{axis line style=opaque,
          ticklabel style=opaque,
          tick style=opaque,
          grid=none}
        \pgfplotsdrawaxis}
      }
    }

    \begin{tikzpicture}
        \scriptsize
        \begin{groupplot}[group style={group size=4 by 1,
                     horizontal sep=0pt},
                     xmode=normal,ymode=normal,
                     enlargelimits=false,
                     no markers, width=0.25\linewidth,
                     height=0.15\linewidth,
                     xlabel={Pixel intensity$(e^-) - B(e^-)$},
                     scale only axis, axis line on top,
                     area legend, legend cell align=left,
                     legend style={draw=none, fill=none,
                                   at={(0.895\linewidth,0.085\linewidth)},
                                   anchor=south west, font=\scriptsize}]

          \nextgroupplot[yticklabels={}, xtick={0,100,200},
                         ytickmin=-2, ytickmax=-1]
            \addplot [const plot, fill=myblue]
                     table [x expr=\thisrowno{0}-10000, y index=1]
                           {\inputdir/onelarge_hist.txt}
            \closedcycle;
            \addplot [const plot, fill=mygreen, fill opacity=0.6]
                     table [x expr=\thisrowno{0}-10000, y index=1]
                           {\inputdir/onelarge_chist.txt}
            \closedcycle;

            \addplot [thick, color=myblack]
                     table [x expr=\thisrowno{0}-10000, y index=1]
                     {\inputdir/onelarge_ccfp.txt};

            \draw [thin]
                  ({axis cs:-12.41,0}|-{rel axis cs:0,1})--
                  ({axis cs:-12.41,0}|-{rel axis cs:0,0});
            \draw [thin, dotted]
                  ({axis cs:100,0}|-{rel axis cs:0,1})--
                  ({axis cs:100,0}|-{rel axis cs:0,0});
          \legend{Original, Convolved}

          \nextgroupplot[yticklabels={}, ytickmin=-2, ytickmax=-1]
            \addplot [const plot, fill=myblue]
                     table [x expr=\thisrowno{0}-10000, y index=1]
                           {\inputdir/sensitivity3_hist.txt}
            \closedcycle;
            \addplot [const plot, fill=mygreen, fill opacity=0.6]
                     table [x expr=\thisrowno{0}-10000, y index=1]
                           {\inputdir/sensitivity3_chist.txt}
            \closedcycle;

            \addplot [thick, color=myblack]
                     table [x expr=\thisrowno{0}-10000, y index=1]
                     {\inputdir/sensitivity3_ccfp.txt};

            \draw [thin]
                  ({axis cs:-13.47,0}|-{rel axis cs:0,1})--
                  ({axis cs:-13.47,0}|-{rel axis cs:0,0});
            \draw [thin, dotted]
                  ({axis cs:100,0}|-{rel axis cs:0,1})--
                  ({axis cs:100,0}|-{rel axis cs:0,0});

          \nextgroupplot[yticklabels={}, ytickmin=-2, ytickmax=-1]
            \addplot [const plot, fill=myblue]
                     table [x expr=\thisrowno{0}-10000, y index=1]
                           {\inputdir/sensitivity2_hist.txt}
            \closedcycle;
            \addplot [const plot, fill=mygreen, fill opacity=0.6]
                     table [x expr=\thisrowno{0}-10000, y index=1]
                           {\inputdir/sensitivity2_chist.txt}
            \closedcycle;

            \addplot [thick, color=myblack]
                     table [x expr=\thisrowno{0}-10000, y index=1]
                     {\inputdir/sensitivity2_ccfp.txt};

            \draw [thin]
                  ({axis cs:-11.81,0}|-{rel axis cs:0,1})--
                  ({axis cs:-11.81,0}|-{rel axis cs:0,0});
            \draw [thin, dotted]
                  ({axis cs:100,0}|-{rel axis cs:0,1})--
                  ({axis cs:100,0}|-{rel axis cs:0,0});

          \nextgroupplot[yticklabels={}, xlabel={Pixel intensity
              ($e^- s^{-1}$)$\times10^5s$}, ytickmin=-2, ytickmax=-1]
            \addplot [const plot, fill=myblue]
                     table [x expr=\thisrowno{0}*100000, y index=1]
                           {\inputdir/4_hist.txt}
            \closedcycle;
            \addplot [const plot, fill=mygreen, fill opacity=0.6]
                     table [x expr=\thisrowno{0}*100000, y index=1]
                     {\inputdir/4_chist.txt}
            \closedcycle;

            \addplot [thick, color=myblack]
                     table [x expr=\thisrowno{0}*100000, y index=1]
                     {\inputdir/4_ccfp.txt};

            \draw [thin]
                  ({axis cs:-63.23,0}|-{rel axis cs:0,1})--
                  ({axis cs:-63.23,0}|-{rel axis cs:0,0});

        \end{groupplot}

        \node[anchor=south west] at (0.005\linewidth,0.125\linewidth)
                 {\normalsize a};
        \node[anchor=south west] at (0.255\linewidth,0.125\linewidth)
                 {\normalsize b};
        \node[anchor=south west] at (0.505\linewidth,0.125\linewidth)
                 {\normalsize c};
        \node[anchor=south west] at (0.755\linewidth,0.125\linewidth)
                 {\normalsize d};

    \end{tikzpicture}
        \else
        \includegraphics[width=\linewidth]{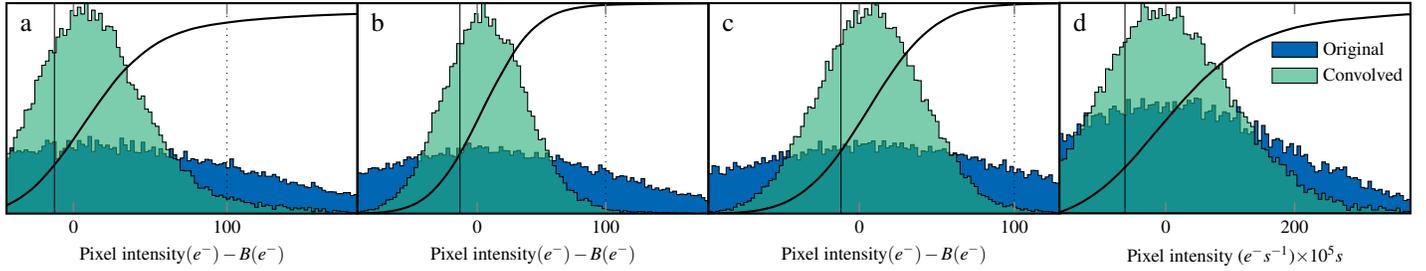}
        \fi
        \caption{\label{NCthresh} Thresholds used in the demonstration
          examples of Figure \ref{det} with the same labels. The blue
          histograms belong to the actual image (column 2 in Figure
          \ref{det}). Only the range that overlaps with the green
          histogram is plotted here. The green histogram is the
          histogram of the convolved image (column 3 of Figure
          \ref{det}). The thick curve shows the cumulative frequency
          of the convolved image. The thin solid vertical lines show
          the $\qthresh$ quantile of the convolved image which is used
          to determine the threshold for it; this value is used to
          make column 4 of Figure \ref{det}. The dotted lines in the
          first three examples (mock galaxies) show the ideal
          $\sigma_{sky}$ of the input image (blue histogram), which is
          $\backstd{}e^-$. $B$ is the background count, see Section
          \ref{sky}.}
      \end{figure*}

    \subsubsection{Thresholding}\label{thresh}
      Pixels with values far below the inherent background or measured
      Sky values (see Section \ref{sky}) can harbor valuable
      information (see Figure \ref{dataandnoise}). After accurate Sky
      subtraction, a real object will always have a positive total
      count. However, the standard deviation of the noise is always
      much larger than the smallest detectable positive count value of
      the real objects. For example, assume that the Gaussian noise
      standard deviation is $30e^-$. After accurate Sky subtraction,
      $\sim37\%$ of\footnote{The cumulative distribution function of a
        normal distribution with mean of 10$e^-$ and standard
        deviation of 30$e^-$ calculated at 0$e^-$.} the pixels that
      harbor $10e^-$ of real counts will have negative recorded pixel
      value. Once a large number of such pixels are found and the Sky
      value is accurately subtracted, the random effects of noise will
      be decreased and useful information can be extracted from such
      pixels.

      Therefore it is wrong to assume that the Sky value is the lowest
      possible threshold. To accurately detect the faintest pixels of
      an object, the threshold has to be less than the Sky value. This
      is very important in accurate measurement of the Sky value as
      the average of undetected pixels.

      In \textsf{NoiseChisel} the threshold is defined using the
      cumulative distribution of the pixels in the convolved image. To
      ensure that the threshold is below the Sky value, the quantile
      has to be less than $0.5$, or the median. For the examples in
      this paper, the quantile threshold is set to \texttt{\small
        -{}-qthresh=\qthresh}, which means that $\invqthresh\%$ of the
      pixels in the convolved image will have a value above it. The
      results of applying a threshold to the convolved image can be
      seen in Figure \ref{det}, column 4, where black pixels are the
      ones that had a count above this threshold. The thresholds can
      be seen relative to the actual and convolved image histograms in
      Figure \ref{NCthresh}. Notice that the threshold is safely below
      the background value which is $0e^-$ for the mock images. In
      Section \ref{lithresh} the complete procedure for finding the
      threshold on a large image is elaborated.

    \subsubsection{Erosion}\label{erosion}
      Applying a threshold to the image (Section \ref{thresh}) results
      in a binary image with pixels either having a count above the
      threshold or below it. The former are known as foreground pixels
      which are black in Figure \ref{det}, column 4. The latter,
      called background pixels, are displayed in white in the same
      figure. Because of this extremely low threshold, it is clear
      from column 4 of Figure \ref{det} that the regions harboring
      true signal in them cover a 2D contiguous (non-porous) region of
      connected pixels, while regions where no data is present form a
      porous structure in the foreground. The figure shows that when
      signal is present, even if it is very close to the background
      value, it will augment the region that it covers. The
      augmentation can be seen through smaller and fewer white holes
      in the black regions that harbor data. The first step in
      \textsf{NoiseChisel} after applying the threshold is thus to
      exploit this porous structure and expand the holes through
      erosion of the foreground to separate the augmented regions.

      Erosion is an operation in morphological image processing
      \citep{starckmurtagh, gonzalezwoods}. By eroding the foreground,
      it is implied that all foreground pixels that \emph{neighbor} a
      background pixel are carved off\footnote{Hence the name
        \textsf{NoiseChisel}: a tool for carving off noise, similar to
        what wood chisels or stone chisels do on their respective
        materials.}  the foreground and changed to a background
      pixel. The neighborhood of, or connected pixels to, a pixel is
      defined based on a \emph{structuring element} and generally has
      two types: 4 and 8 connectivity (see Figures \ref{StructElem}(a)
      and (b)).  Eroding based on 8 connectivity will remove more
      pixels. Therefore, in this step, \textsf{NoiseChisel} uses a 4
      connected structuring element to erode the foreground. The
      statistical properties of this and other operations in
      mathematical morphology are reviewed in \citet{dougherty92}.

      Eroding the foreground expands the holes (white regions in
      Figure \ref{det} column 4) and connects them to each other. The
      best number of erosions (\texttt{\small -{}-erode}) and the type
      of connectivity (\texttt{\small -{}-erodengb}) applied to an
      image is intertwined with \texttt{\small -{}-qthresh}. The
      results of respectively setting them to \texttt{\small \erode}
      and \texttt{\small \erodengb} can be seen in Figure \ref{det},
      column 5. If a set of foreground pixels had the shape of the
      non-white pixels of Figure \ref{StructElem}(c), after applying
      this erosion the pixels with the two brighter shades of gray
      would be carved off and only the central 9 pixels would remain.

    \subsubsection{Opening}\label{opening}
      Dilation is the inverse of erosion in morphological image
      processing. When dilating the foreground, if a background pixel
      touches a foreground pixel, it will be changed to the
      foreground. Applying these two steps in order, namely, first
      eroding then dilating the foreground, is known as
      \emph{opening}.

      Opening is very useful for separating regions of the foreground
      that are only connected through a thin thread of pixels. It also
      has the benefit of removing extremely small objects
      \citep[see][for such an application in astronomy]{vollmer13}. If
      the image is first eroded $n$ times and then dilated the same
      $n$ times, then $n$ is referred to as the ``depth'' of the
      opening. Note that the erosion discussed here is part of opening
      and separate from the one in Section \ref{erosion}. The depth
      can be set in \textsf{NoiseChisel} through the parameter
      \texttt{\small -{}-opening} and the type of connectivity can be
      set through \texttt{\small -{}-openingngb}. In the examples
      shown in Figure \ref{det}, they are set to \texttt{\small
        \opening} and \texttt{\small \openingngb}, respectively. The
      opened image can be seen in column 6 of Figure
      \ref{det}. Compared with column 5, it is clear that, as
      expected, opening has separated the foreground into separately
      connected regions or islands without bridges.

      \begin{figure}[t]
        \ifdefined\makeeps
        \begin{tikzpicture}
    \draw[step=0.0666\linewidth](0, 0.133\linewidth)
    grid (0.2001\linewidth, 0.3335\linewidth);

    \draw[step=0.0666\linewidth](0.266\linewidth,0.133\linewidth)
    grid (0.4669\linewidth, 0.3335\linewidth);

    \draw[step=0.0666\linewidth](0.532\linewidth,0)
    grid (\linewidth, 0.4669\linewidth);

    \draw[black, fill=black] (0.0666\linewidth, 0.1996\linewidth)
                   rectangle (0.1332\linewidth, 0.2662\linewidth);
    \draw[black, fill=gray] (0.0666\linewidth, 0.2662\linewidth)
                  rectangle (0.1332\linewidth, 0.3328\linewidth);
    \draw[black, fill=gray] (0.0666\linewidth, 0.1330\linewidth)
                  rectangle (0.1332\linewidth, 0.1996\linewidth);
    \draw[black, fill=gray] (0               , 0.1996\linewidth)
                  rectangle (0.0666\linewidth, 0.2662\linewidth);
    \draw[black, fill=gray] (0.1332\linewidth, 0.1996\linewidth)
                  rectangle (0.1998\linewidth, 0.2662\linewidth);

    \node[anchor=south, inner sep=1pt]
         at (0.0999\linewidth,-0.5cm) {a};

    \draw[black, fill=black] (0.3330\linewidth, 0.1996\linewidth)
                   rectangle (0.3996\linewidth, 0.2662\linewidth);
    \draw[black, fill=gray] (0.3330\linewidth, 0.2662\linewidth)
                  rectangle (0.3996\linewidth, 0.3328\linewidth);
    \draw[black, fill=gray] (0.3330\linewidth, 0.1330\linewidth)
                  rectangle (0.3996\linewidth, 0.1996\linewidth);
    \draw[black, fill=gray] (0.2664\linewidth, 0.1996\linewidth)
                  rectangle (0.3330\linewidth, 0.2662\linewidth);
    \draw[black, fill=gray] (0.3996\linewidth, 0.1996\linewidth)
                  rectangle (0.4662\linewidth, 0.2662\linewidth);
    \draw[black, fill=gray] (0.2664\linewidth, 0.1330\linewidth)
                  rectangle (0.3330\linewidth, 0.1996\linewidth);
    \draw[black, fill=gray] (0.3996\linewidth, 0.1330\linewidth)
                  rectangle (0.4662\linewidth, 0.1996\linewidth);
    \draw[black, fill=gray] (0.2664\linewidth, 0.2662\linewidth)
                  rectangle (0.3330\linewidth, 0.3328\linewidth);
    \draw[black, fill=gray] (0.3996\linewidth, 0.2662\linewidth)
                  rectangle (0.4662\linewidth, 0.3328\linewidth);

    \node[anchor=south, inner sep=1pt]
         at (0.3663\linewidth,-0.5cm) {b};

    \draw[black, fill=black] (0.7326\linewidth, 0.1996\linewidth)
                  rectangle (0.7992\linewidth, 0.2662\linewidth);

    \draw[black, fill=black!75]
    (0.7326\linewidth, 0.2662\linewidth) rectangle
    (0.7992\linewidth, 0.3328\linewidth);
    \draw[black, fill=black!75]
    (0.7326\linewidth, 0.1330\linewidth) rectangle
    (0.7992\linewidth, 0.1996\linewidth);
    \draw[black, fill=black!75]
    (0.6660\linewidth, 0.1996\linewidth) rectangle
    (0.7326\linewidth, 0.2662\linewidth);
    \draw[black, fill=black!75]
    (0.7992\linewidth, 0.1996\linewidth) rectangle
    (0.8658\linewidth, 0.2662\linewidth);
    \draw[black, fill=black!75]
    (0.6660\linewidth, 0.1330\linewidth) rectangle
    (0.7326\linewidth, 0.1996\linewidth);
    \draw[black, fill=black!75]
    (0.7992\linewidth, 0.1330\linewidth) rectangle
    (0.8658\linewidth, 0.1996\linewidth);
    \draw[black, fill=black!75]
    (0.6660\linewidth, 0.2662\linewidth) rectangle
    (0.7326\linewidth, 0.3328\linewidth);
    \draw[black, fill=black!75]
    (0.7992\linewidth, 0.2662\linewidth) rectangle
    (0.8658\linewidth, 0.3328\linewidth);

    \draw[black, fill=black!50]
    (0.6660\linewidth, 0.0666\linewidth) rectangle
    (0.7326\linewidth, 0.1332\linewidth);
    \draw[black, fill=black!50]
    (0.7326\linewidth, 0.0666\linewidth) rectangle
    (0.7992\linewidth, 0.1332\linewidth);
    \draw[black, fill=black!50]
    (0.7992\linewidth, 0.0666\linewidth) rectangle
    (0.8658\linewidth, 0.1332\linewidth);
    \draw[black, fill=black!50]
    (0.6660\linewidth, 0.3328\linewidth) rectangle
    (0.7326\linewidth, 0.3994\linewidth);
    \draw[black, fill=black!50]
    (0.7326\linewidth, 0.3328\linewidth) rectangle
    (0.7992\linewidth, 0.3994\linewidth);
    \draw[black, fill=black!50]
    (0.7992\linewidth, 0.3328\linewidth) rectangle
    (0.8658\linewidth, 0.3994\linewidth);
    \draw[black, fill=black!50]
    (0.5994\linewidth, 0.1332\linewidth) rectangle
    (0.6660\linewidth, 0.1996\linewidth);
    \draw[black, fill=black!50]
    (0.5994\linewidth, 0.1996\linewidth) rectangle
    (0.6660\linewidth, 0.2662\linewidth);
    \draw[black, fill=black!50]
    (0.5994\linewidth, 0.2662\linewidth) rectangle
    (0.6660\linewidth, 0.3328\linewidth);
    \draw[black, fill=black!50]
    (0.8658\linewidth, 0.1332\linewidth) rectangle
    (0.9324\linewidth, 0.1996\linewidth);
    \draw[black, fill=black!50]
    (0.8658\linewidth, 0.1996\linewidth) rectangle
    (0.9324\linewidth, 0.2662\linewidth);
    \draw[black, fill=black!50]
    (0.8658\linewidth, 0.2662\linewidth) rectangle
    (0.9324\linewidth, 0.3328\linewidth);

    \draw[black, fill=black!25]
    (0.6660\linewidth, 0) rectangle
    (0.7326\linewidth, 0.0666\linewidth);
    \draw[black, fill=black!25]
    (0.7326\linewidth, 0) rectangle
    (0.7992\linewidth, 0.0666\linewidth);
    \draw[black, fill=black!25]
    (0.7992\linewidth, 0) rectangle
    (0.8658\linewidth, 0.0666\linewidth);
    \draw[black, fill=black!25]
    (0.6660\linewidth, 0.3994\linewidth) rectangle
    (0.7326\linewidth, 0.4660\linewidth);
    \draw[black, fill=black!25]
    (0.7326\linewidth, 0.3994\linewidth) rectangle
    (0.7992\linewidth, 0.4660\linewidth);
    \draw[black, fill=black!25]
    (0.7992\linewidth, 0.3994\linewidth) rectangle
    (0.8658\linewidth, 0.4660\linewidth);
    \draw[black, fill=black!25]
    (0.5328\linewidth, 0.1332\linewidth) rectangle
    (0.5994\linewidth, 0.1996\linewidth);
    \draw[black, fill=black!25]
    (0.5328\linewidth, 0.1996\linewidth) rectangle
    (0.5994\linewidth, 0.2662\linewidth);
    \draw[black, fill=black!25]
    (0.5328\linewidth, 0.2662\linewidth) rectangle
    (0.5994\linewidth, 0.3328\linewidth);
    \draw[black, fill=black!25]
    (0.9324\linewidth, 0.1332\linewidth) rectangle
    (0.9990\linewidth, 0.1996\linewidth);
    \draw[black, fill=black!25]
    (0.9324\linewidth, 0.1996\linewidth) rectangle
    (0.9990\linewidth, 0.2662\linewidth);
    \draw[black, fill=black!25]
    (0.9324\linewidth, 0.2662\linewidth) rectangle
    (0.9990\linewidth, 0.3328\linewidth);
    \draw[black, fill=black!25]
    (0.5994\linewidth, 0.0666\linewidth) rectangle
    (0.6660\linewidth, 0.1332\linewidth);
    \draw[black, fill=black!25]
    (0.8658\linewidth, 0.0666\linewidth) rectangle
    (0.9324\linewidth, 0.1332\linewidth);
    \draw[black, fill=black!25]
    (0.5994\linewidth, 0.3328\linewidth) rectangle
    (0.6660\linewidth, 0.3994\linewidth);
    \draw[black, fill=black!25]
    (0.8658\linewidth, 0.3328\linewidth) rectangle
    (0.9324\linewidth, 0.3994\linewidth);

    \node[anchor=south, inner sep=1pt]
         at (0.7659\linewidth,-0.5cm) {c};

\end{tikzpicture}
        \else
        \includegraphics[width=\linewidth]{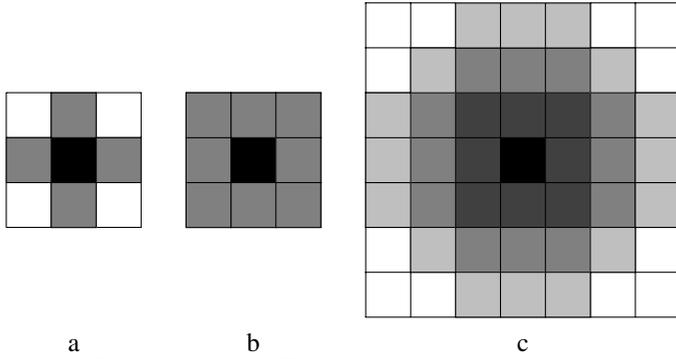}
        \fi
        \caption{\label{StructElem}Structuring elements. The black
          pixel is the pixel under consideration and the gray pixels
          are the neighbors of that pixel. (a) 4 connected
          neighbors. (b) 8 connected neighbors. (c) Non-white pixels
          show the smallest area an object should have to be detected
          with the given two 4 connected erosions (see Section
          \ref{erosion}) and one 8 connected opening (see Section
          \ref{opening}). It is assumed that all the pixels outside
          the box shown are background (white).  The three shades of
          gray show the pixels that are removed in each step
          (brightest first). After the three steps finish, the inner 9
          pixels remain. In processing, white pixels have a value of 0
          and others have a value of 1.}
      \end{figure}

      Multiple erosions and dilations with the same structuring
      element will result in the outer shapes of the foreground
      becoming like a diamond in 4 connectivity or a rectangle in 8
      connectivity. Therefore, since the erosion of Section
      \ref{erosion} was based on a \erodengb{} connected structuring
      element, in this step an \openingngb{} connected structuring
      element was used to compensate for the shape loss in the
      previous step as well as more successfully removing thin
      connections. The composite effect of the two steps of erosion in
      Section \ref{erosion} and the opening done here can be seen in
      Figure \ref{StructElem}(c). Before the dilating section of the
      opening step, only the central black pixel will remain. If a
      region above the threshold is any smaller than the colored
      pixels of Figure \ref{StructElem}(c), it will disappear after
      this step since no pixels remain for dilation.

      The erosion and dilation defined here are based on a binary
      image as discussed in Section \ref{erosion}. It is also possible
      to define such operations on grayscale images where a pixel can
      have a range of values, not just 0 or 1. \citet{perret09} used a
      combination of such grayscale erosion and dilation (hit-or-miss
      transform) for a demonstration on detecting low surface
      brightness galaxies. Since the structuring element is no longer
      a binary, the variation of pixel values over the structuring
      element becomes very important and has to be modeled based on
      specific elliptical parameters, surface brightness profiles, and
      the {\small PSF}; see Figures 4 and 5 of
      \citet{perret09}. Therefore, it is optimal only for profiles
      that satisfy the count distribution of the structuring
      element. Applying it generically to any image with any
      distribution of galaxy morphologies will require a high level of
      customization that may fail for complex galaxy morphologies like
      the real galaxies shown in this paper.

    \subsubsection{Defining and Removing False Detections}
    \label{removenoise}

      \begin{figure*}
        \centering
        \ifdefined\makeeps
        \include{./tex/dettf}
        \else
        \includegraphics[width=\linewidth]{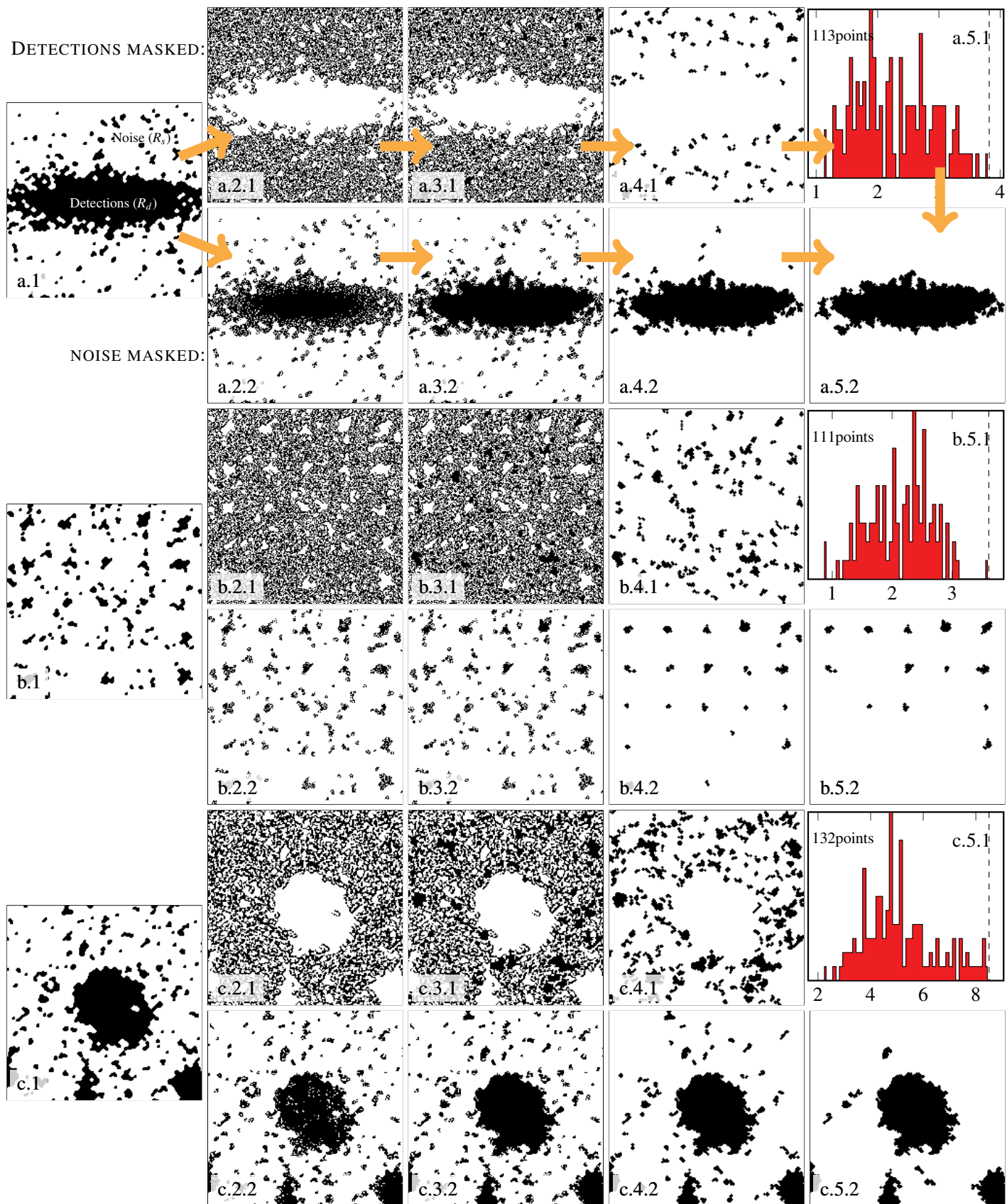}
        \fi
        \caption{\label{dettf}Identifying and removing false
          detections, demonstrated for three of the examples from
          Figures \ref{det}(a), (b) and (d). Column 1 is the same as
          column 6 in Figure \ref{det}. Columns 2 to 4 apply the same
          steps (explained below) on the undetected regions (top) and
          detections (bottom) separately. Column 2: Applying the
          threshold. Column 3: Fill holes surrounded by one 4
          connected region above the threshold. Column 4: Application
          of opening to the previous column and small regions
          removed. Column 5, top: distribution of the total
          signal-to-noise ratio of all the dark connected components
          in column 4 (top). The dashed vertical line in the histogram
          shows the \texttt{-{}-detquant=\detquant} quantile of this
          distribution. This value is used as a threshold to define
          false detections in column 4 (bottom). (a.5.1) is from a
          nearby large mesh (see Section \ref{largeimg}) because the
          number of connected components in (a.4.1) is less than
          \texttt{-{}-minnumfalse=\minnumfalse}. Column 5 (bottom):
          all detections with total signal-to-noise ratio above that
          threshold. The arrows in (a) visualize the steps which are
          schematically shown in the lower box of Figure
          \ref{flowchart}. }
      \end{figure*}

      The foreground is now composed of separately connected objects
      that we define as \emph{initial} detections. However, on the
      mock images (particularly comparing Figure \ref{det}(a.1) with
      (a.6)), it is clear that some detections can be associated with
      true signal while most cannot. These ``false'' detections are
      sets of pixels with random values (above the very low threshold
      on the convolved image) that were randomly positioned close
      enough to each other to pass the erosion and opening steps of
      detection explained above. As the threshold decreases, the
      probability of such random positioning increases. If a larger
      threshold, number of erosions, or number of openings is used,
      false detections will decrease or completely disappear.

      The problem with more stringent detection parameters is that
      besides rejecting more inherently faint objects, they will also
      remove the valuable fainter parts of the final
      detections. Therefore, a classification scheme is defined to
      find and remove false detections very accurately.  This allows
      for keeping the valuable faint pixels of bright detections and
      most of the detections that would have been lost otherwise.

      The bottom group of steps in the flowchart in Figure
      \ref{flowchart} show all the steps used to define and remove
      false detections. The same steps can also be seen in practice in
      Figure \ref{dettf} and elaborated here.  The detections have
      divided the pixels into 2 sets: (1) The undetected sky, $R_s$
      (white regions of column 1 in Figure \ref{dettf}) and (2) the
      detected regions, $R_d$ (black regions in column 1 of Figure
      \ref{dettf}). It can be reasonably assumed that no significant
      contribution from a detectable object exists in $R_s$.  Some of
      the detections in $R_d$ harbor signal and some are localized
      random noise pixels, namely false detections.

      In order to identify the false detections in $R_d$, a secondary
      detection process is applied to the pixels of both $R_s$ and
      $R_d$ independently. The detections in the former will provide a
      scale on which the certainty of the detections in the latter can
      be defined. Let $S_a$ and $\sigma_s$ be the average and standard
      deviation of the count in $R_s$, respectively. The second
      threshold is defined by these two parameters. Note that this
      $S_a$ is not the final Sky value because some of the high
      valued, localized noise pixels (false detections) have
      systematically been removed from it; it is just used here as an
      approximation.

      The user is free to set this second threshold based on $S_a+$
      \texttt{\small dthresh} $\times\sigma_s$. In the examples here,
      \texttt{\small -{}-dthresh=\dthresh}.  The result of applying
      this threshold to the detected and blank regions can be seen in
      Figure \ref{dettf}, column 2.  Note that as discussed in Section
      \ref{conv}, absolute values such as $S_a$ and $\sigma_s$ come
      from the input image. Therefore, while the pixel \emph{indices}
      comprising $R_s$ and $R_d$ are defined from column 6 of Figure
      \ref{det}, which used the convolved image, the pixel
      \emph{values} for finding and applying this second absolute
      threshold come from the actual image. The pixels that were above
      this threshold are marked in black in column 2 of Figure
      \ref{dettf}.

      As discussed in Section \ref{mydetection}, a significant
      advantage of data over noise detections is that the data augment
      a \emph{contiguous} region. In order to exploit this contiguous
      property of true detections, all the holes that are engulfed in
      4 connected foreground regions will be filled. Note that holes
      are white pixels in Figure \ref{dettf}, column 2. The result of
      filling such holes can be seen in column 3 of Figure
      \ref{dettf}. Such holes are most probably due to noise since the
      4 connectivity of all the surrounding pixels is a very strong
      constraint (compare Figure \ref{dettf}(b.2.2) and
      (b.3.2)). Combined with the next step (opening), filling holes
      in this manner significantly enhances the ability to identify a
      faint contiguous signal.

      To remove the thin connections between thicker regions and
      obtain individual connected components, one level of 4 connected
      opening is applied. The result can be seen in column 4 of Figure
      \ref{dettf}.  The two images of column 4 in Figure \ref{dettf}
      are now composed of separate connected components
      (pseudo-detections) that were created in exactly the same manner
      but on different regions of the input image. The parameter used
      to quantify the definition of a false detection is the total
      {\small S/N} ({\small S/N}$_T$) of each pseudo-detection which
      is a combination of its total count and area.

      Let $F$ be the average count in each pseudo-detection and $N$ be
      its area. The areas vary widely, ranging from one pixel to
      thousands of pixels. Therefore in order to get a reasonable
      average count comparison, all pseudo-detections with an area
      smaller than \texttt{\small -{}-detsnminarea} are removed from
      the analysis in both $R_s$ and $R_d$. In the examples in this
      section, it is set to \texttt{\small \detsnminarea}. Note that
      since a threshold approximately equal to the Sky value is used,
      this is a very weak constraint. For each pseudo-detection,
      {\small S/N}$_T$ can be written as,

      \begin{equation}
        \label{tSNeq}
        \mathrm{S/N}_T=\frac{NF-NS_a}{\sqrt{NF+N\sigma_S^2}}
        =\frac{\sqrt{N}(F-S_a)}{\sqrt{F+\sigma_S^2}}.
      \end{equation}

      \noindent
      See Section \ref{SNeqmodif} for the modifications required when
      the input image is not in units of counts or has already been
      Sky subtracted. The distribution of {\small S/N}$_T$ from the
      objects in $R_s$ for the three examples in Figure \ref{dettf}
      can be seen in column 5 (top) of that figure. Image processing
      effects, mainly due to shifting, rotating, and re-sampling the
      images for co-adding, on the real data further increase the size
      and count, and hence, the {\small S/N} of false detections in
      real, reduced/co-added images. A comparison of scales on the
      {\small S/N} histograms between the mock ((a.5.1) and (b.5.1))
      and real (c.5.1) examples in Figure \ref{dettf} shows the effect
      quantitatively. In the histograms of Figure \ref{dettf}, the bin
      with the largest number of false pseudo-detections respectively
      has an {\small S/N} of $\onelargedettfmax$,
      $\sensitivitycdettfmax$, and $\fourdettfmax$.

      The {\small S/N}$_T$ distribution of detections in $R_s$
      provides a very robust and objective scale that can be used to
      select or reject a given detection in $R_d$. The level of
      certainty for true detections can be input by the user and is
      defined based on \texttt{-{}-detquant} which is the desired
      quantile in the S/N$_T$ distribution of $R_s$. In the flowchart
      shown in Figure \ref{flowchart}, this is displayed with
      $D_j>s_q$. The accepted detections from the second thresholding
      can be seen in the bottom row of column 5 in Figure
      \ref{dettf}. Any initial detection that overlaps with at least
      one of the accepted pseudo-detections
      ($B_j\cap{}A_i\neq\emptyset$ in Figure \ref{flowchart}) is
      defined as a true detection, while those that don't are
      considered a false detection and removed. The successful
      detections can be seen in the last two columns of Figure
      \ref{det} (after a final dilation; see Section \ref{dilation}).

      Outside of astronomical data analysis, the technique proposed
      here to separate true from false detections (and later clumps;
      see Section \ref{clumpseg}) can be considered a form of anomaly
      detection in data mining. An anomaly or outlier is defined as
      ``an observation which deviates so much from other observations
      as to arouse suspicion that it was generated by a different
      mechanism'' \citep[page 1]{hawkins80}. This is a very general
      and qualitative definition applicable to any form of anomaly
      \citep[see][for a review]{chandola09}.

      The {\small S/N} of the connected components in $R_s$ (after
      applying the threshold, filling holes, and opening) gives a
      scale to define the outliers in $R_d$ (true
      pseudo-detections). However, a quantile and not the histogram is
      used (the histogram is only displayed in Figure \ref{dettf} for
      demonstration). Therefore based on the classification of
      \cite{chandola09}, the true/false classification technique
      proposed here can be considered a non-parametric, ``quantile
      based'' statistical anomaly detection technique.

      This technique for determining an ``anomaly'' (or a real object
      within noise) has some similarities to the Higher Criticism
      statistic \citep{highercriticism}. It was used in detecting
      non-Gaussianities in the cosmic microwave background
      \citep{cayon05} with the modification of using
      $p$-values. However, the higher criticism statistic is derived
      from and applied to the same set of observations, while here the
      {\small S/N} threshold is found using a completely separate set
      of pixels (those in $R_s$). It is then applied to another set of
      pixels ($R_d$). Also, while the Gaussian/Poisson distribution is
      assumed for the noise, there is no such assumption for the
      distribution of the {\small S/N}.

    \subsubsection{Final Dilation}\label{dilation}
      The final step is to restore, through dilation, the pixel layers
      of each remaining initial detection that were removed by erosion
      in Section \ref{erosion}. Dilating \texttt{\small -{}-erosion}
      times would presumably restore the object pixels out to the
      initial \texttt{\small -{}-qthresh} quantile. The user can set
      the number of final dilations with the \texttt{\small
        -{}-dilate} option.  Images of astronomical objects never have
      a strong cutoff. Even if they have a physical cutoff, the
      {\small PSF} will soften any sharp boundary except cosmic rays
      which are independent of the {\small PSF}. Therefore it can be
      assumed that the object extends beyond the initial threshold. To
      exploit this fact, this final dilation is only based on 8
      connectivity and in the examples of this paper; it is set to
      \texttt{\small -{}-dilate=\dilate}.

  \subsection{Segmentation}\label{mysegmentation}

    The detection thresholds used are extremely low in
    \textsf{NoiseChisel} (see Section \ref{thresh}). Combined with the
    fact that galaxies have no clear cutoff, it often happens that
    several apparently nearby objects on the image will be detected as
    one region. There are two approaches to separating a detected
    region into potentially several sub-detections or to find
    substructure: segmentation and deblending.

    Segmentation is the procedure that assigns each pixel to one
    sub-component. Deblending, on the other hand, identifies the
    contribution in each pixel from different blended
    sources. Therefore in deblending, each pixel can belong to more
    than one object. Deblending is more realistic: the sum of the
    light profiles of multiple overlapping objects specifies the final
    pixel value. However, to be accurately done, deblending will
    require parametric analysis and fitting algorithms. Segmentation,
    on the other hand, is a very low-level measurement and only goes
    so far as to assign each pixel to the sub-component that
    contributes most to it. Therefore segmentation provides the best
    starting point for the higher level deblending if it is
    needed.\footnote{We will be working on such a fitting tool for
      deblending as part of the {\scriptsize GNU} Astronomy Utilities
      to start using the outputs of \textsf{NoiseChisel}.}
    \textsf{NoiseChisel} will not do deblending but only performs
    segmentation.

    \begin{figure*}[tbp]
      \centering
      \ifdefined\makeeps
      \input{./tex/overseg}
      \else
      \includegraphics[width=\linewidth]{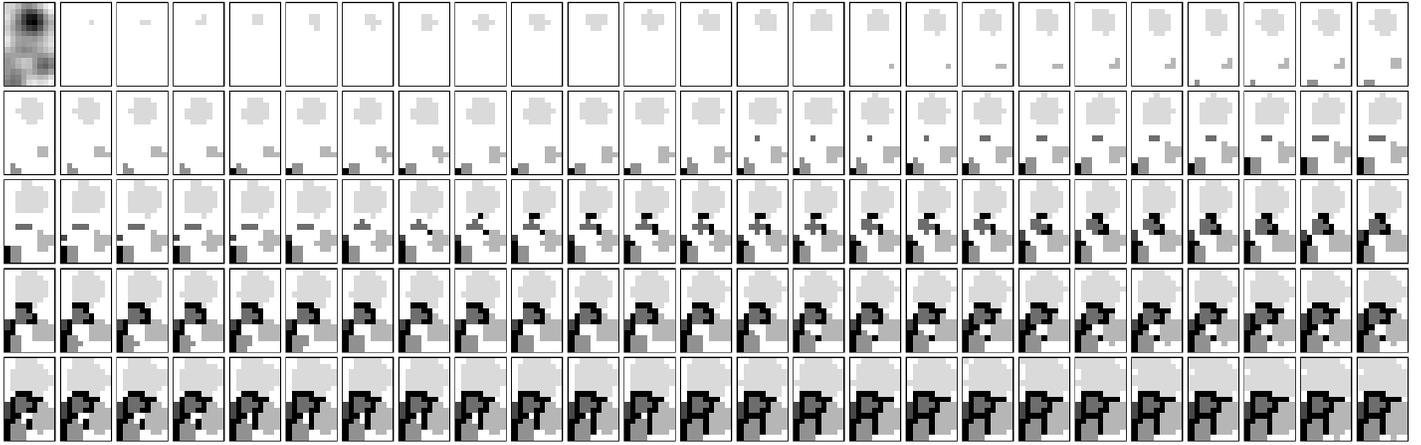}
      \fi
      \caption{\label{overseg} Pixel-by-pixel demonstration of
        oversegmentation (see Section \ref{clumpseg} for a complete
        explanation of the algorithm). The order is from the top left
        to the bottom right. First: a small region in the center of
        Figure \ref{segtf}(b.2). The rest show how pixels are labeled
        in order of their value. The shades of gray become darker as
        the labels increase. The darkest shade of gray represents
        river pixels.}
    \end{figure*}

    In order to segment each detection into sub-components,
    \emph{clumps} (Section \ref{clumpseg}) and \emph{objects} (Section
    \ref{ObjSeg}) are defined. A detected region might contain one or
    more objects and anything from none to multiple ``true'' clumps.
    Finding the clumps in an object or detected region is a low-level
    measurement that provides us with a multitude of false clumps
    mixed with possibly true clumps. True clumps can be found robustly
    by studying the noise properties in the vicinity of the detected
    regions. This process is very similar to how true/false detections
    were separated in Section \ref{removenoise}. In Section
    \ref{clumpseg} the proposed definition for a clump and the
    true/false classification is explained. The segmentation of a
    detected region to objects is a high-level measurement thoroughly
    discussed in Section \ref{ObjSeg}.

    \begin{figure*}[t]
      \centering
      \ifdefined\makeeps
      \input{./tex/segtf}
      \else
      \includegraphics[width=\linewidth]{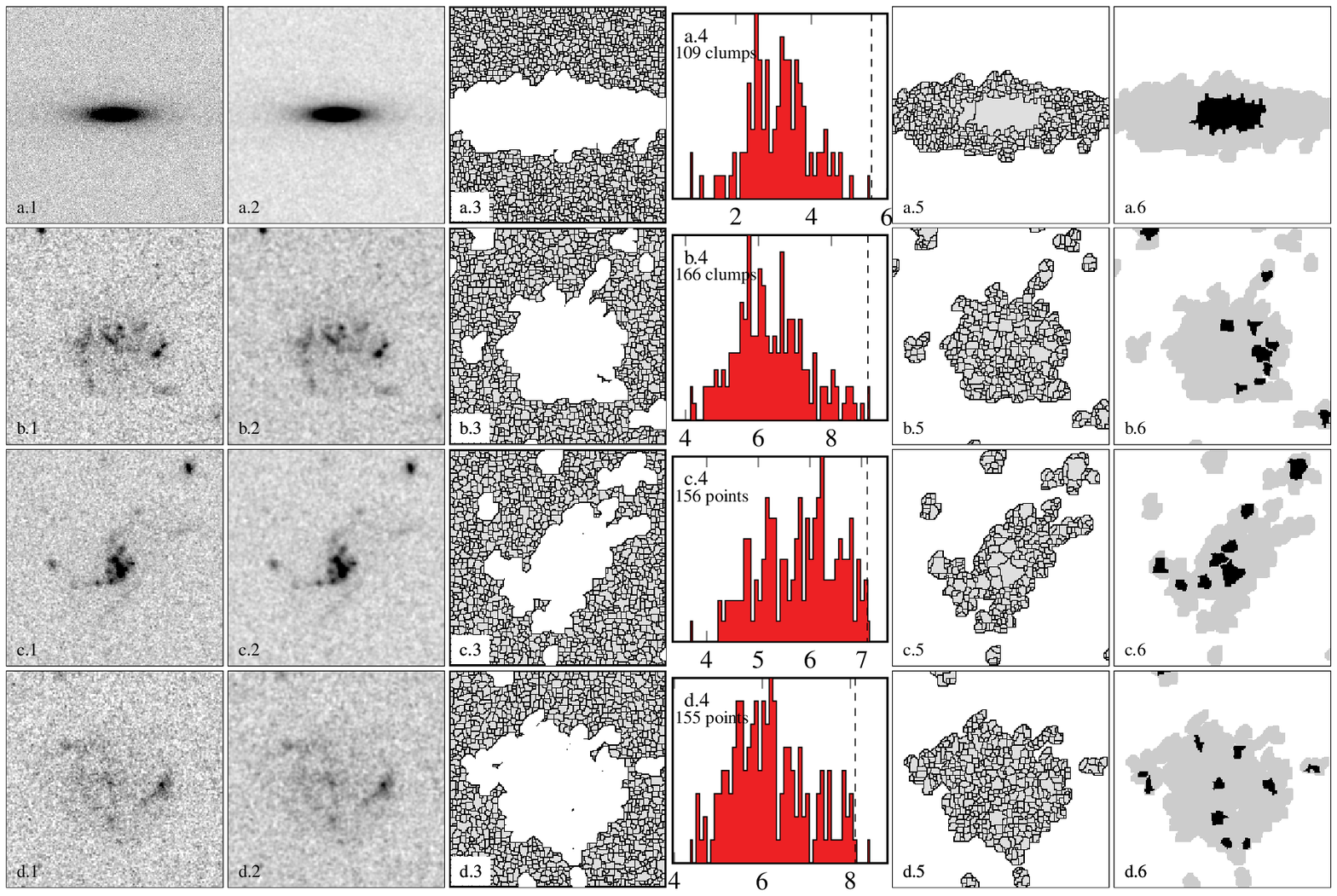}
      \fi
      \caption{\label{segtf} Finding clumps through
        oversegmentation. Column 1: input images. Column 2: convolved
        image (Section \ref{conv}). Column 3: oversegmentation applied
        to non-detected regions on the convolved image. Column 4:
        signal-to-noise ratio histogram (Equation (\ref{clumpSNeq}))
        of all segments in column 3 with an area larger than
        \texttt{-{}-segsnminarea=\segsnminarea}. The dashed line shows
        the position of the \texttt{-{}-segquant=\segquant}
        quantile. Column 5: oversegmentation applied to the central
        object. Column 6: clumps in all the detected objects with
        signal-to-noise ratio larger than the threshold of column
        4. Note that some detections have no \emph{true} clumps.}
    \end{figure*}

    \subsubsection{Finding ``True'' Clumps}\label{clumpseg}
      Pixels at a local maximum are defined as those whose value is
      larger than their 8 connected neighbors. The 8 connected region
      around each local maximum whose pixels all have a lower count
      than the peak is defined as a \emph{clump}. Note that the clump
      can be much larger than 9 pixels.

      The approach introduced here for finding the clumps associated
      with all the local maxima of an image is conceptually very
      similar to and inspired by the method proposed by
      \citet{vcwatershed}, without using layers because of the biases
      they produce (see the discussion in Section
      \ref{SEdeblend}). Figure \ref{overseg} shows the process for 124
      pixels of the central region of Figure \ref{segtf}(b.2).

      A zero valued array with the size of the image is first created
      to store the labels of the different clumps. All the pixels of
      each object are ordered based on their value. Starting from the
      brightest pixel and in decreasing order, the pixels are labeled
      based on the following conditions.
      \begin{enumerate}
      \item If the pixel is not 8 connected to any other labeled
        pixels, it is a local maximum and is assigned a new
        label. This can be seen as new labels (darker shades of gray)
        are added in Figure \ref{overseg}.
      \item If its 8 connected neighbors all have the same label, the
        pixel is given that label as well. In Figure \ref{overseg},
        this occurs most often (when the colored areas expand).
      \item If its neighbors have different labels, like a river
        flowing between two mountains,\footnote{As \citet[page
            583]{vcwatershed} put it, ``In the field of image
          processing and more particularly in mathematical morphology,
          grayscale pictures are often considered as topographic
          reliefs. In the topographic representation of a given image
          $I$, the numerical value (that is, the gray tone) of each
          pixel stands for the elevation at this point. Such a
          representation is extremely useful since it allows one to
          better appreciate the effect of a given transformation on
          the image under study.'' In this analogy, the noisy image
          can be considered as a mountain range and when rain comes,
          the water gathers in the local minima of those mountain
          ranges to form rivers.} the pixel is a local minimum. The
        \emph{river} pixels act as separators of each mountain or
        clump. River pixels are the darkest pixels in Figure
        \ref{overseg}. The pixel-by-pixel construction of the rivers
        can be seen from the middle of the third row in Figure
        \ref{overseg} as the growing regions start touching. In this
        setup the rivers are 4 connected. If 4 connectivity was chosen
        for checking neighbors, the rivers would be 8 connected.
      \end{enumerate}

      \begin{figure*}[t]
        \centering
        \ifdefined\makeeps
        \input{./tex/objseg}
        \else
        \includegraphics[width=\linewidth]{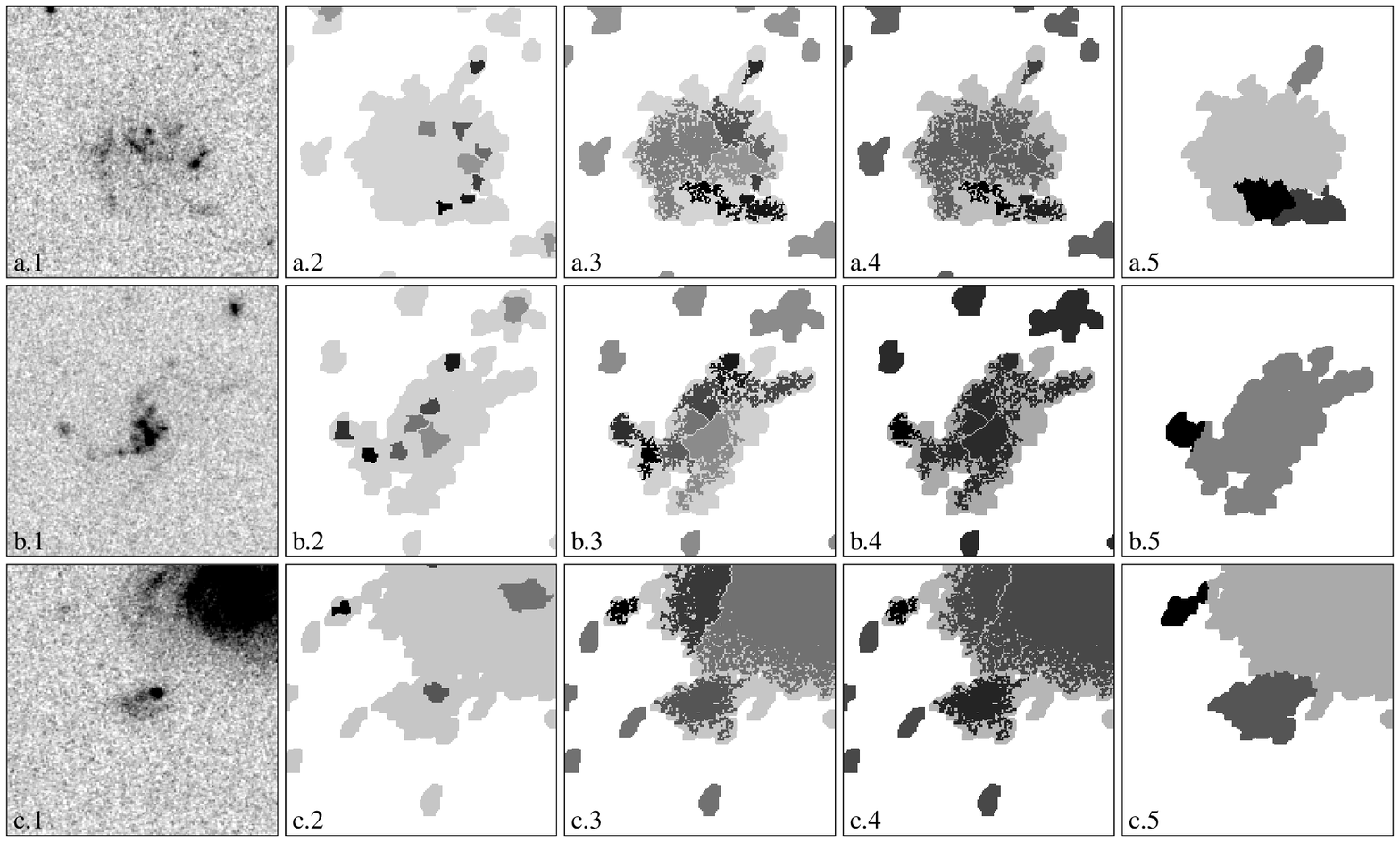}
        \fi
        \caption{\label{objseg} Segmenting objects. Column 1: actual
          images. Column 2: true clumps in the detections (see Section
          \ref{clumpseg}). Each clump is color-coded based on its
          label. The underlying light gray region shows the detection
          region. Column 3: the clumps are grown until
          \texttt{-{}-gthresh=\gthresh} (see Section
          \ref{ObjSeg}). Column 4: all objects with boundary
          signal-to-noise ratio larger than
          \texttt{-{}-objbordersn=\objbordersn} are given the same
          label. Column 5: the remaining pixels of the region are
          filled by growing the labels of column 4. In the last column
          only the central segmented objects are shown for clarity.}
      \end{figure*}

      The application of this simple algorithm to the non-detected
      regions of the image can be seen in column 3 of Figure
      \ref{segtf} and on the detected regions in column 5. Similar to
      our discussion in Section \ref{removenoise}, the region of no
      detection is used as a basis or calibrator to find the true
      clumps over the detected regions. This algorithm for finding the
      clumps uses their relative values. Therefore as discussed in
      Section \ref{conv}, the convolved image is used to find the
      index of pixels in each clump. The small convolution kernel is
      beneficial for this step because the spatial resolution is very
      important for an accurate result. This pixel-by-pixel expansion
      process provides the ultimate usage of the dynamic range of the
      convolved image.

      A very large fraction of the clumps found over a detected object
      (column 5 in Figure \ref{segtf}) are due to background and
      correlated noise. To identify the false clumps, the segmentation
      results on the noise regions of the image (column 3 of Figure
      \ref{segtf}) are used as a reference. As discussed in Section
      \ref{mydetection}, it can be assumed with high certainty that
      there is no significant signal (from a true physical object) in
      the noise regions. Furthermore, the noise regions in the
      vicinity of the object would contain the same undetected
      astronomical objects and instrumental and data analysis biases
      as the object pixels. Therefore this region is the most accurate
      and objective reference available to judge between a true and a
      noise clump.

      On the original unconvolved and not Sky subtracted image, $F_i$
      and $F_o$ are respectively defined as the average count in a
      clump and on all river pixels surrounding it. The average value
      of the river pixels around each clump approximately show the
      base elevation or count if the clump was not present. Let $N_i$
      be the total area or number of pixels inside a clump. The
      {\small S/N} of each clump can be written as,

      \begin{equation}
        \label{clumpSNeq}
        \mathrm{S/N}=\frac{N_iF_i-N_iF_o}{ \sqrt{N_iF_i+N_iF_o}} =
        \frac{ \sqrt{N_i}(F_i-F_o) }{ \sqrt{F_i+F_o} }.
      \end{equation}

      See Section \ref{SNeqmodif} for how this equation can be
      modified for images that are not in units of counts or have
      already been Sky subtracted.  The {\small S/N} distribution of
      all the clumps in the blank Sky can be seen for all the examples
      of Figure \ref{segtf} in column 4 of that figure. The
      \texttt{\small -{}-segquant=\segquant} quantile of the
      distribution is used to find the {\small S/N} threshold to
      accept or reject a clump over the detected object. This
      threshold is thus completely independent of the particular
      object in which the potential clumps are embedded. Note that the
      criteria in \textsf{SExtractor} and some other existing tools
      for deblending an object is the fraction of counts in each clump
      to that in the parent detection (see Section \ref{SEdeblend} for
      a discussion of the resulting biases). Since no layering is
      necessary and the clump counts are not compared to the object
      anymore, the clumps in various galaxies can now be objectively
      compared with each other.

      Since the average count in each clump is very important in the
      {\small S/N} calculation, only clumps having an area larger than
      \texttt{\small -{}-segsnminarea} are considered in the
      comparison. Any clump that is smaller than this area will be
      discarded as noise. For the examples in this paper it is set to
      \texttt{\small \segsnminarea}. This is larger than
      \texttt{\small -{}-detsnminarea}, because the areas there were
      found by applying a threshold to the input image, while the
      clumps here are found on the convolved (smoothed) image where
      the areas become larger. Clumps smaller than this area add very
      strong skewness to the {\small S/N} distributions.

    \subsubsection{Object Segmentation}\label{ObjSeg}

      The very low thresholds that were used (see Section
      \ref{thresh}) cause objects that are close enough to each other
      on the image to be blended in one detection. Three real examples
      are displayed in the first column of Figure \ref{objseg}. The
      basis for separating potential objects in a given region is the
      true clumps which have been found in that detection.  Therefore,
      if a detected region has none or only one clump, it is
      considered to be only one object. If there is more than one
      clump in a detected region, the clumps are grown until a certain
      threshold of the input image. If the Sky value (average of
      non-detected pixels) and its standard deviation are labeled with
      $S$ and $\sigma_s$, then the threshold to stop clump growth is
      $S+$\texttt{\small gthresh}$\times\sigma_s$. In the examples of
      Figure \ref{objseg} it is set to \texttt{\small
        -{}-gthresh=\gthresh}.

      The process of growing the clumps is very similar to the
      algorithm in Section \ref{clumpseg} (Figure \ref{overseg}) with
      the exception that no new labels are added. If a pixel has no
      labeled neighbors, it is kept in a queue to be checked on a next
      loop. The loop continues until no new pixel can be labeled. The
      grown segments of each clump of the examples in Figure
      \ref{objseg} can be seen in column 3.

      Objects are defined based on the average {\small S/N} of the
      river pixels between the grown clumps of Figure \ref{objseg},
      column 3. Two grown clumps are defined as separate objects if
      the river between them is below a user-defined {\small S/N}. If
      it is larger, then the connection between the grown clumps is
      too strong to be regarded as separate objects. Therefore the
      grown clumps are considered as parts of one object.  Let
      $F_{ij}$ represent the average count on the river between the
      grown clumps $i$ and $j$. The average {\small S/N} of the river
      between these two clumps is defined as:

      \begin{equation}\label{bordsneq}
        \mathrm{S/N}_{ij}=\frac{F_{ij}-S}{\sqrt{ F_{ij}+\sigma_s^2 }}.
      \end{equation}

      If {\small S/N}$_{ij}<$\texttt{\small objbordersn}, then these
      two grown clumps are considered separate objects and if not,
      they are defined as belonging to the same object. In the
      examples of Figure \ref{objseg} this is set to \texttt{\small
        -{}-objbordersn=\objbordersn}. {\small S/N}$_{ij}(=${\small
        S/N}$_{ji})$ is calculated for all the grown clumps of each
      detection. Finally, all the clumps that are connected to each
      other within a detection are given one label (see column 4 of
      Figure \ref{objseg}). Note that if two grown clumps have {\small
        S/N}$_{ij}<$\texttt{\small objbordersn}, and they are both
      connected to a third clump with {\small
        S/N}$_{ik}>$\texttt{\small objbordersn} and {\small
        S/N}$_{jk}>$\texttt{\small objbordersn} then all three grown
      clumps will be given the same label.  The rivers between two
      grown clumps are only one pixel wide. Therefore in the
      calculation of $F_{ij}$, the count of each pixel is taken as the
      average of that pixel and its 8 connected neighbors. Hence, in
      practice the rivers between two grown clumps are 3 pixels wide.

      Because of the absolute nature of the two parameters for object
      definition (as opposed to the relative nature in the case of
      detection and clumps), namely the user directly providing
      \texttt{\small -{}-gthresh} and \texttt{\small -{}-objbordersn},
      the object definition is not robust and objective like that of
      detections and clumps. In fact \texttt{\small -{}-objbordersn}
      is the only user given {\small S/N} value in
      \textsf{NoiseChisel}.

      The definition of objects made here is purely based on the one
      image input to \textsf{NoiseChisel}. The example above shows
      that this definition of objects is fundamentally subjective
      (directly determined by the particular data-set and the user and
      not necessarily based on physical reality). Two galaxies at
      vastly different redshifts might be sufficiently close to each
      other, on the same line of sight, that they are each defined as
      clumps in one larger object. However, due to surface brightness
      limits, the connection (for example, through a spiral arm or a
      tidal tail) of a real star-forming region of a galaxy might not
      be detected. Therefore that region will be identified as a
      separate object. This also applies to rings and filaments. In
      order to solve these issues, the information in one data set
      (image in this case) is not sufficient, and ancillary data, for
      example, from spectroscopic data or images in other wavelengths,
      will be necessary. Therefore the physical reality of the
      `objects' defined here is beyond the scope of any analysis based
      solely on one image.

  \subsection{S/N Equation Modifications}
    \label{SNeqmodif}

    The standard deviation of the non-detected regions ($\sigma_s$),
    incorporates all sources of error: Poisson noise, read-out noise,
    etc. Hence, if the image is in units of counts and the backgrounds
    are not already subtracted once, then Equations (\ref{tSNeq}) --
    (\ref{bordsneq}) can be used no matter if the noise is
    background-dominated or read-out-noise-dominated. However, input
    images do not necessarily satisfy this condition. For example, the
    \emph{\small HST}/{\small ACS} images shown above are processed
    (already Sky-subtracted) and distributed in units of counts per
    second.

    When the image is in units of counts s$^{-1}$, and if the noise is
    background-dominated, all the terms in the equations have to be
    multiplied by a constant in units of time. In finding true
    detections and clumps that depend on the quantile of the {\small
      S/N} distribution, the absolute value of this constant makes
    absolutely no difference. However, in identifying objects, it is
    important, because the user specifies an absolute {\small S/N}
    value for rivers. \textsf{NoiseChisel} will automatically detect
    if an image is in units of counts or counts s$^{-1}$ using the
    minimum standard deviation in all the meshes (see Section
    \ref{largeimg}). If this minimum is larger than unity, then it is
    assumed that the image is in units of counts. If not, the inverse
    of this minimum (which has units of time) will be used as the time
    constant discussed above.

    If the input images are processed, then Sky subtraction has
    already been applied. Therefore the error in the average count $F$
    in the denominators should be $F+\sigma_s^2$, not $F$ alone as is
    currently present in the equations. It is up to the user to
    specify this condition through the \texttt{\small
      -{}-skysubtracted} parameter (see the manual for more details).

\section{Large (real) Images}\label{largeimg}
\begin{figure*}[t]
      \centering
      \ifdefined\makeeps
      \input{./tex/largeimage}
      \else
      \includegraphics[width=\linewidth]{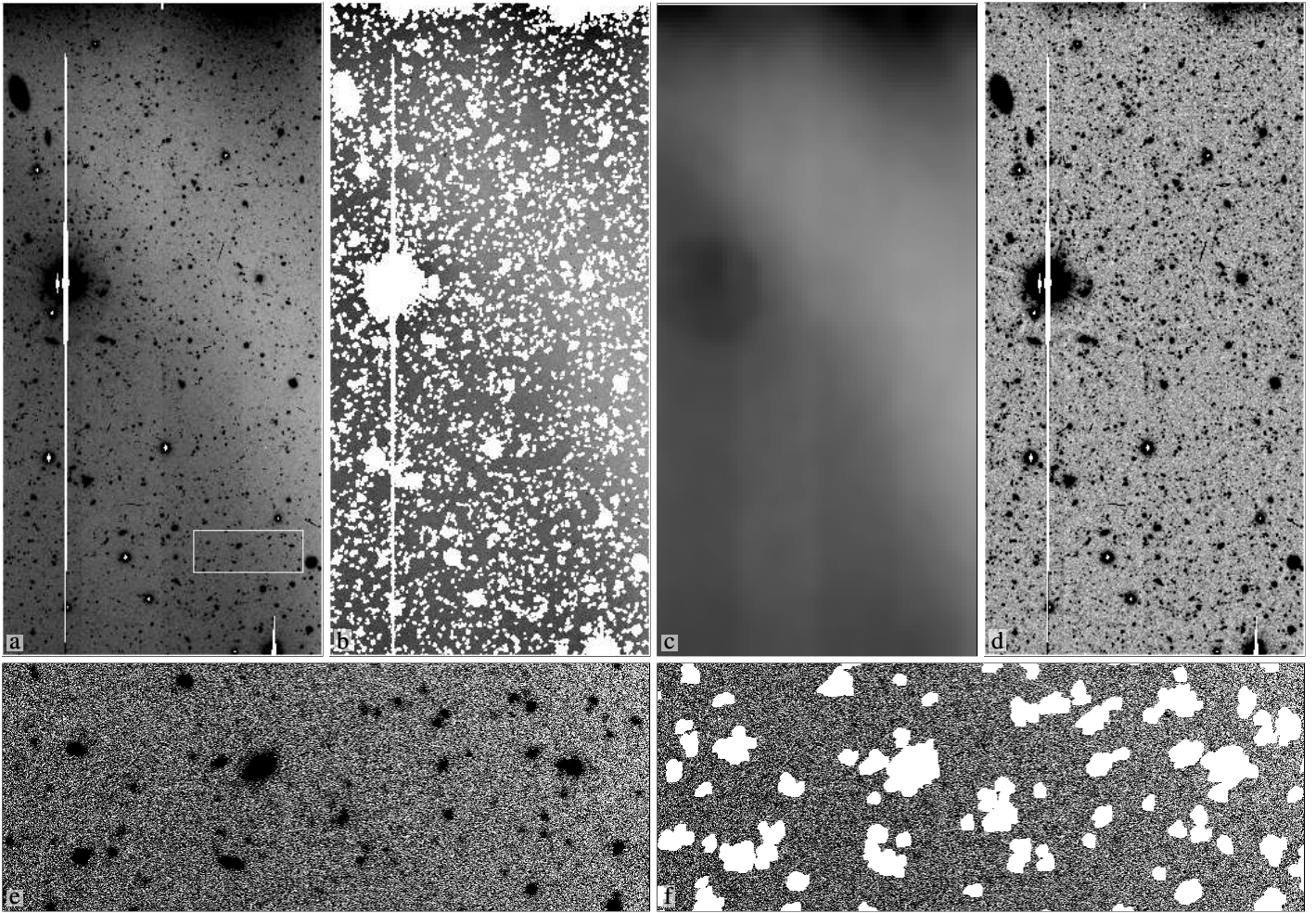}
      \fi
      \caption{\label{largeimage} Applying \textsf{NoiseChisel} on a
        raw image prior to Sky subtraction from Subaru SuprimeCam. The
        image is $\scccdnaxisa\times\scccdnaxisb$ pixels. (a) The raw
        image after bias subtraction, dark subtraction and flat
        fielding have been applied. (b) All the detected regions in
        the image have been removed to expose the Sky. (c) The
        interpolated and smoothed Sky value defined as the average of
        undetected regions over a mesh grid. (d) The Sky value on each
        mesh is subtracted from the respective region of the
        image. Notice how the gradients in the input image have been
        mostly removed. (e) Enlarged image of white box in (a). (f)
        Zoom-in of the same region in (b). The color truncation values
        in (a)--(c), (e), and (f) are exactly the same; for (d) they
        were set such that the bright star has the same apparent
        diameter. The white pixels in (a) and (d) are masked pixels.}
    \end{figure*}

  Real astronomical images are not as small and clean as the examples
  in the figures in Section \ref{noisechisel}. They are often much
  larger or might contain masked pixels. The former often causes
  variations in background and noise (see Figure
  \ref{largeimage}). These gradients do not necessarily have to be
  created by the background, but they might be due to processing, for
  example, bad fittings in flat fielding. Even if the image is very
  clear, and the targets are already accurately defined, the
  statistics when using such small postage stamps will not be
  accurate. By statistics we mean the number of false detections and
  clumps in the non-detected region that go into calculating the
  {\small S/N} quantile; see Section \ref{removenoise} and
  \ref{clumpseg}. Therefore much larger postage stamps, for example,
  from surveys, have to be used. As discussed in Section
  \ref{aboutimages}, the input {\small FITS} images of all the figures
  so far were much larger than those shown here.

  The image is considered to be a grid of meshes. All the steps
  explained in Section \ref{noisechisel} can be applied independently
  on each mesh. Since the operation on each mesh is completely
  independent of the rest, parallel processing can be applied to
  significantly improve processing time. This is only possible because
  all the methods in \textsf{NoiseChisel} are non-parametric.

  Some procedures are directly linked to the potential gradients in
  the image, for example, convolution (Section \ref{liconv}),
  thresholds (Section \ref{lithresh}), and Sky subtraction (Section
  \ref{lisky}), while others are only defined after the gradients have
  been removed and need a large area for statistical accuracy, for
  example, removing false detections (Section \ref{removenoise}) and
  finding clumps (Section \ref{clumpseg}). To be able to adequately do
  both jobs, two mesh sizes are defined: a small mesh specified by
  \texttt{\small -{}-smesh} for the former class of operations and a
  large mesh specified by \texttt{\small -{}-lmesh} for the
  latter. Each specifies the sides of a square mesh. As explained in
  Section \ref{aboutimages}, in the examples of Section
  \ref{noisechisel}, \texttt{\small -{}-lmeshsize=\lmeshsize} (so one
  mesh covers the displayed postage stamps) and for Figure
  \ref{largeimage} it is set to \texttt{\small \scccdlmesh}. In all
  the examples of this paper, \texttt{\small
    -{}-smeshsize=\smeshsize}.

  \subsection{Convolution}\label{liconv}

    Image convolution is mostly done using the discrete Fourier
    transform in the frequency domain
    \citep[see][]{gonzalezwoods}. When the convolution kernels are
    large, for example, when the image {\small PSF} is used, this
    technique provides significant performance benefits. However, it
    fails on the edges of the image. Spatial domain convolution, on
    the other hand, can be corrected to account for the edge effects
    or masked pixels and when the kernels are small it can be even
    faster. The kernels in \textsf{NoiseChisel} are small so added to
    its extra capability, spatial domain convolution is used (see the
    Convolve section of the {\small GNU} Astronomy Utilities manual
    for a complete explanation).

  \subsection{Interpolation and Smoothing}\label{interpsmooth}
    Some meshes will not be able to provide an input in the grid, for
    example, because of a large object that is larger than the
    mesh. Therefore interpolated values will be used for unsuccessful
    meshes, and finally, the mesh grid is smoothed. We will discuss
    the case for the quantile threshold (Section \ref{lithresh}) and
    Sky value (Section \ref{lisky}) more accurately here.

    Following the non-parametric nature of \textsf{NoiseChisel},
    functional interpolation techniques like spline or cubic
    interpolation will not be used. They can result in strong outliers
    on the edges or corners like those in Figure \ref{SEback}(a)--(c)
    and (l). To interpolate over each blank mesh, the median value of
    the nearest \texttt{\small -{}-numnearest} acceptable meshes is
    used. The nearest non-blank neighbors are found efficiently
    through a breadth-first search strategy in graph theory. In such a
    median interpolated grid, the result will not be smooth. Hence an
    average filter\footnote{The value of each mesh is replaced with
      the average of it and its 8 connected neighbors (see Figure
      \ref{StructElem}(b)).} is applied to the interpolated grid to
    make smoother variations between neighboring meshes. The final
    interpolated grid for the Sky value can be seen in Figure
    \ref{largeimage}(c).

    \subsubsection{Threshold}\label{lithresh}
      The quantile threshold of Section \ref{thresh} can only have a
      comparable value between all meshes, if the signal in the mesh
      is not significant compared to the noise. Otherwise, with more
      signal contributing to the mesh, the pixel distribution in the
      mesh will become more skewed and thus the quantile value found
      will also shift and not be comparable between different
      meshes. In order to find the threshold, the novel algorithm of
      Appendix \ref{findingmode} is used to find the mode of the
      image. The mode acts as a gauge for the amount of data that is
      mixed with the noise in that mesh.

      With more signal contributing to the mesh, the average, median,
      and mode will shift (with decreasing rates) to the positive (see
      Figure \ref{dataandnoise} and Appendix
      \ref{findingmode}). Exploiting this property, the significance
      of signal in noise of a mesh regardless of its morphology can be
      assessed using the difference between the median and the mode
      such that the closer the mode is to the median, the less
      significant signal there is in the mesh. Through the parameter
      \texttt{\small -{}-minmodeq}, the user can set the minimum
      acceptable quantile for the mode. In this paper it is
      \texttt{\small \minmodeq}. Note that the skewness induced by
      convolution, further skews the distribution, or, distances the
      mode and median (see Section \ref{conv} and Figure
      \ref{convolution}). Hence the \texttt{\small \minmodeq} value is
      a very strong constraint.

    \subsubsection{Sky and $\sigma_{sky}$}\label{lisky}
      As defined in Section \ref{sky}, the Sky value is the average of
      undetected pixels in each mesh and its error is their standard
      deviation.  Figure \ref{largeimage}(a) shows an example Subaru
      Telescope SuprimeCam image that has undergone initial processing
      after flat fielding and prior to Sky subtraction. The visible
      gradients in the input image are most probably due to problems
      in flat fielding rather than actual variation in the
      Sky. Finding the cause of this gradient is very important, but
      beyond the scope of this paper. No matter how it was created, a
      good detection algorithm has to be as resilient to such
      gradients as possible. This particular image was intentionally
      chosen to demonstrate \textsf{NoiseChisel}'s ability to account
      for such gradients.

      Figure \ref{largeimage}(b) shows the detected objects removed
      from the input image. The image is covered with a mesh grid of
      size \texttt{\small -{}-smesh}. The Sky on each mesh is found
      from the average of the undetected pixels in that mesh. Only
      those meshes with a sufficiently large fraction of undetected
      pixels are used. If too much of the mesh area is covered with
      detected objects, that mesh will not be used because of the
      potential fainter wings penetrating to the undetected regions;
      compare Figures \ref{det}(a.1) and (a.8). The fraction can be
      set by \texttt{\small -{}-minbfrac}, in this paper it is set to
      \texttt{\small \minbfrac}.

      Cosmic rays can be a problem for determining the average and
      standard deviation. For example, long exposures in raw
      \emph{\small HST}/{\small ACS} images are heavily peppered with
      cosmic rays (see Figure \ref{CRinnoise} for a normal
      example). Since cosmic rays are very sharp and can be very
      small, some will not be detected by the detection
      algorithm. Therefore a simple calculation of the average and
      standard deviation will be significantly biased if cosmic rays
      are present. To remove the effect of cosmic rays,
      $\sigma$-clipping with the same convergence-based approach as
      \textsf{SExtractor} is used (see Section
      \ref{sigmaclipping}). Note that by this step the mode and median
      are approximately equal, so that $\sigma$-clipping is just used
      to remove the effect of cosmic rays on the average and standard
      deviation.

    \begin{table*}[t]
      \footnotesize
        \begin{tabularx}{\textwidth}{L{0.16\linewidth} C{0.05\linewidth} C{0.05\linewidth} C{0.12\linewidth} C{0.12\linewidth} C{0.12\linewidth} C{0.12\linewidth} C{0.1\linewidth}}
    \hline\hline

    Figure Label & Median & Mode & $\sigma$-clip & $\sigma$-clip &
    \textsf{SExtractor} & \textsf{NoiseChisel} & \textsf{NoiseChisel}
    \Tstrut\\

    & & & Converge & (4,5) & Average of & Average of & $N_f$
    (Section \ref{A_purity})\\

    & & & & & Undetected & Undetected &  \\

    \hline

    \ref{NCsensitivity}(a) and \ref{SEsensitivity}(a) & $\senamed$ &
    $\senamode$ & $\senascconv$ & $\senascfixed$ & $\senasemean$ &
    $\senancmean$ & $\senanfalse$ \\

    \ref{NCsensitivity}(b) and \ref{SEsensitivity}(b) & $\senbmed$ &
    $\senbmode$ & $\senbscconv$ & $\senbscfixed$ & $\senbsemean$ &
    $\senbncmean$ & $\senbnfalse$ \\

    \ref{NCsensitivity}(c) and \ref{SEsensitivity}(c) & $\sencmed$ &
    $\sencmode$ & $\sencscconv$ & $\sencscfixed$ & $\sencsemean$ &
    $\sencncmean$ & $\sencnfalse$ \\

    \ref{NCsensitivity}(d) and \ref{SEsensitivity}(d) & $\sendmed$ &
    $\sendmode$ & $\sendscconv$ & $\sendscfixed$ & $\sendsemean$ &
    $\sendncmean$ & $\sendnfalse$ \\

    \ref{NCreal}(a) and \ref{SEreal}(a) & $\realamed$ & $\realamode$ &
    $\realascconv$ & $\realascfixed$ & $\realasemean$ &
    $\realancmean$ & {\small N/A} \\

    \ref{NCreal}(b) and \ref{SEreal}(b) & $\realbmed$ & $\realbmode$ &
    $\realbscconv$ & $\realbscfixed$ & $\realbsemean$ &
    $\realbncmean$ & {\small N/A} \\

    \ref{NCreal}(c) and \ref{SEreal}(c) & $\realcmed$ & $\realcmode$ &
    $\realcscconv$ & $\realcscfixed$ & $\realcsemean$ &
    $\realcncmean$ & {\small N/A} \\

    \ref{NCreal}(d) and \ref{SEreal}(d) & $\realdmed$ & $\realdmode$ &
    $\realdscconv$ & $\realdscfixed$ & $\realdsemean$ &
    $\realdncmean$ & {\small N/A} \\

    \ref{NCreal}(e) and \ref{SEreal}(e) & $\onedgemed$ & $\onedgemode$ &
    $\onedgescconv$ & $\onedgescfixed$ & $\onedgesemean$ &
    $\onedgencmean$ & $\onedgenfalse$ \\

    \ref{SEthresh}(a) & $\onelargemed$ & $\onelargemode$ &
    $\onelargescconv$ & $\onelargescfixed$ & $\onelargesemean$ &
    $\onelargencmean$ & $\onelargenfalse$  \\

    \ref{SEthresh}(b) & S/A & S/A & S/A & S/A & $\onelargetonesemean$
    & S/A & S/A\\

    \ref{SEthresh}(c) & S/A & S/A & S/A & S/A & $\onelargethalfsemean$
    & S/A & S/A\\

    \ref{SEthresh}(d) & S/A & S/A & S/A & S/A &
    $\onelargettenthsemean$ & S/A & S/A\\

    \hline
  \end{tabularx}

      \caption{\label{analysistable} \footnotesize Sky value on the
        figures in units of counts. counts $s^{-1}$ data multiplied by
        $10^4$s. For mock images, the background ($=\backcount{}e^-$)
        was subtracted. $LS$: low symmetricity, see Appendix
        \ref{findingmode}. S/A: same as above. N/A: not applicable.}
    \end{table*}

      The interpolated and smoothed (see Section \ref{interpsmooth})
      Sky map can be seen in Figure \ref{largeimage}(c), each pixel in
      this image represents a $\smeshsize\times\smeshsize$ pixel
      mesh. Finally in Figure \ref{largeimage}(d), the Sky value for
      each mesh is subtracted from the pixels in that mesh. The
      gradients and differences between the different {\small CCD}
      chips have been significantly suppressed. Notice that even with
      this small mesh size some of the strong top right gradient still
      remains.

\section{Analysis}\label{analysis}

  \textsf{NoiseChisel} is based on detecting an object deeply buried
  in noise using its fainter parts (see Section \ref{noisechisel}). In
  contrast, existing techniques rely on detecting the brightest
  regions and growing those (based on a priori models) to include the
  fainter parts (see Appendix \ref{existingmethods}). Hence these
  methods are fundamentally different. The detection accuracy is
  analyzed here using three fundamental and basic statistics that can
  be done with mock profiles in mock noise. Comparing the Sky and
  background value when the latter is accurately known (in a mock
  image) is the most basic test to evaluate the success of a detection
  algorithm.  This comparison is done in Section \ref{A_sky}. Purity
  is a quantitative measure to assess the contamination of false
  detections that remain in the final result. In Section
  \ref{A_purity} purity is defined and used for an analysis of false
  detections in the faintest limits. Finally, in Section
  \ref{A_magdisp}, the faint end magnitude dispersion is studied. The
  statistical significance of the imposed requirements on the objects
  detected as true is discussed in Section \ref{restrictions}.

  While being demonstrative for a ``proof-of-concept'' paper like this
  one, tests using mock objects or noise can never simulate all real
  circumstances accurately.  Thus \textsf{NoiseChisel} is tested using
  only real data without any modeling in a companion paper by
  M. Akhlaghi et al. (2015, in preparation).  Its accuracy is further
  evaluated, for example completeness, purity, number counts and etc
  on real data in that work. The details of the real data set,
  necessary modifications to the existing pipeline, tests, and an
  analysis of their results is beyond the scope of this definition or
  ``proof-of-concept'' paper.

  \subsection{Sky Value}\label{A_sky}

    The Sky is the average of undetected regions. Therefore the
    success of a detection algorithm can be assessed with the effect
    of undetected objects on the Sky value. The more successful the
    detection algorithm is, the closer the Sky value will be to the
    background value, which is $\backcount{}e^-$ for all the mock
    images (see Section \ref{sky}). There are two types of ``faint''
    pixels that can be left undetected. (1) When the brightest pixels
    of an object are faint, for example, Figure \ref{det}(c.2). (2)
    When the object is bright enough to be detected but has wings that
    penetrate into the noise very slowly (for example Figure
    \ref{det}(a.2)). The failure to detect both will be imprinted into
    the average of undetected regions of the image.

    Table \ref{analysistable}, has all the relevant data for this
    comparison. Except for the last column, all columns in Table
    \ref{analysistable} show the pixel statistics of images shown in
    the figures of column 1. In these statistics, only the pixels
    within the $200\times200$ pixel box are used and not the whole
    image (see Section \ref{aboutimages}). As $\sigma$-clipping is
    still extensively used in existing pipelines (see Section
    \ref{sigmaclipping}) in Table \ref{analysistable} the results of
    the two major $\sigma$-clipping strategies are also included, by
    convergence and (\issigclipmultip,\issigclipnum). In
    \textsf{SExtractor}, which uses the former, the average is used as
    the Sky value and in the latter, the median is used, so we take
    the same approach here.

    The $\sigma$-clipping results show that due to the skewness caused
    by galaxies (which do not have a sharp cutoff), none are
    significantly different from the median. Since the final average
    is used in the former, its reported values can be even larger than
    the initial median. The correlated noise, which acts like a
    convolution and skews the distribution, causes this difference
    with the mode to be much larger in the real (processed) images of
    Figures \ref{NCreal} and \ref{SEreal} (a)--(d) than the mock
    images.

\begin{figure}[t]
      \centering
      \ifdefined\makeeps
      \input{./tex/NCsensitivity}
      \else
      \includegraphics{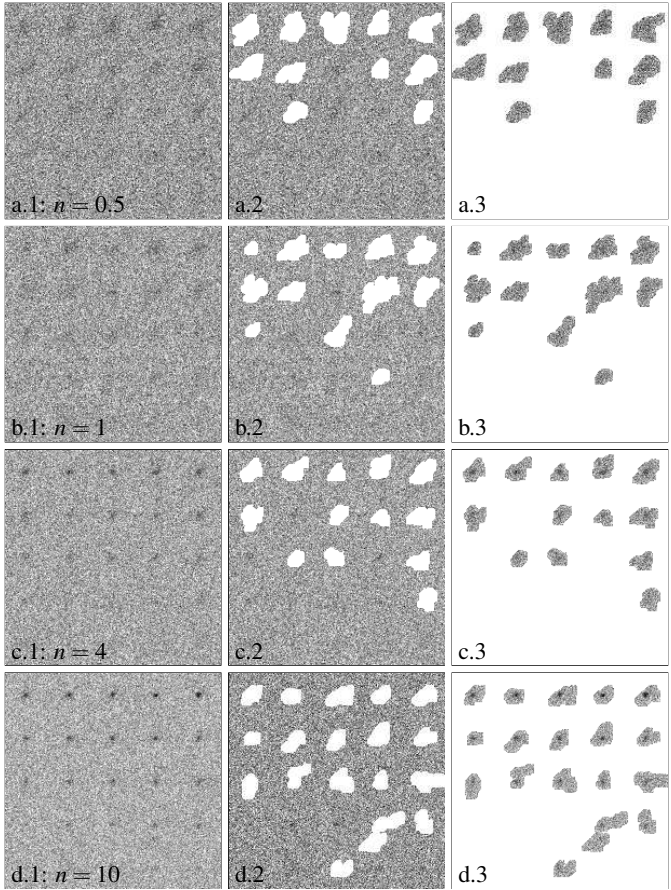}
      \fi
      \caption{\label{NCsensitivity} \textsf{NoiseChisel} sensitivity
        test. In each image there are 25 mock profiles, equally spaced
        (see Figure \ref{det}(b.1) and (c.1) for the images prior to
        adding noise for the cases of $n=\sennc$ and
        $n=\sennb$). Except for their total magnitude, all profiles in
        each image have the same parameters, $r_e=\senre$ pix,
        $\theta=\sentheta^\circ$, $q=senq$. The S\'{e}rsic index in
        each input image is different. The profiles on the bottom left
        are the faintest with $\senfaintmag$ magnitude while those on
        the top right are the brightest with $\senbrightmag$
        magnitude. The second column shows the regions of the detected
        objects masked from the input image. The third column is the
        inverse of the second.}
    \end{figure}

\begin{figure}[t]
      \centering
      \ifdefined\makeeps
      \input{./tex/NCreal}
      \else
      \includegraphics{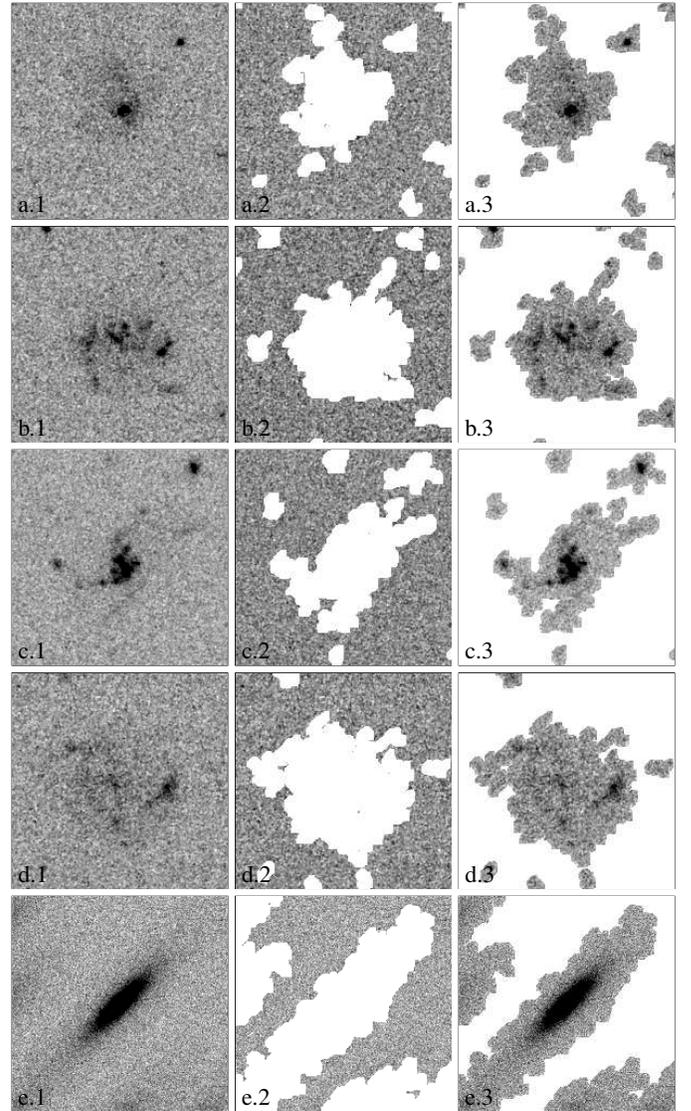}
      \fi
      \caption{\label{NCreal} \textsf{NoiseChisel} applied to
        realistic conditions. (a)--(d) Four real galaxies. (e) Six
        mock profiles with the same parameters of Figure
        \ref{dataandnoise}(b) and position angle of
        $\theta=\onedgepa^\circ$. Five of the profiles are centered
        $\onedgedist$ pixels outside the bottom and left edges of the
        image; see Section \ref{aboutimages}. The columns are the same
        as those in Figure \ref{NCsensitivity}. The truncation count
        (where black pixels are defined) in (e.1) is half that of
        Figure \ref{dataandnoise}(b) to emphasize the faint wings of
        the profiles outside the image edge. }
    \end{figure}

    In order to check the sensitivity of \textsf{NoiseChisel}, it is
    applied to a set of mock and real images of Figures
    \ref{NCsensitivity} and \ref{NCreal} with the default parameters
    discussed in Section \ref{noisechisel} and Section \ref{largeimg}
    and fully listed in Appendix \ref{NCconfig}. The values for the
    S\'{e}rsic index in Figure \ref{NCsensitivity} ($n=\senna, \sennb,
    \sennc$ and $\sennd$) were chosen to approximately cover the range
    of observed galaxy profiles \citep[for example,
      see][]{lackner12}. The objects in Figure \ref{NCreal} are
    \emph{\small HST}/{\small ACS}, {\small F814W} images from the
         {\small COSMOS} survey of what are thought to be three
         $z\sim1.5$ star-forming galaxies, so that they are imaged in
         rest-frame {\small UV}. The same input images are given to
         \textsf{SExtractor} (Figures \ref{SEsensitivity} and
         \ref{SEreal} in Appendix \ref{existingmethods} with the
         configuration file shown in Appendix \ref{SEconfig}) and the
         results are reported here. Some studies use the average of
         undetected pixels when \textsf{SExtractor}'s detections have
         been removed. Therefore in column 6 of Table
         \ref{analysistable} the average pixel value when all
         \textsf{SExtractor} detections have been masked to
         $\rkron{}r_k$ is reported.

    The mock images (with a number in the last column) show that
    compared to $\sigma$-clipping, removing \textsf{SExtractor}'s
    detected objects has been more successful in approaching the real
    background value ($0e^-$), but the average of the undetected
    regions are still a significant overestimation. The last 4 rows of
    Table \ref{analysistable}, which successively mask more of the
    bright profile, show that this is not only due to undetected
    objects, but also the fainter parts of bright objects. The last
    row shows that when too much of the image is masked, the average
    of non-detected pixels goes below the real Sky value. This is
    reasonable because localized high valued noise pixels are
    systematically removed.

    Over-estimating the Sky value results in a systematic
    underestimation of the total count of the detected objects. In the
    most diffuse and low surface brightness cases, a severe Sky
    overestimation can even result in a negative total count for the
    targets. Some of the individual pixels of a real object can be
    below the Sky value (Section \ref{thresh}) but if the Sky value is
    accurately found, the total count of a real detection can never be
    negative. If it occurs, it can only be due to the overestimation
    of the Sky. This problem can only occur in the existing signal
    based approach to detection where any pixel above the threshold is
    assumed to be the top of a profile. Negative total counts have
    been reported and extensively used in the {\small SDSS} survey,
    for example. However, only the problem of not being able to
    measure the logarithm of a negative value (for conversion to a
    magnitude) was addressed \citep[see][]{lupton99}. The source of
    the problem, which is an overestimation of the Sky, was not
    sought. In \textsf{NoiseChisel}, if the numerator of any of the
    {\small S/N} measurements (Equations
    (\ref{tSNeq})--(\ref{bordsneq})) becomes zero or negative, that
    detection, clump, or river is discarded.

    For the mock galaxies, where the background is accurately known,
    when the area covered by a diffuse source is the greatest (Figures
    \ref{NCsensitivity}(a), \ref{NCreal}(e) and \ref{SEthresh}(a) and
    (b)), \textsf{NoiseChisel}'s detection algorithm is
    $\senasencfrac$, $\onedgesencfrac$, $\onelargettsencfrac$, and
    $\onelargetosencfrac$ times more successful than
    \textsf{SExtractor}.

  \begin{figure*}[t]
    \ifdefined\makeeps
    \input{./tex/purity}
    \else
    \includegraphics[width=\linewidth]{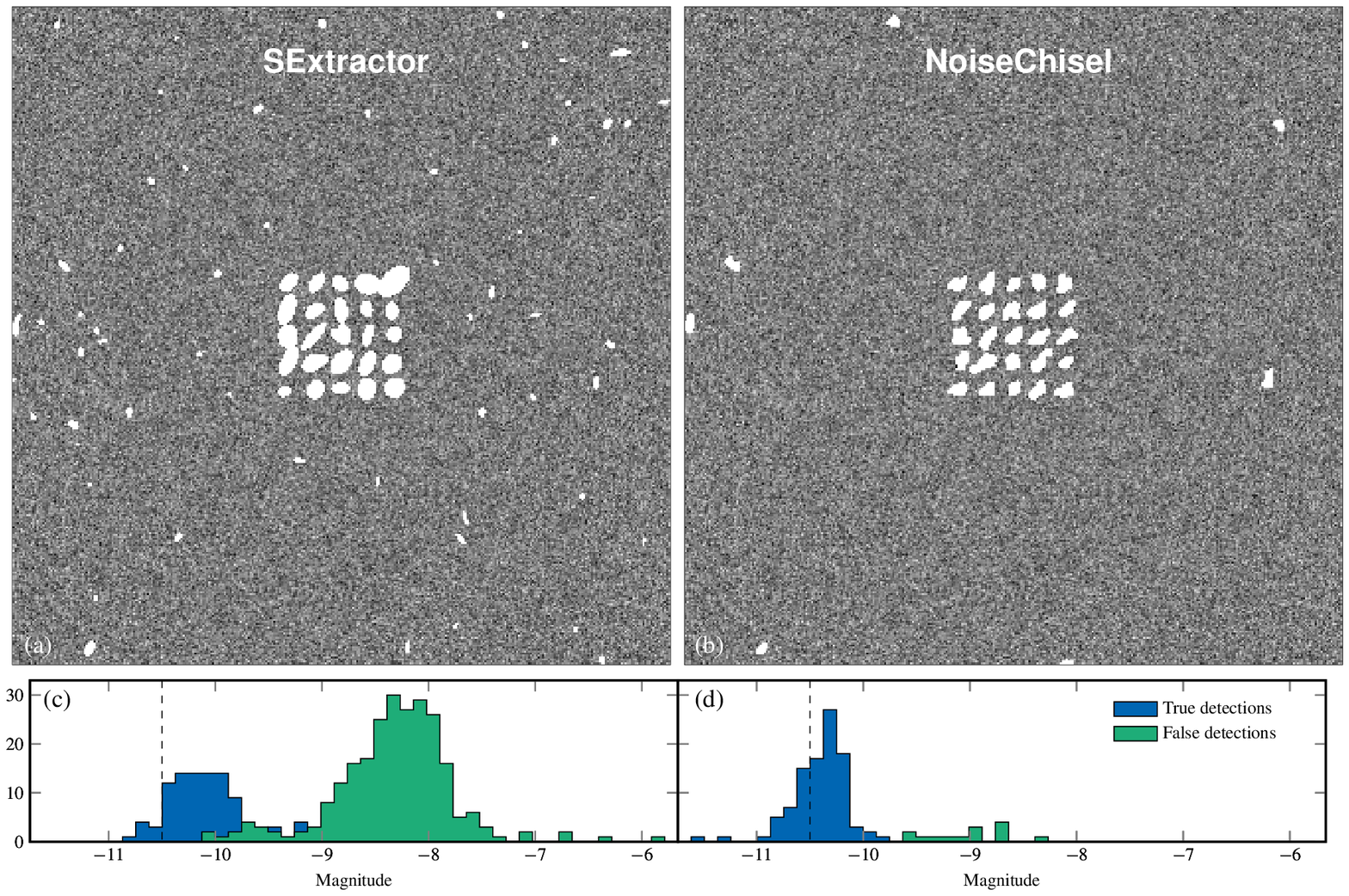}
    \fi
    \caption{\label{purity} Comparison of purity (false detections)
      and dispersion in measured magnitude when completeness is 1 for
      identical and very faint $n=\purityn$ profiles. (a) and (b) 25
      profiles are located in the central $200\times200$ region, so
      any detection outside of it is false, see Section
      \ref{aboutimages}. Completeness is constrained to 1 for both
      techniques. \textsf{SExtractor} and \textsf{NoiseChisel} have
      $\sepuritydemonfalse$ and $\ncpuritydemonfalse$ false detections
      respectively. The two histograms (c) and (d) show the
      distribution of magnitude (with zeropoint magnitude of 0) for
      the true and false detections for the respective technique. The
      dashed line shows the true magnitude of all the profiles. The
      histograms are derived from four runs (100 true detections) on
      an identical no-noised image but with different simulated noise
      seeds. (a) and (b) had the most number of false detections for
      each method.}
  \end{figure*}

    For Figures \ref{NCreal}(e) and \ref{SEreal}(e), Table
    \ref{analysistable} shows the edge and crowded field effects on
    both \textsf{SExtractor} and \textsf{NoiseChisel}. Compared to the
    other mock images, \textsf{NoiseChisel}'s Sky value is
    systematically higher in this case. This is not due to the edge
    but due to the six large $n=\onelargen$ profiles that exist in
    this image (see Figure \ref{dataandnoise} (b.1) for the complete
    pixel coverage of one of those six profiles). If \texttt{\small
      -{}-smesh} was set to a larger size to find the threshold (see
    Section \ref{lithresh}), the meshes covered by the objects would
    be ignored and \textsf{NoiseChisel}'s final Sky value would be
    lower. However, Figure \ref{largeimage}(d) shows that even such a
    small \texttt{\small -{}-smesh} was not enough to completely
    remove some strong gradients. Therefore this higher Sky value can
    be corrected if there are no gradients in the image. Only one
    value for the full displayed region was discussed here. In
    practice, \textsf{SExtractor} interpolates over such regional
    values to find a background value for each pixel.  Section
    \ref{SEbackinterp} discusses the process and shows the histogram
    of the pixel by pixel Sky values used.

  \subsection{Purity}\label{A_purity}

    Assume that we have made mock profiles in mock noise where the
    number and positions of the mock objects buried in the noise are
    already known. By construction we know there is no other source of
    signal in such an image. Take $N_f$ to be the number of false
    detections, $N_t$, the the number of true detections, $N_{T}$, the
    total number of detections, and $N_{I}$ the total number of input
    mock profiles. Therefore $N_f+N_t=N_{T}$. The completeness, $C$,
    and purity, $P$, of a given detection algorithm with a given set
    of parameters are defined as

    \begin{equation}
      C\equiv\frac{N_t}{N_{I}},\quad\quad
      P\equiv\frac{N_t}{N_T}=1-\frac{N_f}{N_{T}}.
    \end{equation}

    Completeness is commonly measured with mock profiles, for example,
    mock point sources or mock galaxies. The results from such mock
    objects are not realistic. This is because, as discussed
    throughout this paper, real galaxies display very diverse
    morphologies and do not generically satisfy the clean elliptical
    and radial profiles that can be modeled. Therefore any result
    based on mock objects will overestimate $C$. When the detection
    algorithm also depends on profiles having simple shapes with a
    clearly defined center and radially decreasing profile, the
    systematic bias is exacerbated. Purity, on the other hand, deals
    with false detections which don't require modeling.

    Without true detections, purity will be $0$ since $N_t=0$ or
    $N_f=N_T$. In general, for a given set of parameters in any
    detection algorithm, $C$ and $P$ are anti-correlated: as purity
    increases, completeness decreases and vice versa. Therefore
    neither can be studied independently. For example, in
    \textsf{SExtractor} decreasing the threshold in Figure
    \ref{SEsensitivity} will allow the detection of more of the faint
    mock profiles therefore increasing completeness. Simultaneously,
    however, the number of false detections will increase (see Figure
    \ref{SEthresh}, for example) and thereby decreasing the purity of
    the output.

    The methodologies of \textsf{NoiseChisel} and \textsf{SExtractor}
    are fundamentally different, therefore comparing their purity
    should be done carefully. To have a reasonably objective measure
    of purity in this case, the only way is to constrain completeness
    between the two. Therefore a mock image that has $C=1$ with the
    default parameters of \textsf{NoiseChisel} is created. All the
    parameters of \textsf{SExtractor} are fixed to Appendix
    \ref{SEconfig}, except \texttt{\small DETECT\_THRESH} which is
    decreased until $C=1$ is achieved in \textsf{SExtractor} too. This
    strategy is chosen based on the advice of \textsf{SExtractor}'s
    manual to keep the sensitivity only related to \texttt{\small
      DETECT\_THRESH}. The result will not be independent of the mock
    profiles used to define completeness. The sharper the profiles,
    the fewer faint pixels will be needed to reach $C=1$. Hence an
    ideal test would be on the flattest profiles or those with
    $n=\purityn$, which both \textsf{SExtractor} and
    \textsf{NoiseChisel} show as their lowest completeness (see
    Figures \ref{NCsensitivity} and \ref{SEsensitivity}).

    To get an accurate measure of purity, a large area of the image
    has to have no data. The mock image is very similar to the input
    image of Figures \ref{NCsensitivity}(a.1). Recall that $96\%$ of
    the area of all the mock images in this paper is blank noise
    (Section \ref{aboutimages}). The only difference is that all 25
    profiles have the same total $\puritymag$ magnitude. This is done
    to decrease the effect of random noise positioning. The fixed
    magnitude of the mock profiles is set to be such that with the
    default parameters of Appendix \ref{NCconfig},
    \textsf{NoiseChisel} detects all 25 in the image. To make the test
    more robust, the mock noisy image is created four times, such that
    only the noise differs. The magnitude for all $25\times4=100$
    profiles was the faintest magnitude that \textsf{NoiseChisel}
    would give $C=1$ in all four images.

    \textsf{SExtractor} could not detect all the objects without
    flagging some of their magnitudes as unreliable. So ignoring the
    flags, the largest threshold that \textsf{SExtractor} also gives
    $C=1$ for all thresholds below it and for all four mock images is
    found to be \texttt{\small DETECT\_THRESH=\puritysethresh}. The
    masked results of one of the images can be seen in Figure
    \ref{purity} once masked with \textsf{SExtractor}'s detections and
    once with \textsf{NoiseChisel}'s. In total, in the four images,
    100 mock galaxies are present and in all four, completeness is
    1. Therefore the result of all four images can be reported in one
    purity value for \textsf{SExtractor} ($P_s$) and
    \textsf{NoiseChisel} ($P_n$).

    \begin{displaymath}
      P_s=\frac{25+25+25+25}
               {\sepuritya+\sepurityb+\sepurityc+\sepurityd}=\sepurity
    \end{displaymath}

    \begin{displaymath}
      P_n=\frac{25+25+25+25}
               {\ncpuritya+\ncpurityb+\ncpurityc+\ncpurityd}=\ncpurity
    \end{displaymath}

    Even though the threshold used in \textsf{NoiseChisel} is so much
    lower than the smallest possible threshold to \textsf{SExtractor}
    (which is the Sky value), this result shows that the purity of
    \textsf{NoiseChisel} is about three times higher in detecting
    real, faint objects with the parameters used (see Appendices
    \ref{SEconfig} and \ref{NCconfig}). In the last column of Table
    \ref{analysistable} the number of false detections ($N_f$) in all
    the mock images is reported. Recall that like Figure \ref{purity},
    all mock images are also covered by Gaussian noise for $96\%$
    their area. With exactly the same input parameters, the reported
    $N_f$ values agree well with those reported in this test.

    As discussed in Sections \ref{removenoise} and \ref{clumpseg},
    \textsf{NoiseChisel} is very robust in dealing with correlated
    noise that processed real images also have. However,
    \textsf{SExtractor} and all existing signal-based detection
    methods in general treat the pixels above the threshold as the top
    of a known profile; therefore they are not immune to the effects
    of correlated noise, resulting in lower purity when applied to a
    real processed image.

    \subsection{Magnitude Dispersion}\label{A_magdisp}

    All the profiles in Figure \ref{purity} are identical, however,
    Figure \ref{purity}(a) displays a large dispersion in the shapes
    and areas in the detections. This results in the roughly 2
    magnitude rage that is visible in \textsf{SExtractor}'s true
    detections (see Figure \ref{purity}(c)). It is very important to
    note that the bimodality that is observed in Figure
    \ref{purity}(c) (ignoring color) is due to the fact that all mock
    profiles were identical. In a real image, the faintest true
    detections can have a variety of inherent profiles. Therefore the
    distribution will be unimodal with negative skew. The net result
    is a common plot in existing catalogs (for example see Figure 10
    in \citet{XDFsurvey}). On the contrary, \textsf{NoiseChisel}'s
    true detections all have roughly the same area in Figure
    \ref{purity}(b) and thus the range in the measured magnitude for
    true detections in Figure \ref{purity}(d) is about half that of
    \textsf{SExtractor}. The two \textsf{NoiseChisel} detections with
    magnitudes brighter than $-11$ are the result of four of the
    profiles being detected as two detections.

  \subsection{Imposed Restrictions on the Signal}\label{restrictions}

    The proposed algorithm imposes some restrictions on the objects it
    will detect as true. Therefore it is not ``absolutely'' noise
    based or independent of \emph{any} signal parameter. Here we will
    discuss its methodological limits.  With the chosen parameters in
    this paper (listed in Appendix \ref{NCconfig}) the target has to
    satisfy the following conditions.
    \begin{enumerate}
    \item To be detected, an object has to have an area equal to or
      larger than Figure \ref{StructElem}(c), above the initial
      quantile threshold (which is below the Sky value; see Section
      \ref{thresh}) on the convolved image.
    \item To be classified as a true detection, it has to contain at
      least one connected component larger than $\detsnminarea$ pixels
      above the Sky threshold with a sufficiently large {\small S/N}
      specified from the ambient noise (see Section \ref{removenoise})
      on the actual image.
    \end{enumerate}

    As the methodological limits or requirements imposed on the target
    by a detection technique decrease, the detection algorithm becomes
    more successful because it becomes more generic. The ``success''
    of a detection algorithm can be defined as how accurate its Sky
    measurement can approximate a known background, the purity of its
    result and the scatter in its photometry as discussed above. It
    was shown that because of far fewer constraints on the target
    objects, this technique was significantly more ``successful''
    compared to \textsf{SExtractor} (with the parameters of Appendix
    \ref{SEconfig}).

    In astronomical data analysis, an {\small S/N} above five is
    usually considered as a ``solid'' result, any measurement with a
    lower {\small S/N} is usually included in an analysis with caution
    \citep[Section 1.5.2]{eimginastro}. Using ideal simulated noise we
    showed that objects and clumps with an {\small S/N} as low as
    $\dettfsmallestsnqnt$ (Figure \ref{dettf}) can be accurately
    ($\detquant\%$) detected with this technique using this set of
    parameters. Note that the Sky value used in those plots was a
    first (under-estimated) approximation, therefore the counts and
    hence the {\small S/N}s are overestimated. Hence the true limiting
    {\small S/N} is slightly less than what is reported in those
    plots.

    Through setting fewer constraints (if a similar purity level can
    be achieved) it would be possible to decrease the {\small S/N} of
    true detections even further. The extreme case would be to detect
    objects with an {\small S/N} of unity. However, the statistical
    significance of any resulting measurement (for example, the effect
    of that detection on the measured Sky, its total count, its central
    position, or any other measurements on the object) will
    correspondingly decrease to values that are no longer significant
    in any scientific/statistical measurement and analysis.

\section{Discussion}\label{discussion}
  A new method for detecting extremely faint objects and fainter parts
  of brighter objects in noise is introduced. Unlike the existing
  approach to detection where the signal is the basis, this approach
  is based on operations and calculations on the noise with
  insignificant requirements for the signal. This method is also
  non-parametric, in other words, it does not involve a functional
  regression analysis. It is therefore ideal for the study of
  nebulous/amorphous objects that are heavily immersed in noise, for
  example, galaxies.

  Galaxies can have rich dynamic histories involving internal
  processes as well as external interactions with the halo or other
  galaxies. This creates very complicated, diverse morphologies. The
  fact that we cannot image them in their full 3D glory further
  complicates their final observed image.  Therefore, imposing any a
  priori model on their shape or radial light profile during the
  detection process is a self-inflicted systematic bias in their
  study.  Modeling the light profiles of galaxies is vital to our
  understanding of the galaxies and testing our models, but it should
  be done \emph{after} detection is complete. Mixing detection and
  modeling will only bias both.

  One major argument in favor of elliptical-based detection and
  photometry techniques like the Kron and Petrosian radii as opposed
  to isophotal methods (including \textsf{NoiseChisel}) is that they
  are impervious to $(1+z)^4$ dimming and K-correction and for any
  profile, they are ``well defined.'' By well defined it is meant that
  if the same galaxy (with the same morphology) exists in multiple
  redshifts, then such methods will yield the same fraction of total
  light for all the redshifts. Ideally, when the apertures used extend
  to infinity (see \cite{grahamdriver05}), and the galaxies have no
  morphological evolution, the result would be independent of the
  noise and thus the threshold used. However, this is not necessarily
  the case in the real world.

  The interpretation of signal-based detection results depends on the
  morphology of the galaxies under study. To demonstrate this point,
  consider the simple mean in a 1D distribution as an analogy. Note
  that the Kron radius is a weighted mean. The mean is a well defined
  statistic for most distributions because it gives a unique and
  unambiguous value.  Regardless of whether the underlying
  distribution is a Gaussian or a log-normal distribution, for
  example, an unambiguous mean can be \emph{defined}, but for
  \emph{interpreting} the mean, the distributions should not be
  ignored; otherwise the analysis will be biased when such different
  distributions are involved.

  The same applies to the Kron or Petrosian radii. For any given
  surface brightness (morphology), a unique and unambiguous radius can
  be found, rendering them well defined. However, when it comes to
  interpreting what that radius physically means (in kiloparsecs or
  fraction of total light, for example), the interpretation will be
  biased when the morphology (spatial and count distribution of pixels)
  differs between the sample of galaxies. Therefore, comparing Kron
  magnitudes for the galaxies of a general population can be
  problematic. For example, all the galaxies in a (noisy) survey or
  image which can simultaneously contain irregulars, spirals, and
  elliptical galaxies.

  The Kron or Petrosian radii satisfy the redshift independence in
  studies of galaxies that are \emph{already known} to have
  approximately the same morphology; in other words, a morphologically
  selected sample where the sample has been detected independently of
  the shape of the galaxies. For example, unlike the references in
  Section \ref{intro}, galaxies might indeed have similar morphologies
  across redshifts \citep[for example, see][]{Lee2013}. However, this
  claim, or any other claims, including those in Section \ref{intro},
  about the morphology of detected sources can only be tested when the
  detection technique does not depend on the morphology as an
  assumption. In other words, if the detection technique depends on
  galaxies having a similar morphology at similar or various
  redshifts, it should not come as a surprise if galaxies with similar
  morphologies are detected.

  Isophotal detection methods like \textsf{NoiseChisel} do have the
  limitation that they use a different threshold for objects at
  different redshifts, or $(1+z)^4$ dimming. However, it should be
  considered that the Sky value and its standard deviation in the
  input image are also completely independent of such issues. These
  vital image statistics, which propagate in all subsequent
  measurements, only depend on the arriving photons in each pixel
  regardless of the physical origin of the source. Therefore imposing
  such high-level (related to physical interpretation and not the
  data) constraints on low-level measurements (to do with the data,
  regardless of later physical interpretations) is a source of
  systematic bias. This noise-based approach to detection was designed
  to create an accurate low-level measurement tool with as few
  assumptions as possible so high-level interpretations can be less
  biased.

  High-level interpretations of the outputs of low-level measurements
  can easily be done after the low-level procedure is complete. For
  example, to account for $(1+z)^4$ dimming, an individual threshold
  can be defined based on the measured redshift (a high-level product)
  for each object in a study. In the 1970s, when the Kron and
  Petrosian radii were defined, the processing power to achieve this
  level of customization was not available. Therefore such high-level
  constraints had to be imposed on the techniques that originated
  decades ago. However with the very fast computers cheaply available
  today, there is no more need for such self-inflicted biases.

  Through the particular definition of threshold (in Figure
  \ref{flowchart}, Section \ref{thresh} and \ref{lithresh}) the cycle
  of the Sky estimate depending on detection and detection depending
  on Sky are significantly weakened. In the proposed technique, the
  detection threshold and initial detection process are defined
  independently of the Sky value. An independent initial approximation
  of the Sky is found by averaging the initial undetected regions and
  is only used in classifying and removing false initial detections
  (see the flowchart of Figure \ref{flowchart}).

  Due to all the novel, low-level and non-parametric methods employed
  in this noise based detection, fainter profiles can be detected more
  successfully than the existing signal-based detection technique (at
  least with the parameters used here). The purity measurements of
  Section \ref{A_purity} show how successful this algorithm and its
  implementation have been in removing false detections, while
  detecting the very faint true profiles and the fainter parts of
  large profiles. Section \ref{A_magdisp} then showed the
  significantly smaller scatter in the photometry of those very
  diffuse and faint galaxies.

  A new approach to segment a detected region into objects that are
  heavily immersed in noise is also developed. In this approach, the
  true clumps are first found using the global properties of the
  noise, not based directly on a user input value or the particular
  object (compare Section \ref{clumpseg} with Section
  \ref{SEdeblend}). By expanding the clumps over the light profile of
  the detected region, possible objects are found and the detected
  region is segmented.

  One of the design principles of \textsf{NoiseChisel} was to decrease
  the dependency of the output on user input. To ensure that the
  results are more dependent on the image than the user's subjective
  experience. To achieve this, the true/false criteria were defined
  based on the image and not an absolute hand input value (see Section
  \ref{removenoise} and Section \ref{clumpseg}). This approach
  accounts for local variations in the noise properties or correlated
  noise that becomes a significant issue in processed data
  products. Note that by using relative measures the dependence on
  user input has significantly decreased, but not disappeared.

  This noise based approach to detection allows for the input
  parameters to require minimal customization between images of
  different instruments and in different processing stages. For
  example, in this paper, all the outputs for mock images with ideal
  noise, raw ground-based images with real noise and \emph{\small
    HST}/{\small ACS} drizzled space-based images with correlated
  noise were processed with nearly identical input parameters.

  \textsf{NoiseChisel} is designed with efficiency in mind. Therefore
  it is also very fast in processing a large image. As an example the
  image of Figure \ref{largeimage} was processed in $\nconscccdtime$
  seconds using a $\cpughz${\small GH}z, $\cpunumcore$ physical core
  {\small CPU} on a desktop computer. Until now the design and
  applicability of the concepts introduced here was the primary
  concern, but in time more efficient algorithms and methods might be
  found to increase its efficiency. For example including {\small GPU}
  functionality can significantly decrease the running time. The
  detection and segmentation techniques introduced through
  \textsf{NoiseChisel} are easily expandable to the detection of any
  signal in noise, for example, to 3D data in radio studies similar to
  \textsf{clumpfind} \citep{williams94} and 1D data like spectroscopic
  data.

  The technique proposed here for detection and segmentation along
  with the ancillary \textsf{NoiseChisel} program provides a
  fundamentally fresh and new approach to astronomical data
  analysis. The limited preliminary tests on mock profiles and noise
  in this paper show that it significantly increases the detection
  ability, with a reasonable purity and very small photometric
  scatter, to detect the fainter parts of bright galaxies with any
  shape or to find small faint objects that will be undetected with
  the existing signal based approach to detection. Both of these new
  abilities can play a major role in most branches of galaxy evolution
  or a astronomy research as whole (some applications are reviewed in
  Section \ref{intro}). Along with the new instruments such as
  \emph{James Webb Space Telescope}, the Large Synoptic Survey
  Telescope ({\small LSST}), and the Thirty Meter Telescope, which
  will be available in the next few years, this new approach to
  detection will play a very important role in current and future
  astronomical discoveries.

\ifdefined\acknowledgesection
\section*{Acknowledgments}
\else
\acknowledgements
\fi

  We are most grateful to John Silverman for providing very
  constructive comments and suggestions, particularly on the
  scientific questions that led to the creation of this
  technique. Robert Lupton's very critical assessment of this work
  during several discussions in various stages of its development were
  very constructive and we are most grateful to him. Anupreeta More
  provided some very critical and useful comments on testing,
  particularly in Section \ref{A_purity}. Paul Price kindly supplied
  us with Figure \ref{largeimage} along with his valuable opinions on
  this work. Alan Lefor reviewed this text and provided very useful
  writing style comments for which we are most grateful; he also
  patiently helped in the initial portability tests of the {\small
    GNU} Astronomy Utilities. The ApJS Scientific Editor (Eric
  Feigelson) provided valuable comments on the layout and
  presentation. The critical comments, and references introduced, by
  the anonymous referee were also very important for the final
  presentation. Mohammad-reza Khellat, Henry McCracken, Masashi Chiba,
  Marcin Sawicki, Surhud More, James Gunn, Emanuelle Daddi, Steve
  Bickerton, and Claire Lackner also provided valuable critical
  suggestions throughout the duration of this work. M.A. is supported
  by a Ministry of Education, Culture, Sports, Science and Technology
  in Japan (MEXT) scholarship. This work has been supported in part by
  a Grant-in-Aid for Scientific Research (21244012, 24253003) from
  MEXT and by the Global Center of Excellence (GCOE), Young Scientist
  Initiative A.

\appendix

\section{Existing Sky Calculation Methods}\label{existingsky}

  In existing algorithms for detection, the first step of detection,
  namely setting a threshold, is based on the Sky value (see Appendix
  \ref{existingmethods}). To approximate the Sky value prior to
  detection, researchers use multiple techniques, such as the mode,
  $\sigma$-clipping, etc. A critical analysis of the most common
  techniques for approximating the Sky value is given here.

  \subsection{Local Mode}\label{localmode}

    In a symmetric random distribution, for example, pure noise in an
    astronomical image,\footnote{The photon noise actually comes from
      a Poisson distribution which is not symmetric especially for
      very low mean ($\lambda$) values, but with the very high mean
      values ($\lambda=B=\backcount{}e^-$ from Figure
      \ref{dataandnoise}), the Poisson distribution can safely be
      considered approximately equal to a Gaussian distribution with
      $\mu=\lambda$ and $\sigma=\sqrt{\lambda}$. Images taken from the
      \emph{HST} commonly have much smaller Sky values. In such cases,
      the main source of noise is noise produced by the
      instrumentation, for example, read-out noise which can
      out-weight the non-symmetric low-$\lambda$ Poisson noise.} the
    mean, median, and mode are equal within their statistical errors.
    With the addition of signal (astronomical objects, yellow
    histogram in Figure \ref{dataandnoise}), a skewed distribution is
    formed (red histogram in Figure \ref{dataandnoise}).  The mean is
    the most highly dependent on the skewness since it shifts toward
    the positive much faster. The median is less affected since it
    will be smaller than the mean for a given positively skewed
    distribution. Finally, the mode of the distribution is the least
    affected statistic of the three.  This argument has thus led many
    to take the mode of a noisy image as the Sky value.

    There are several approaches to finding the mode of the image:
    \citet{bijaoui80} finds a functional form to the image histogram
    in the vicinity of the mode using Bayes's theorem.  \citet{kron}
    finds the mode of the pixel distribution in a
    $50\mathrm{pixel}\times50\mathrm{pixel}$
    ($20^{\prime\prime}\times20^{\prime\prime}$) box about the peak of
    each object based on the mean of the 7 bins in the histogram with
    the largest number of pixels and considered that as the Sky value.
    \citet{beard90} use a more elaborate approach in the search to
    find the mode.  The histogram is first smoothed with a moving box
    filter.  The highest peaks are found and used in a cubic fit to
    find the mode which they consider to be the Sky.

    \textsf{DAOPHOT} \citep{daophot} also uses a very similar method:
    a circular annulus with an inner radius several times the stellar
    {\small FWHM} is taken from a nearby part of the image,
    sufficiently close to the target and the mode of the pixels in
    that region is considered as the Sky value. The mode is
    approximated based on the following relation between the mean and
    median: $3\times\mathrm{median}-2\times\mathrm{mean}$. This
    particular relation between the mode, median and mean, can only
    exist when one assumes a particular pixel count probability
    distribution function ({\small PDF}) or histogram for the image
    pixels.

    \textsf{SExtractor} can be considered a hybrid, first relying on
    $\sigma$-clipping, and then on finding the mode to find the Sky
    value. The former will be discussed in Section
    \ref{sigmaclipping}. In the latter, like \textsf{DAOPHOT},
    \textsf{SExtractor} uses the following relation to find the mode:
    $2.5\times\mathrm{median}-1.5\times\mathrm{mean}$. In a recent
    re-analysis of the {\small SDSS} stripe 82 (the {\small SDSS} deep
    field), \citet{stripe822014} use a similar but more accurate
    hybrid method. The main difference is that they remove
    \textsf{SExtractor} detections first---with \texttt{\small
      DETECT\_THRESH=2} and \texttt{\small DETECT\_MINAREA=4} (see
    Section \ref{SEdetection}). If the mean is not smaller than the
    median, the mode is assumed to be found based on a similar
    relation to \textsf{DAOPHOT} between the mean and mode.

    Therefore existing methods to use the mode as a proxy for the Sky
    either employ the parametric approximations of \citet{bijaoui80},
    \textsf{DAOPHOT}, \textsf{SExtractor}, and \citet{stripe822014}
    based on an assumed function for the pixel count {\small PDF} or a
    non-parametric one \citep{kron, beard90}. It is clear from Figure
    \ref{dataandnoise} that the histogram, which can be considered a
    binned {\small PDF} can take any form of skewness depending on the
    type and positioning of astronomical data in the image. Therefore
    by assuming a fixed generic functional form for all possible
    {\small PDF}s, significant systematic errors will be induced.

    The existing non-parametric methods reviewed here rely on the
    image histogram. The results from a histogram depend on the
    bin-width and the bin positioning, especially in the vicinity of
    the mode. For example note how the peaks near the mode vary in
    Figure \ref{mirrordemo} only due to bin positioning. The method of
    \citet{kron} for finding the mode (explained above) for the
    histogram in Figure \ref{dataandnoise}(b.1), yields
    $\onelargekronmean{}e^-\pm\onelargekronstd{}e^-$, where the
    histogram is made from 200$\times$200 pixels; recall that
    \citet{kron} used a 50$\times$50 pixels box, and the area here
    with no signal is much larger than the object. Different bin
    widths would change this result, but since no generic distribution
    can be assumed for all images, any generic bin width choice would
    ultimately be arbitrary. A very accurate, non-parametric method
    for finding the mode of a distribution that does not rely on the
    histogram, but on the cumulative frequency plot of the pixels is
    introduced in Appendix \ref{findingmode}.

    Regardless of how the mode is found, such attempts at using the
    mode of a distribution as a proxy for the Sky fail to consider the
    fact that the mode of the distribution also shifts significantly
    from the actual image background depending on the data that is
    embedded in the noise. Figure \ref{dataandnoise} shows that like
    the mean and median, the mode is a biased estimator for the Sky
    value due to the data. As the fainter parts of large objects or a
    large number of faint objects cover a larger fraction of the
    image, the mode of the distribution shifts to the positive. Figure
    \ref{dataandnoise} shows how this shift increases, with mode
    values of $\txtrandomsmallmode{}e^-$ (symmetricity of
    $\txtrandomsmallmodesym$, see Appendix \ref{findingmode}),
    $\txtonelargemode{}e^-$ ($\txtonelargemodesym$) and,
    $\txtseverallargemode{}e^-$ ($\txtseverallargemodesym$),
    respectively. Recall that the background value is $0e^-$ in all
    three. See Section \ref{A_sky} for further discussion on the mode
    as compared to the true Sky.

    Reversing the argument above (that data cause a difference between
    the mode and median), we conclude that if the mode and median are
    approximately equal, there is no significant contribution of
    signal or data. In fact this argument is very important for
    \textsf{NoiseChisel} (see Section \ref{lithresh}). Note that this
    argument is only possible when the mode is found independently
    from the median as in Appendix \ref{findingmode}.

    Based on the idea in the previous paragraph we have also created
    \textsf{SubtractSky}, which is also distributed as part of the
           {\small GNU} Astronomy Utilities. It can be used for cases
           where only Sky subtraction is desired. Note that the
           argument above ``detects'' signal through its effect on the
           distance of the mode and median, but only based on pixel
           values and independently of the signal's spatial
           distribution. See the manual for more information; its
           operation is very similar to that explained in Section
           \ref{largeimg} with small modifications.

  \subsection{Removing Detections}\label{medianfilter}

    If the detection algorithm is independent of noise, this method of
    finding the Sky is the most accurate (see Section
    \ref{sky}). However, the existing detection algorithms depend on
    knowing the Sky before they run (see Section
    \ref{SEdetection}). Even so, this method is used by some
    researchers to find the Sky value, for example, \citet{sdssedr}
    used this technique over a grid to estimate the Sky in the SDSS
    survey. Bright objects, which are defined as those with at least
    one pixel above $200\sigma$, are first subtracted from the
    image. They are all assumed to have power-law wings which are
    modeled and subtracted from the image and finally a median filter
    is applied to the image in a box of side
    $100^{\prime\prime}\approx252$ pixels.  Increasing the box size
    will be a serious burden on the computational process of the
    pipeline and there will always be objects that are, collectively
    or individually, large enough to bias the result.

    \citet{mandelbaum05} reported a systematic decrease in the number
    of faint galaxies near $<90^{\prime\prime}$ of the center of
    bright objects.  It was also observed that this method
    underestimates the total count of the brighter galaxies \citep[for
      example,][]{lauer07}.  The issue was finally addressed in
    \citet{sdssdr8}. The major solution was to change the threshold to
    detect bright sources to $51\sigma$. Detected objects were
    deblended, and a linear combination of the best exponential and de
    Vaucouleurs models were subtracted.

    While the decreased threshold and new fitting functions do
    increase the number of objects that are masked, they still fail in
    the following respects. (1) Subtraction is done based on
    parametric model fitting which can potentially be severely wrong
    in the fainter outer parts of the galaxies that are the primary
    source of bias.  Such cases are when the galaxy cannot be
    characterized by a simple ellipse or when it is on the edge of the
    image (see Section \ref{rkron}).  Other cases are when the galaxy
    might not be a simple $n=1$ or $n=4$ profile.  (2) The bright and
    extended objects that do not have a pixel above $51\sigma$ are
    still a cause of systematic bias.  For example, none of the bright
    profiles in Figure \ref{dataandnoise}((c), untruncated) have a
    pixel above that extremely high threshold. The brightest pixel in
    Figure \ref{dataandnoise}(c.1) is only $\severallargemaxpix\sigma$
    above the background.

  \subsection{$\sigma$\emph{-clipping}}\label{sigmaclipping}

\begin{figure}
      \centering
      \ifdefined\makeeps
      \input{./tex/CRinnoise}
      \else
      \includegraphics{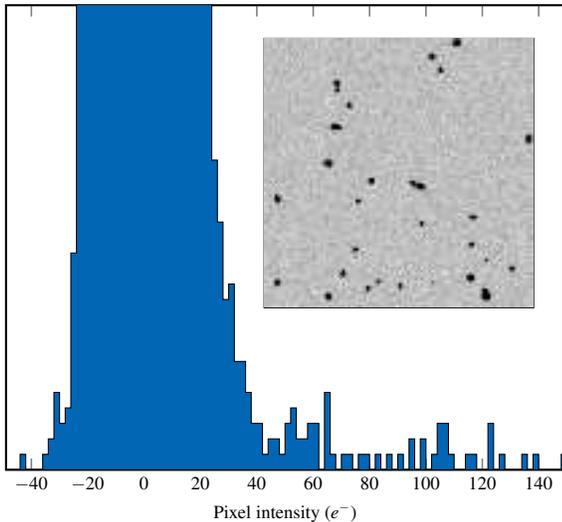}
      \fi
      \caption{\label{CRinnoise} Effect of cosmic rays on the
        histogram in a small postage stamp ($100\times100$ pixels) of
        a raw Sky subtracted \emph{\small HST}/{\small ACS}
        (\texttt{flt.fits}) image. The histogram is vertically
        truncated and the color range in the displayed image has the
        same range as the histogram. $\crinnoisenumhigher$ cosmic ray
        pixels have values above $\crinnoisemax{}e^-$ and are not
        shown in the histogram. The outlying cosmic rays can easily be
        distinguished in the histogram due to their sharp boundaries
        in contrast to the fading boundaries of galaxies, for example,
        Figure \ref{dataandnoise}(b).}
    \end{figure}

    When the outliers of a data set are sufficiently distinct from the
    majority of the (noisy) distribution, $\sigma$-clipping can be
    very useful in removing their effect.  Assuming $m$ is the
    distribution median, any data point lying beyond $m\pm a\sigma$
    can be clipped (removed) and this process can be repeated on the
    smaller data set $n$ times.  The parameters to define
    $\sigma$-clipping can thus be written as $(a, n)$.  Such
    $\sigma$-clipping is only useful if the objects (outliers) have a
    very high {\small S/N} with very sharp boundaries, for example
    cosmic rays (see Figure \ref{CRinnoise}).

    Some prominent uses of this approach in finding the Sky value in
    an image are: the \emph{\small HST} image processing pipeline
    \citep{drizzlepac} with $(4,5)$ and SDSS Stripe 82
    \citep{stripe82} with $(3,5)$ and again \citep{stripe822014} with
    $(\_$\footnote{\citet{stripe822014} do not to mention the first
      parameter.}$,20)$. As its primary tool, \textsf{SExtractor} also
    relies on this approach though the termination criterion is not a
    fixed number of times, it is the convergence of
    $\sigma$. Asymmetric clipping was proposed by
    \citet{ratnatunga84}. \citet{drizzlepac} obtains the value
    globally and uses it for the whole image; others use this
    technique to find the local Sky value and use interpolation over
    the whole image. The five vertical lines in all three examples of
    Figure \ref{dataandnoise} show the position of each iteration's
    $\sigma$-clipped median with $(4,5)$ on the mock images.  Due to
    their extreme proximity, they might not be resolved.

    Once $\sigma$ converges, \textsf{SExtractor} uses the mean as the
    Sky value. For the displayed area of the three examples of Figure
    \ref{dataandnoise}, the mean, when \textsf{SExtractor}'s
    convergence is achieved, is, respectively,
    $\dnintxtrandomsmallscconv{}e^-$, $\dnintxtonelargescconv{}e^-$,
    and $\dnintxtseverallargescconv{}e^-$. Prior to $\sigma$-clipping,
    the median of each image was: $\dnintxtrandomsmallmed{}e^-$,
    $\dnintxtonelargemed{}e^-$, and
    $\dnintxtseverallargemed{}e^-$. Hence, a simple median (no
    $\sigma$\--clipping) was a better approximation to the true
    background value ($0\pm\backstd{}e^-$). The medians of the
    $(\issigclipmultip,\issigclipnum)$ $\sigma$-clipping technique are
    shown on Figure \ref{dataandnoise}. It is clear that because the
    profiles of astronomical objects go slowly into the noise, the
    correction provided by $\sigma$-clipping is very insignificant for
    the mean and median. For the standard deviation, however,
    $\sigma$-clipping is very useful. The initial standard deviation
    of these three examples was $\dnintxtrandomsmallscfstd{}e^-$,
    $\dnintxtonelargescfstd{}e^-$ and
    $\dnintxtseverallargescfstd{}e^-$ (see Section \ref{A_sky} for a
    more complete comparison).

    \begin{figure*}[t]
      \centering
      \ifdefined\makeeps
      \include{./tex/SEback}
      \else
      \includegraphics{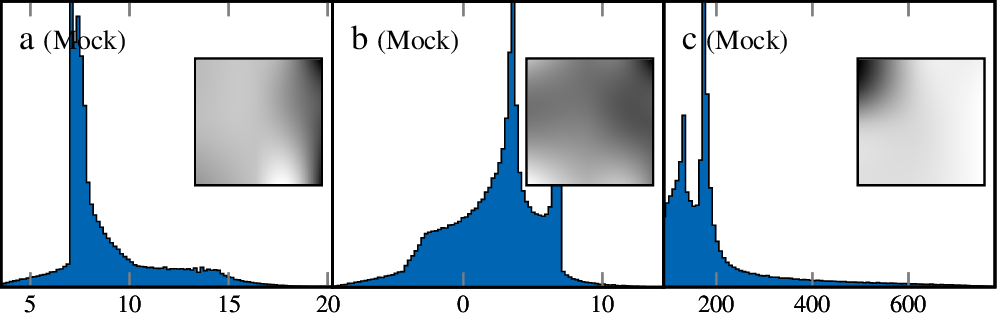}\\
      \includegraphics{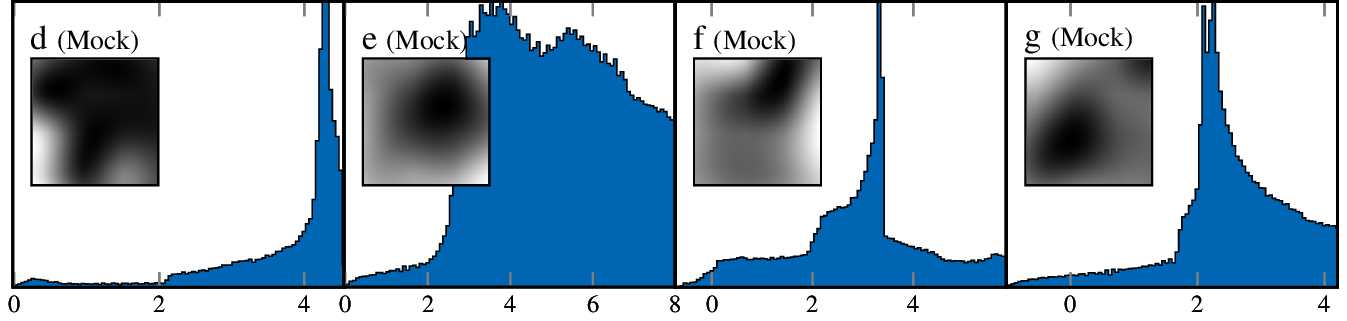}\\
      \includegraphics{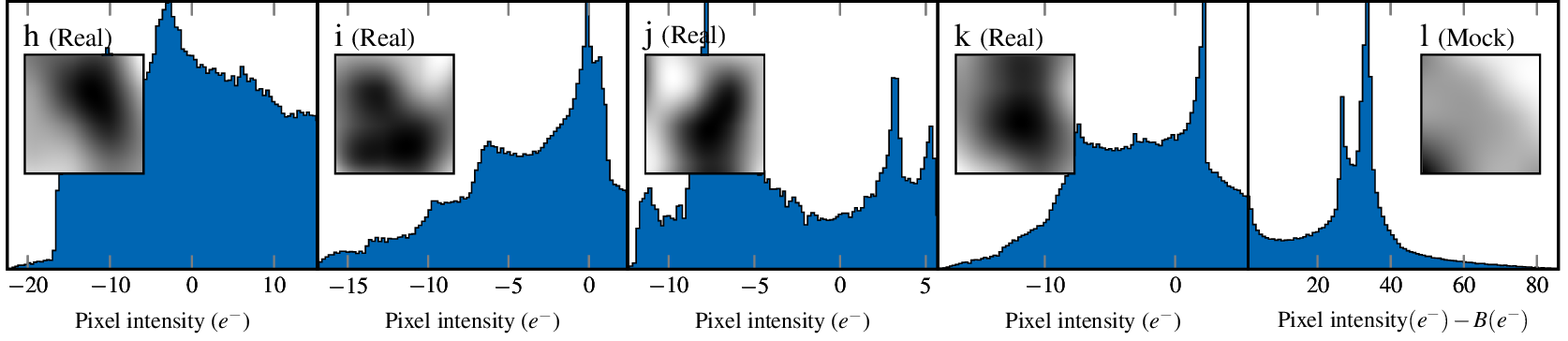}
      \fi
      \caption{\label{SEback} Sky value check images and their
        histograms, produced by \textsf{SExtractor} with the input
        parameters of Appendix \ref{SEconfig} applied to the images in
        some of the figures in this paper. First row (a)--(c): the
        three mock images of Figure \ref{dataandnoise}. Only for these
        three, the displayed $200\times200$ region was input to
        \textsf{SExtractor}. Second row (d)--(g): the four mock images
        of Figure \ref{SEsensitivity}. Third row: (h)--(l) the four
        real and one mock image of Figure \ref{SEreal}. Failure to
        interpolate accurately on the corners of the images are
        clearly visible in (a)--(c) and (l) where the image edge can
        be seen. Real images are in units of counts s$^{-1}$, so that
        the horizontal axis of their histograms are multiplied by
        $10^5s$. The true background ($B$) value of all mock images is
        subtracted in the histograms.}
    \end{figure*}

    Figure \ref{dataandnoise} shows how through their very gradual
    penetration into the noise, stars (the {\small PSF}), galaxies,
    and undetected objects in astronomical images are very different
    from cosmic rays. Therefore $\sigma$-clipping will not suffice
    when such objects are present. In other words, no matter how much
    the brightest pixels of objects in an image are clipped, their
    faint wings penetrate very deep into the noise. Therefore when
    astronomical objects are present in the image, a generic
    $\sigma$-clip is a very crude overestimation of the Sky value.

  \subsection{Radius of Flat Profile}\label{flatprofile}

    Another possible solution to finding the Sky was presented
    recently as a proxy for the local Sky background behind an
    object. \citet{galapagos} has proposed finding the
    $\sigma$-clipped mean value on elliptical annuli after removing
    all known source contributions, using \textsf{SExtractor}.  As
    long as the $\sigma$-clipped mean is very high above the noise,
    the slope of the mean pixel value on an annulus as a function of
    radius will be negative. However, as it approaches the noise, due
    to the scatter caused by noise, the slope will become repeatedly
    positive and negative.  They consider the position of the second
    positive to be the annulus on which they find Sky value for each
    object.

    The drawbacks to this method are that it assumes that the profiles
    can be represented as an ellipse with a well-defined center. It is
    also based on the Kron radius (Section \ref{rkron}) which can
    underestimate the full extent of an object when \texttt{\small
      DETECT\_THRESH>1} (see Section \ref{SExthresh} for a complete
    discussion). The most important issue with this approach is the
    area (or number of pixels) which is used to estimate the Sky
    value.  As \citet{daophot} noted, to reach a statistically
    acceptable precision, the area used to find the Sky value has to
    be sufficiently larger than the object itself. However, the
    average value on an annulus is based on a smaller number of pixels
    than that spanning the parent object especially when all
    neighboring objects are removed (see their Figure 5). Therefore
    the random scatter due to the small number of pixels used will
    limit the ability to measure the true Sky level with
    accuracy. This is further exacerbated by the fact that the growth
    of the elliptical annuli is halted based on scatter in the
    measured Sky of each.

  \subsection{Median Value of Dithered Images}\label{meddithered}

    Median values of dithered images can only be used in images prior
    to co-adding.  This method exploits the dithering, where the field
    of view of the telescope is shifted slightly between different
    exposures \citep[for example in][]{kaj2011}.  In order to find the
    Sky value between several temporally close observations, the
    median pixel value of those observations is taken as the Sky value
    on each pixel.

    The weaknesses of this method are: (1) the Sky value might change
    between separate observations, particularly in infrared imaging.
    (2) The main objects in the image have to be far smaller than the
    dithering length.  (3) The field should not be crowded. (4) It
    assumes perfect bias, dark, and flat fielding since it depends on
    operations that are done prior to them.

  \subsection{Background Interpolation}\label{SEbackinterp}

    \textsf{SExtractor} and most other algorithms do not find one Sky
    value for the whole image. Instead, the Sky values are found on a
    mesh grid. In \textsf{SExtractor} the sizes of the meshes are set
    to \texttt{\small BACK\_SIZE} pixels. It then uses bicubic-spline
    interpolation to find a the Sky value for each pixel. Figures
    \ref{SEback}(a)--(c) show \textsf{SExtractor}'s Sky value
    calculated for each pixel for the three mock images of Figure
    \ref{dataandnoise} along with the images in Figures
    \ref{SEsensitivity} and \ref{SEreal}. Recall that the background
    value for all mock images is zero. It is clear that as the area
    covered by the faint parts of mock galaxies has increased from (a)
    to (c) the background values calculated by \textsf{SExtractor}'s
    $\sigma$-clipping and mode approximation have been shifted very
    strongly to the positive. In this figure the problems with using
    model based interpolation techniques are also evident. Note how
    the corners of the images have extremely high and low values
    compared to the centers of the images in Figure \ref{SEback}
    (a)--(c) and (l). Note that the other cases of Figure \ref{SEback}
    do not include image corners.

    In \textsf{SExtractor}, through the parameter \texttt{\small
      BACK\_SIZE} (see Appendix \ref{SEconfig}), a user can set a
    smaller grid on the image. \textsf{SExtractor} then finds the Sky
    in those sub grids, removes the outliers with median filtering,
    and interpolates over them. Therefore for each particular image of
    a particular target, a good choice of \texttt{\small BACK\_SIZE}
    will slightly correct the very large overestimations discussed
    here. However, for any \texttt{\small BACK\_SIZE}, there might be
    objects in the image that cause a similar or even larger
    bias. Therefore, unless the best \texttt{\small BACK\_SIZE} is
    found for each particular input image---depending on the objects
    and their position in the input image---there will be a
    bias. Such customized application is not practical because of the
    sheer number of galaxies studied in most papers. If the user
    chooses a custom background for each image of a large set of
    targets, comparison between the different images cannot be
    accurate because of the different statistical properties of the
    background. Therefore, such significant systematic overestimations
    will plague any astrophysical analysis or hypothesis testing that
    is based on them.

\section{Existing Detection and Segmentation/Deblending Methods}
\label{existingmethods}

  The detection technique used in astronomy up until now can generally
  be classified as a signal-based approach with various
  implementations. It is impractical to review all implementations and
  their minor differences here. Thus in this appendix we use
  \textsf{SExtractor} \sextractorversion{} \citep{sextractor} while
  mentioning the important alternative implementations during the
  steps.  \textsf{SExtractor} was chosen here because it can easily be
  installed and has a relatively complete manual. Because of this it
  is by far the most commonly used tool by most authors in generating
  large catalogs or source extraction of known targets and is thus
  widely recognized by the community. Its techniques are also
  conceptually very similar and inherited from other packages, for
  example, \textsf{DAOPHOT} \citep{daophot}, \textsf{clumpfind}
  \citep{williams94} and \citet{kron}.

  To extract sources from a noisy image, \textsf{SExtractor} first
  needs to find the noise characteristics or the Sky value and its
  error for every pixel. A comprehensive review of the common methods
  for finding the Sky value are thus provided in Appendix
  \ref{existingsky}. Based on the noise, a threshold is applied and
  regions above noise are chosen as detections (Section
  \ref{SEdetection}). The detections are then deblended (Section
  \ref{SEdeblend}). The deblended detections are extended beyond the
  threshold using the Kron radius concept that is critically analyzed
  in Section \ref{rkron}. In Section \ref{combdetseg} a discrete
  hierarchical Markov image modeling method for simultaneous detection
  and segmentation of objects in an image is discussed. Since it also
  requires growing the objects afterwards, it is discussed before
  Section \ref{rkron}. The images are outputs of
  \textsf{SExtractor}. The interpretations of how it has operated are
  based on its manual or corresponding paper. The configuration file
  used to run \textsf{SExtractor} is shown in Appendix \ref{SEconfig}.

  \subsection{Detection}\label{SEdetection}

    After finding the Sky value, \textsf{SExtractor} uses it to define
    a threshold (see Section \ref{SExthresh}). The threshold is
    applied to a smoothed image (see Section \ref{SEconv}). Of the
    various regions found above the threshold, those with an area
    smaller than a user defined value are excluded as a false
    detection (Section \ref{SEminarea}) and the remaining objects
    (true detections) are deblended (Section \ref{SEdeblend}). Using
    the \citet{kron} radius concept, the deblended detections are
    grown on a fixed ellipse to include pixels below the threshold
    (see Section \ref{rkron}).

    \subsubsection{Convolution}\label{SEconv}

      A convolution kernel comparable with the {\small PSF} has been
      the best choice for this purpose. A qualitative argument in
      support of this choice is that it is the widest convolution
      kernel that can be used such that no object is smoothed out
      since no object can be smaller than the image {\small PSF}
      \citep[see][for a detailed quantitative
        discussion]{bijaoui70}. Based on this argument, for the
      example runs of \textsf{SExtractor} on the mock images, the same
      {\small PSF} (Moffat function with $\beta=\moffatbeta$ and
      {\small FWHM}=\mockfwhm{} pixels) that was used to create the
      input image was used. In \textsf{SExtractor} the convolution
      kernel is specified by the parameter \texttt{\small
        FILTER\_NAME}.

    \subsubsection{Threshold}\label{SExthresh}
      In order to detect objects above the noise, a threshold needs to
      be applied to the convolved image. Only pixels with values above
      this threshold are considered for object detection. A higher
      threshold will decrease the probability of false detections,
      because fewer noise pixels will have a count above it. However,
      it will miss fainter objects and the fainter parts of brighter
      objects.

      In \textsf{SExtractor}, the threshold is defined in terms of the
      approximated Sky ($s$, see Appendix \ref{existingsky}) and its
      standard deviation ($\sigma$): $s+$\texttt{\small
        DETECT\_THRESH}$\times\sigma$ which are measured on the
      original input image (prior to convolution). The values set for
      \texttt{\small DETECT\_THRESH} are different in various
      studies. The recently released {\small 3D}-\emph{\small HST}
      {\small WFC3}-selected Photometric Catalogs in the five {\small
        CANDELS}/{\small 3D}-\emph{\small HST} Fields
      \citep{skelton14} set this parameter to $1.8$. In the
      K$_{s}$-selected Catalog in the {\small COSMOS}/{\small
        ULTRAVISTA} field \citep{muzzin13}, it is set to $1.7$. As a
      final example, in the Subaru/{\small MOIRCS} deep survey of the
      {\small GOODS-N} field \citep{kaj2011}, it is set to $1.3$.

      In order to go deeper in the noise, \citet{rix04} proposed a
      two-stage detection process. In their technique, objects are
      detected in two runs \citep[values from][]{rix04}: (1) a
      ``cold'' run where \texttt{\small DETECT\_THRESH} is set to a
      very high value of 2.3, and only the brightest regions of the
      brightest objects are detected and (2) a ``hot'' run where it is
      set to a lower value of 1.65 to detect the fainter objects.  The
      two resulting catalogs are then compared and ``hot'' objects
      that lie over the ``cold'' ones are removed from the final
      catalog. In their ``hot'' run, such techniques use the lowest
      possible threshold they can consider. Working on deeper images,
      \citet{leuthaud07} used $2.2$ and $1$ for their cold/hot
      detection thresholds, respectively.

      The threshold is calculated based on the original image, but it
      is applied to the convolved image. Therefore it no longer has
      the statistical properties that any multiple of the standard
      deviation is expected to have, for example, that $\sim95\%$ of
      the noise pixels have a value below $2\sigma$. This occurs
      because of the decreased dynamic range after convolution (see
      Section \ref{conv} and Figure \ref{convolution}). In Figure
      \ref{convolution}, $1\sigma$, which is the standard deviation of
      the un-convolved image, corresponds to 100$e^-$. However, when
      applied to the convolved image of Figure \ref{convolution}(b)
      only $\convsampbnumhigher$ (of the 40,000) pixels are above this
      threshold. Note that the kernel used in Figure
      \ref{convolution}(b) is the same {\small PSF} that was used to
      make the mock image. In Figure \ref{convolution}(c)--(e),
      100$e^-$ far exceeds the brightest convolved pixel.

      Based on this idea, \citet{galametz2013} adopt a different
      approach to the cold and hot detection technique. \texttt{\small
        DETECT\_THRESH} is approximately similar between the two (0.75
      and 0.7), but the convolution kernels (see Section \ref{SEconv})
      used are significantly different such that the cold run has a
      much wider kernel (see their Appendix A for the respective
      \textsf{SExtractor} parameters).

      A different approach to finding the threshold of an image,
      called the ``False-Discovery Rate,'' was introduced in astronomy
      by \citet{miller01}. In this approach the average fraction of
      false pixel detections to true pixel detections is held fixed
      when assuming a fixed null hypothesis (background and its
      noise). It was implemented as a detection technique in
      \citet{hopkins02}. All the pixels in the region must be sorted
      and a $p$ value has to be calculated from an approximated mean
      and standard deviation. Thus this thresholding technique also
      needs the Sky and its error prior to the actual detection
      process.

    \subsubsection{Detection: True/False}\label{SEminarea}

    \begin{figure*}[t]
      \centering
      \ifdefined\makeeps
      \include{./tex/SEthresh}
      \else
      \includegraphics[width=\linewidth]{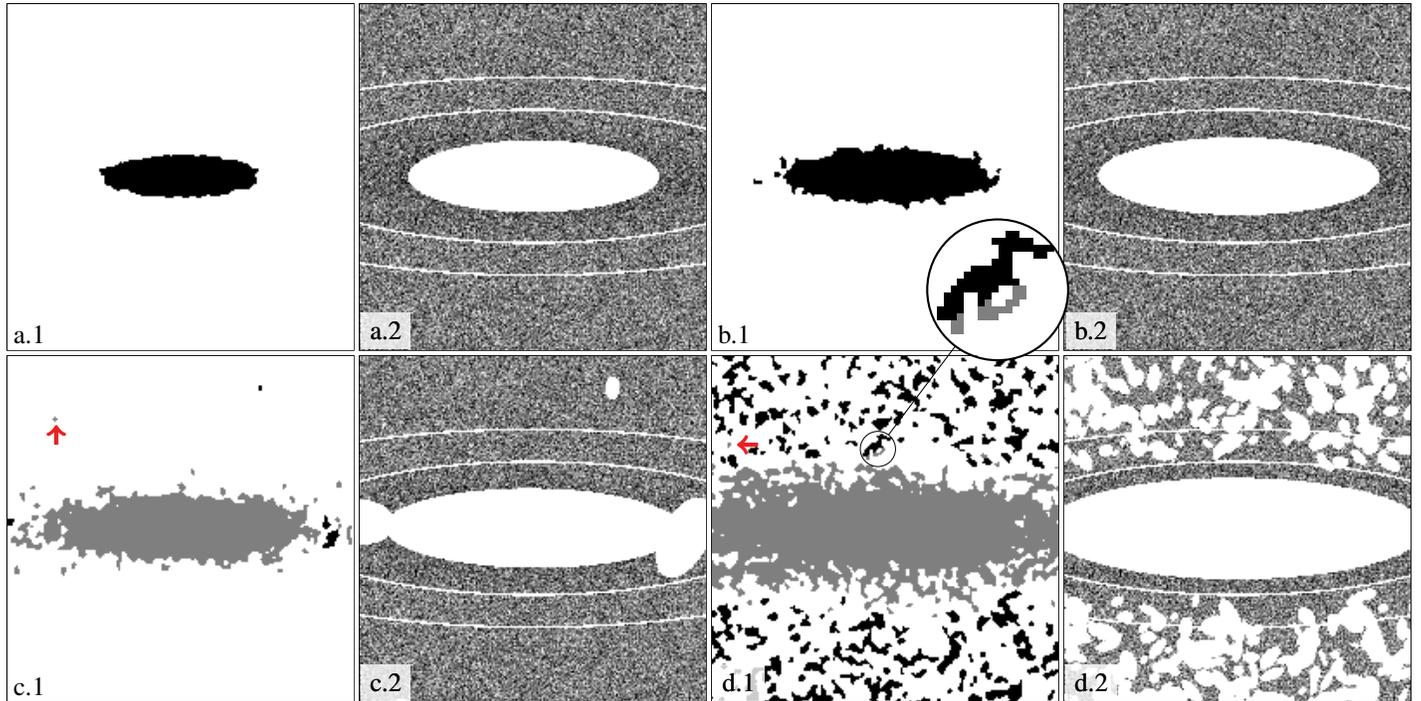}
      \fi
      \caption{\label{SEthresh} \textsf{SExtractor} results of various
        thresholds with configuration parameters from Appendix
        \ref{SEconfig} except for \texttt{DETECT\_THRESH}. The input
        image only has the one profile of Figure \ref{dataandnoise}(b)
        (see Section \ref{aboutimages}). The threshold values for
        \texttt{DETECT\_THRESH} are (a) $2$, (b) $1$, (c) $0.5$, (d)
        $0.1$.  For each threshold there are two images. Left: the
        segmentation map. Right: the elliptical region within three
        times the Kron radius ($r_k$, see Section \ref{rkron}) is
        masked (white) from the original image. Notice the darker
        profile regions surrounding the elliptical mask that have not
        been included in the $\rkron{}r_k$ elliptical aperture. The
        smaller elliptical border marks the aperture containing
        $\sim90\%$ of the counts in the profile. The larger elliptical
        border shows the region containing $\sim95\%$ of the
        counts. The segmentation maps in (c.1) and (d.1) have
        different color codes for the main object (gray) and any other
        detections (black).  The marked regions show how connectivity
        is not a criteria in \textsf{SExtractor}'s deblending
        algorithm. }
    \end{figure*}

\begin{figure}[tb]
      \centering
      \ifdefined\makeeps
      \input{./tex/SEsensitivity}
      \else
      \includegraphics{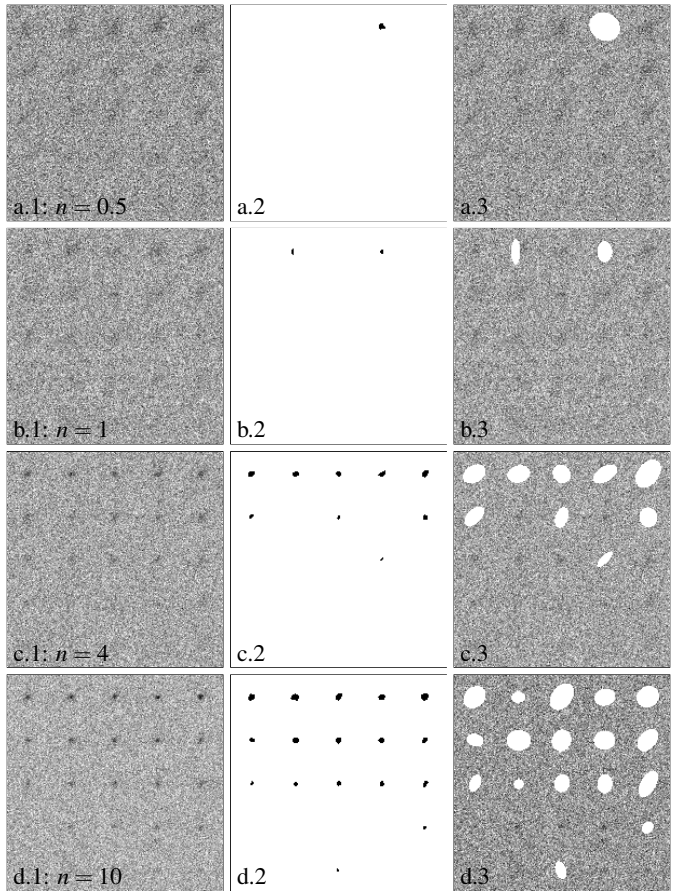}
      \fi
      \caption{\label{SEsensitivity} \textsf{SExtractor} sensitivity
        test. Column 1: the input images, which are the same as those
        in Figure \ref{NCsensitivity}. Column 2: \textsf{SExtractor}'s
        segmentation maps. Column 3: \textsf{SExtractor}'s detections
        extended to three times the Kron radius. The
        \textsf{SExtractor} configuration parameters can be seen in
        Appendix \ref{SEconfig}.}
    \end{figure}

      Once the threshold of Section \ref{SExthresh} is applied to the
      smoothed image of Section \ref{SEconv}, the image is divided
      into two groups of pixels: those that are above and below the
      threshold. The non-white pixels in Figure \ref{SEthresh}
      (a.1)--(d.1) and column 2 in Figures \ref{SEsensitivity} and
      \ref{SEreal} show this image with a threshold applied (they are
      all \textsf{SExtractor}'s segmentation maps).  In Figure
      \ref{SEthresh}, \textsf{SExtractor} is run on the mock image of
      Figure \ref{dataandnoise}(b.2) with the configuration file of
      Appendix \ref{SEconfig}.  All of \textsf{SExtractor}'s
      parameters in all four cases are fixed to those in Appendix
      \ref{SEconfig} except for \texttt{\small DETECT\_THRESH}, which
      is set to \texttt{\small 2}, \texttt{\small 1}, \texttt{\small
        0.5}, and \texttt{\small 0.1}, respectively.  The first
      observation after examining Figure \ref{SEthresh} is that as the
      detection threshold decreases, the number of false detections
      increases: $\setwothresh$, $\seonethresh$, $\sehalfthresh$ and
      $\setenththresh$, (in the $1000\times1000$ images, see Section
      \ref{aboutimages}) respectively.

      The major tool available in \textsf{SExtractor} to define and
      thus remove such false detections is \texttt{\small
        DETECT\_MINAREA}. With this parameter the user can specify the
      minimum area that a true detection should have. As the threshold
      decreases, more and more pixels become connected.  As an example
      when \texttt{\small DETECT\_THRESH=0.1} is fixed but
      \texttt{\small DETECT\_MINAREA} is set to \texttt{\small 10},
      \texttt{\small 25}, \texttt{\small 50}, \texttt{\small 75}, and
      \texttt{\small 100} pixels, $\setenminarea$,
      $\setwofiveminarea$, $\sefiftyminarea$, $\sesevenfiveminarea$
      and $\sehundredminarea$ false detections are found. The larger
      \texttt{\small DETECT\_MINAREA} will also result in not being
      able to detect many compact bright objects that will also be
      present in a real image.  The \textsf{SExtractor} manual
      therefore correctly suggests setting \texttt{\small
        DETECT\_MINAREA} to small values ($1$--$5$) so that the
      convolution kernel and \texttt{\small DETECT\_THRESH} define the
      sensitivity.

      Figure \ref{SEsensitivity} shows the outputs of
      \textsf{SExtractor}, configured with Appendix \ref{SEconfig},
      for extremely faint mock galaxies; see Sections
      \ref{mydetection} and \ref{A_sky}. \textsf{SExtractor}'s results
      on a bright and large profile are already analyzed in Figure
      \ref{SEthresh}. With \texttt{\small DETECT\_THRESH=1}, only the
      top few pixels of the brightest and sharpest profiles have been
      successfully detected. Note that as discussed in Section
      \ref{SExthresh}, most surveys use much larger thresholds.

      When a large {\small S/N} for detections is required for
      accurate fittings, for example, in photometric redshifts or
      {\small SED} fittings, some researchers might opt for discarding
      such faint detections. However, their detection in the image is
      nevertheless very important because if they are not detected and
      removed, they will cause an overestimation in the Sky value,
      hence systematically underestimating the total count of the
      brighter, high {\small S/N}, objects (see Section \ref{A_sky}).

  \subsection{Segmentation/Deblending}\label{SEdeblend}

    Segmentation and deblending are defined in Section
    \ref{mysegmentation}. An interesting deblending algorithm is
    proposed by \citet{SDSSdeblender} and is employed in the {\small
      SDSS}, Subaru Telescope Hyper Suprime-Cam, and {\small LSST}
    pipelines. In one image (one color) real peaks are defined as
    those that are at least $3\sigma_{sky}$ above the saddle point or
    valley that separates it from neighboring peaks in terms of
    count. True peaks also have to be at least one pixel distant from a
    neighboring peak. Once the true ``child'' peaks are found over a
    detected region, a circular template is built for the child by
    using the smaller value of two pixels on the opposite sides of the
    peak. A weight is given to each child and a cost function is
    defined and minimized to deblend the sources \citep[see Figure 1
      of][for a 1D demonstration]{SDSSdeblender}.

    In choosing a true peak, the area belonging to the peak and the
    count in it is ignored and a peak pixel is defined as true only if
    that pixel alone is sufficiently above its immediate valley or
    saddle point (compare to the discussion in Section
    \ref{clumpseg}). The very important technical issue of how to find
    the saddle point count is not discussed in that report. One
    drawback is that this selection criteria will miss real peaks that
    are below this large threshold but are not due to noise, for
    example with a very flat profile. In this approach, deblending is
    completely independent of segmentation. Such pure deblending will
    also suffer when the detected regions cannot be fitted (the cost
    function minimized) with the combinations of templates over the
    detection.

    \textsf{SExtractor} uses a concept very similar to contour maps
    known as \emph{multi-thresholding}. The count range of the detected
    region is divided into several layers. Then the separate objects
    in those layers are analyzed and ``real'' peaks are modeled to
    find the separate local maxima and their contribution to the total
    count \citep[see Figure 2 in][]{sextractor}. \textsf{clumpfind}
    \citep{williams94} also uses a very similar approach. Therefore in
    this approach, first a segmentation map is derived and used to
    model the objects, and then deblending is done.

    In \textsf{SExtractor}'s implementation, a fixed number of layers
    is defined for all objects, specified by \texttt{\small
      DEBLEND\_NTHRESH} in Section \ref{SEconfig}. Therefore brighter
    objects, with a very large difference between their peak count and
    the Sky, will receive layers that are much more widely separated
    than fainter objects. This will oversegment the fainter objects
    where the spacing between the layers is less and undersegment the
    brighter ones.  The scale on which the count is divided will also
    significantly alter the result. In \textsf{SExtractor},
    \texttt{\small DETECTION\_TYPE} specifies the scale. If set to
    \texttt{\small CCD}, an exponential scale is used and setting it
    to \texttt{\small PHOTO} will use a linear scale. Each will output
    significantly different results.

    \begin{figure}[!t]
      \centering
      \ifdefined\makeeps
      \input{./tex/SEreal}
      \else
      \includegraphics{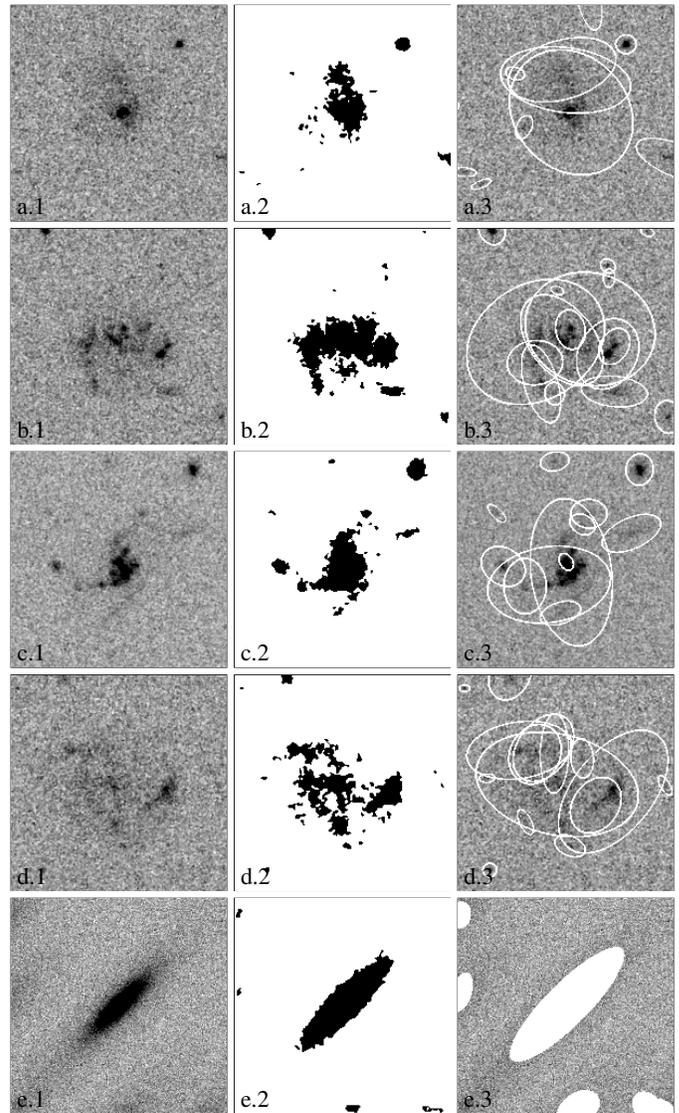}
      \fi
      \caption{\label{SEreal} Tests of \textsf{SExtractor} on the same
        input images as Figure \ref{NCreal}. First column: input
        image.  Second column: \textsf{SExtractor}'s segmentation
        maps. Third column: $\rkron{}r_k$ elliptical apertures (see
        Section \ref{rkron}). In the real images, because of their
        complexity, only the borders of the elliptical apertures are
        shown. In the mock image, the elliptical apertures are
        filled. Most of the central object detections in a.3, b.3 and
        d.3 were flagged by \textsf{SExtractor} for possibly biased
        magnitude measurements. }
    \end{figure}

    Once peaks are found between the layers, they are accepted as a
    true detection or rejected as false based on the parameter
    \texttt{\small DEBLEND\_MINCONT} which specifies the fraction of
    the total intensity in that peak, or clump, compared to the total,
    undeblended, detection. Similar to the layer spacing problem
    mentioned above, this parameter is by definition, biased toward
    detecting more clumps in fainter objects. Since the threshold to
    accept a clump differs from one detection to another (based on the
    total count within each parent detection), a fixed peak will be
    detected in a fainter object and discarded in a bright
    one. Therefore when this selection criteria is adopted for a
    comparison of clumps in galaxies, added with the layering bias
    mentioned above, it will bias the results by preferentially
    detecting more segments (or clumps) in objects with a lower total
    count before deblending.  Another consequence of this method in
    finding true clumps is that the result will vary significantly
    depending on \texttt{\small DETECT\_THRESH}. When the threshold is
    high, the detected total count will be less, and therefore more
    peaks will be detected in a given object.

    The highlighted (magnifying glasses and red arrows) regions in
    Figure \ref{SEthresh} (for mock galaxies) and the large fraction
    of flagged detections in Figure \ref{SEreal} (for real images)
    show how \textsf{SExtractor}'s deblending algorithm has failed in
    the low surface brightness regions.  The primary reason for this
    failure is that \textsf{SExtractor} was designed for high
    thresholds. When reliable peaks are found (with a high threshold),
    each pixel is assigned to a peak, assuming all the peaks are a 2D
    Gaussian far larger than the region above the threshold. This is
    useful in cases where the peaks' {\small S/N} and the threshold
    are sufficiently high and all peaks follow a Gaussian profile over
    the whole image.

    In practice, hardly any non-stellar objects of astronomical
    interest has a Gaussian profile and postage stamps are usually
    chosen to be larger than the objects in order to get an accurate
    estimate of the Sky value. Figure \ref{SEthresh} shows two ways
    that \textsf{SExtractor} fails in deblending as the threshold
    decreases: (1) some of the very far detections of Figure
    \ref{SEthresh}(c.1) and (d.1), designated with arrows have been
    identified as belonging to the main object and (2) a connected
    region can be shared between two objects that are not physically
    connected to it as shown in the magnified regions of the same
    figure. These problems might not cause a significant problem when
    the magnitude of the very bright mock galaxy in this figure is
    desired, but when applied to low surface brightness and diffuse
    galaxies like Figure \ref{SEreal} (a)--(d), very serious
    systematic errors can be produced.

    Figure \ref{SEreal} (a.1)--(d.1) shows \textsf{SExtractor}'s
    result for the same galaxies in Figure \ref{NCreal}. Appendix
    \ref{SEconfig} shows the configuration parameters used to run
    \textsf{SExtractor} on these galaxies. The input {\small PSF} for
    these four real galaxies is obtained from the web interface of
    \textsf{TinyTim}\footnote{\url{http://tinytim.stsci.edu/cgi-bin/tinytimweb.cgi}.
      On Chip 1, pixel position (2047, 1048), filter F814W, Spectrum
      value 12, PSF diameter $3^{\prime\prime}$, focus
      0.0. \textsf{SExtractor} does not accept kernels larger than
      $31\times31$ pixels.  Therefore, the central region of this size
      was cropped from \textsf{TinyTim}'s output for input into
      \textsf{SExtractor}.}  \citep{tinytim}. The {\small PSF} of
    \textsf{TinyTim} is the theoretical {\small PSF} produced on the
           {\small CCD}. In order to create the processed images shown
           here, various steps of image processing have been applied
           to these images \citep[see][]{cosmoshstiband,
             massy10}. Therefore, in practice, the actual PSF on the
           image will be slightly different from that produced by
           \textsf{TinyTim} \citep[see][]{vdWel12}. While this
           difference plays a very important role in model fitting,
           for the demonstrative purposes on detection here, the
           \textsf{TinyTim} output is adequate.

    It is clear from Figure \ref{SEreal} that the two more diffuse
    objects in Figure \ref{SEreal}(b) and (d) have been deblended
    into many more separate objects than the two brighter cases of
    Figure \ref{SEreal} (a) and (c). As explained above, this is due
    to the inherent biases of \textsf{SExtractor}'s deblending
    algorithm.

  \subsection{Combining Detection and Segmentation}\label{combdetseg}

    A discrete hierarchical Markov image modeling approach was
    introduced in astronomy recently by \citet{vollmer13}. This is a
    fundamentally different approach to detection and segmentation. It
    is reviewed here because like the more popular methods used today,
    the detected region needs to be grown to complete the detection as
    we will discuss next in Section \ref{rkron}. In this approach
    object detection and segmentation are done simultaneously. The
    basic technique comes from image processing applications in
    particular the quadtree of \citet{laferte2000}.

    Hierarchical Markov models have the advantage that they allow to
    use multi-band images at potentially different pixel resolutions
    without the need to modify the data as most techniques used in
    astronomy currently need to do. This makes them ``scale free.''
    However \citet{vollmer13} acknowledge that their implementation
    ({\small MARSIAA}) does not account for {\small PSF} variation
    between the inputs. In this approach the problem is labeling of
    the image pixels ``which are spatially connected through a
    pre-defined neighborhood system. These labels correspond to
    discrete classes of objects with similar surface
    brightness''\citep[page 3]{vollmer13}. Using the multi-band data
    as input, the procedure iterates along the hierarchies to find a
    distribution for the labels over the image, see
    \citet{laferte2000} for details.

    In this approach there is no specific functional basis to do the
    detection, therefore it is considered as a model-independent
    detection and segmentation technique. However, instead of known
    functions, it classifies the image pixels into ``classes'' which
    are shared by all the objects in the image. For example, noise has
    a class of 0 and as the gradient over the object increases, the
    class also increase; see Figure 9 in \citet{vollmer13} for how the
    class changes toward the center of the object. The number of
    classes is a free parameter specified by the user. The final
    simultaneous detection/segmentation depends on the total number of
    classes and the dynamic range of the image. To detect low surface
    brightness galaxies, \citet{vollmer13} had to truncate or clip the
    image pixels beyond $20\sigma$, therefore this approach cannot
    simultaneously detect/segment very bright and very faint objects
    in one image.

    An interesting consequence of this technique is that it does not
    need to define a count threshold, however, it can only get
    $\sim1.5\sigma$ close to the Sky value. Figures 8--10 of
    \citet{vollmer13} show the results of this technique compared to
    \textsf{SExtractor}'s segmentation maps for three galaxies. In the
    three cases, the detected area of \textsf{SExtractor}'s
    segmentation map with a threshold at $1.5\sigma$ was comparable or
    even larger (better covering the objects). This technique heavily
    relies on iterative processes and thus can be very slow.
    \citet{vollmer13} report it was at least 90 times slower than
    \textsf{SExtractor} for the same data set.

  \subsection{Growing Detections Below the Threshold} \label{rkron}
    The high thresholds that were used to avoid false detections cause
    a very large fraction of the object to not be initially detected
    (see Figure \ref{SEthresh}(a.1)--(d.1)). Failing to detect a
    bright object's complete area will deprive us of all the valuable
    information lying hidden there and cause a systematic under-
    (over-)estimation of the object's magnitude (Sky value). Therefore
    when using such large thresholds, it is necessary to use the
    regions above the threshold as a basis to ``grow'' the detections
    and include their fainter parts.

    The Kron radius, $r_k$ \citep{kron}, is one of the most popular
    methods to grow the detection into regions of the image with a
    count lower than the threshold. It is defined as the light-weighted
    average radius of a profile,

    \begin{displaymath}
      r_k=\frac{\sum rI(r)}{\sum I(r)}.
    \end{displaymath}

    From Figure \ref{SEthresh} it is clear that the elliptical
    parameters and $r_k$ are found based on the region above the
    threshold, therefore the fainter regions of the galaxies that
    often harbor valuable information on the dynamical history of the
    galaxy and might have different morphological parameters cannot be
    directly measured in the existing signal based approach to
    detection.

    It is generally considered that $2r_k$ contains $\geq90\%$ of the
    total count, integrated to infinity ($F_I$), of a galaxy light
    profile \citep[see][]{kron, infante87}. Therefore in the
    literature it is common to use $2.5r_k$ to find the total count of
    a galaxy. In order to test this claim, \textsf{SExtractor}'s
    \texttt{\small FLUX\_AUTO} is divided by $F_I$\footnote{$F_I$ is
      approximated by the total count within $30r_e$. Note that the
      larger ellipse in Figure \ref{SEthresh} which contains $95\%$ of
      the total light is on $7r_e$.} for the four different masked
    $\rkron{}r_k$ apertures shown in Figure \ref{SEthresh}. For this
    profile, $\rkron{}r_k$ contains $\setwokronff\%$,
    $\seonekronff\%$, $\sehalfkronff\%$ and $\setenthkronff\%$ of the
    total count. The area within the $\rkron{}r_k$ has been masked to
    show the extent of the object that has not been detected. The
    elliptical apertures containing $90\%$ and $95\%$ of $F_I$ are
    also shown in Figure \ref{SEthresh} for visual comparison.

    \citet{grahamdriver05} did a thorough analytic study of the
    mathematical (continuous) 1D S\'{e}rsic profile to show how such
    differences from the expected $>90\%$ could occur in the
    calculation of the Kron radius. They showed how dependent the
    total count within the 2.5 times the Kron radius is to the area
    that is used to find the Kron radius. If this area is extended to
    infinity, it is indeed as expected and 2.5 times the Kron radius
    contains $>90\%$ of $F_I$, even for $n=10$ profiles. However,
    integration to infinity is not a realistic condition. In practice
    any measurement of the Kron radius is confined to a certain
    aperture. By considering the effects of the finite aperture that
    is used in real measurements of $r_k$, they show how the values
    reported above, which have also been reported by other authors are
    reasonable \citep[see][]{Bernstein02a, Benitez04}.

    This discussion demonstrates that the statement that an aperture
    $>2r_k$ will contain nearly all ($\geq90\%$) of the light and area
    of a galaxy is not generic and is only practically correct for
    very flat profiles. If ignored and used for all galaxies in a
    survey or image, it can result in a very significant
    underestimation of the total light and boundaries of the sharper
    objects, even if it is a pure ellipse as in Figure
    \ref{SEthresh}. The underestimation will cause an overestimation
    in the Sky value of the image, further compounding the resulting
    systematic bias. The mock galaxy used in this example has a
    S\'{e}rsic index ($n$) of $\onelargen$, while in the local
    universe, galaxies with $n$ as large as $11.84$ have been reported
    \citep{kormendy09}. The case for distant galaxies becomes much
    worse, since their apparent $r_e$ in pixels, due to the angular
    diameter distance, is much smaller with much lower surface
    brightness.

    Figure \ref{SEsensitivity} shows \textsf{SExtractor}'s output for
    mock galaxies, both flat, low S\'{e}rsic index, and sharp or high
    S\'{e}rsic index. Even with \texttt{\small DETECT\_THRESH=1}, only
    a few pixels have been found for each object, the elliptical
    parameters, namely center, axis ratio, position angle and major
    axis length ($r_k$) are all measured using these few pixels. The
    second argument to \texttt{\small PHOT\_AUTOPARAMS} is set to a
    minimum radius such that if $r_k$ is too small, this minimum
    radius is used instead of it. This is why most of the ellipses in
    Figure \ref{SEsensitivity} have similar sizes. However, such hand
    input values could add more bias to the final result (see Section
    \ref{A_magdisp} for a discussion on the dispersion of the
    magnitudes measured in this method).

    By definition, $r_k$ depends on finding the full 1D light profile
    of an object.  In order to do that, it is necessary that the
    center of the profile, its position angle and axis ratio are
    accurately defined.  Furthermore, a good enough sampling of the
    profile at various $(r,\theta)$ is needed in order to approximate
    the count at a certain radius.  These preconditions are only valid
    in the most optimistic cases, like the bright purely elliptical
    mock profile of Figure \ref{dataandnoise}(b). In general only a
    very rare fraction of galaxies display such a simple and bright
    profile, clearly separated from neighbors or image edges.

    Galaxies can have multiple components, for example, the bulge, a
    bar, and a disk.  Each can have a different axis ratio and
    position angle.  The central parts (bulge and a possible bar) are
    much brighter than the outer disk.  Therefore their effect on the
    overall, light-weighted, estimation of ellipticity will completely
    outweigh the fainter outer parts, which might not even be detected
    because of the high thresholds (\texttt{\small DETECT\_THRESH>1})
    that are commonly used.  This compounds the problem mentioned in
    the previous paragraph for a real galaxy's total light because all
    the elliptical parameters in one profile can change in a real
    galaxy due to their rich dynamical history.

    \begin{figure*}
      \centering
      \ifdefined\makeeps
      \newcommand{\inputdir}{\figtexdir/mirrordemo}

\makeatletter \newcommand{\pgfplotsdrawaxis}{\pgfplots@draw@axis} \makeatother

    \pgfplotsset{axis line on top/.style={
      axis line style=transparent,
      ticklabel style=transparent,
      tick style=transparent,
      axis on top=false,
      after end axis/.append code={
        \pgfplotsset{axis line style=opaque,
          ticklabel style=opaque,
          tick style=opaque,
          grid=none}
        \pgfplotsdrawaxis}
      }
    }

    \begin{tikzpicture}
        \scriptsize
        \begin{groupplot}[group style={group size=5 by 1,
                     horizontal sep=0pt},
                     xmode=normal,ymode=normal,
                     enlargelimits=false,
                     no markers, width=0.2\linewidth,
                     height=0.14\linewidth,
                     xlabel={Pixel intensity ($e^-$) - $M_m$},
                     scale only axis, axis line on top,
                     area legend, legend cell align=left,
                     legend style={draw=none, fill=none,
                                   at={(0.5\linewidth,0.08\linewidth)},
                                   anchor=south west, font=\scriptsize}]

          \nextgroupplot[yticklabels={}, ytickmin=-2, ytickmax=-1]
            \addplot [const plot, fill=myblue]
                     table [x expr={\thisrowno{0}*10000}, y index=1]
                           {\inputdir/10hist.txt}
            \closedcycle;
            \addplot [const plot, fill=mygreen, fill opacity=0.6]
                     table [x expr={\thisrowno{0}*10000}, y index=2]
                           {\inputdir/10hist.txt}
            \closedcycle;
            \legend{Original, Mirrored}

            \addplot [very thick, color=myblack]
                     table [x expr={\thisrowno{0}*10000}, y index=1]
                     {\inputdir/10cfp.txt};
            \addplot [thin, color=myblack]
                     table [x expr={\thisrowno{0}*10000}, y index=2]
                     {\inputdir/10cfp.txt};

          \nextgroupplot[yticklabels={}, ytickmin=-2, ytickmax=-1]
            \addplot [const plot, fill=myblue]
                     table [x expr={\thisrowno{0}*10000}, y index=1]
                           {\inputdir/25hist.txt}
            \closedcycle;
            \addplot [const plot, fill=mygreen, fill opacity=0.6]
                     table [x expr={\thisrowno{0}*10000}, y index=2]
                           {\inputdir/25hist.txt}
            \closedcycle;

            \addplot [very thick, color=myblack]
                     table [x expr={\thisrowno{0}*10000}, y index=1]
                     {\inputdir/25cfp.txt};
            \addplot [thin, color=myblack]
                     table [x expr={\thisrowno{0}*10000}, y index=2]
                     {\inputdir/25cfp.txt};

          \nextgroupplot[yticklabels={}, ytickmin=-2, ytickmax=-1]
            \addplot [const plot, fill=myblue]
                     table [x expr={\thisrowno{0}*10000}, y index=1]
                           {\inputdir/50hist.txt}
            \closedcycle;
            \addplot [const plot, fill=mygreen, fill opacity=0.6]
                     table [x expr={\thisrowno{0}*10000}, y index=2]
                           {\inputdir/50hist.txt}
            \closedcycle;

            \addplot [very thick, color=myblack]
                     table [x expr={\thisrowno{0}*10000}, y index=1]
                     {\inputdir/50cfp.txt};
            \addplot [thin, color=myblack]
                     table [x expr={\thisrowno{0}*10000}, y index=2]
                     {\inputdir/50cfp.txt};

          \nextgroupplot[yticklabels={}, ytickmin=-2, ytickmax=-1]
            \addplot [const plot, fill=myblue]
                     table [x expr={\thisrowno{0}*10000}, y index=1]
                           {\inputdir/75hist.txt}
            \closedcycle;
            \addplot [const plot, fill=mygreen, fill opacity=0.6]
                     table [x expr={\thisrowno{0}*10000}, y index=2]
                           {\inputdir/75hist.txt}
            \closedcycle;

            \addplot [very thick, color=myblack]
                     table [x expr={\thisrowno{0}*10000}, y index=1]
                     {\inputdir/75cfp.txt};
            \addplot [thin, color=myblack]
                     table [x expr={\thisrowno{0}*10000}, y index=2]
                     {\inputdir/75cfp.txt};

          \nextgroupplot[yticklabels={}, ytickmin=-2, ytickmax=-1]
            \addplot [const plot, fill=myblue]
                     table [x expr={\thisrowno{0}*10000}, y index=1]
                           {\inputdir/90hist.txt}
            \closedcycle;
            \addplot [const plot, fill=mygreen, fill opacity=0.6]
                     table [x expr={\thisrowno{0}*10000}, y index=2]
                           {\inputdir/90hist.txt}
            \closedcycle;

            \addplot [very thick, color=myblack]
                     table [x expr={\thisrowno{0}*10000}, y index=1]
                     {\inputdir/90cfp.txt};
            \addplot [thin, color=myblack]
                     table [x expr={\thisrowno{0}*10000}, y index=2]
                     {\inputdir/90cfp.txt};
        \end{groupplot}

        \node[anchor=south west] at (0.002\linewidth,0.118\linewidth)
             {\normalsize a: {\small $10\%$}};
        \node[anchor=south west] at (0.202\linewidth,0.118\linewidth)
             {\normalsize b: {\small $25\%$}};
        \node[anchor=south west] at (0.402\linewidth,0.118\linewidth)
             {\normalsize c: {\small $50\%$}};
        \node[anchor=south west] at (0.602\linewidth,0.118\linewidth)
             {\normalsize d: {\small $75\%$}};
        \node[anchor=south west] at (0.802\linewidth,0.118\linewidth)
             {\normalsize e: {\small $90\%$}};

    \end{tikzpicture}
      \else
      \includegraphics[width=\linewidth]{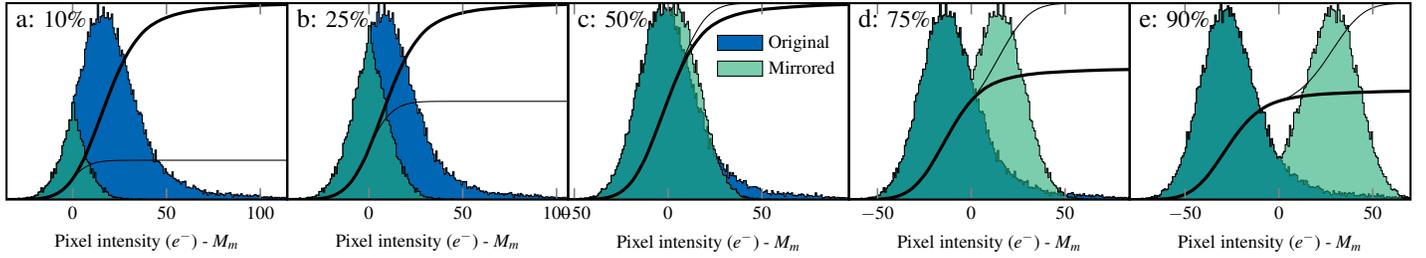}
      \fi
      \caption{\label{mirrordemo} Mirrored distribution compared with
        the original distribution. The data set is the pixels of
        Figure \ref{segtf}(b.2) multiplied by $10^4$ seconds. $M_m$ is
        the value on the mirror point. Blue and green histograms are
        the histograms of the original and mirrored distributions,
        respectively. The thick and thin lines are the cumulative
        frequency plots of the original and mirrored distributions,
        respectively. The percentage value shown after the label of
        each plot shows the percentile of the original distribution on
        which the mirror was placed. It is clear that as the mirror
        point gets closer to the mode of the distribution, the two
        lines diverge more smoothly. The cumulative frequency curve is
        scaled to the tip of the blue histogram. }
    \end{figure*}

    The four real galaxies in Figure \ref{SEreal} visually appear to
    have a distinct region in the noise. Whether they are composed of
    separate objects in the same line of sight or are actually one
    object cannot be judged from one image. In all four examples,
    because the connectivity of the objects is below the threshold
    (\texttt{\small DETECT\_THRESH=1}), \textsf{SExtractor} considers
    them to be spatially separated objects. Furthermore, in none of
    these examples can the connectivity be modeled as a simple ellipse
    to be able to model.

    Figure \ref{SEreal}(a.1) appears to have one bright clump and a
    diffuse host. Because of the relatively flat, diffuse structure,
    the ellipses showing $\rkron{}r_k$ for the three detections have
    become extremely large. When this happens, the final result is
    dependent on the accuracy of Sky subtraction, because a large area
    of sky is also included as part of the object. The large fraction
    of flagged detections in the real galaxies of Figure \ref{SEreal}
    shows how \texttt{\small MAG\_AUTO} has been flagged as unreliable
    for most of their sub-components. A user of \textsf{SExtractor}
    will thus have to either ignore flagged detections or rely on
    \texttt{\small MAG\_ISO}, using only the segmented regions above
    the threshold. If they choose the latter, for such diffuse
    objects, \texttt{\small MAG\_ISO} is going to miss a very large
    fraction of the object's area and photon count.

    Due to the sheer number of galaxies that researchers use for their
    studies, it is not feasible to check each detection by eye to
    check when \texttt{\small MAG\_AUTO} should be used or
    \texttt{\small MAG\_ISO}. \textsf{SExtractor} does provide
    \texttt{\small MAG\_BEST} to automatically choose between the two,
    but it is not commonly used in studies today. The main reason is
    that \texttt{\small MAG\_AUTO} and \texttt{\small MAG\_ISO} are
    not consistent and cannot simply be compared with each other in
    the study of a large number of galaxies.

    In Figure \ref{SEreal}(e), because of the strong gradient created
    by bicubic-spline interpolation in \textsf{SExtractor}'s Sky
    measurement (see Figure \ref{SEback}(l)) no pixel from the the
    profile on the corner of the image has been detected. Looking at
    \textsf{SExtractor}'s detections in Figure \ref{SEreal}(e.3), it
    is clear that for such objects, lying on the edges of images, any
    modeling method of trying to account for their faint regions
    fails.  A very large portion of their total count is left out of
    their designated Kron aperture.  Like the case for clumpy objects,
    all the necessary requirements to define $r_k$ fail when enough of
    the object is beyond the edge of an image.  Such objects have to
    be taken into account, particularly if they are the wings of
    bright objects.  While the total magnitude of such objects would
    always be flagged by software like \textsf{SExtractor}, and not
    included in any scientific work, failing to account for as much of
    their light as possible will systematically increase the Sky value
    that is calculated in such an image.  Such cases are common when
    postage stamp images (for example, Figure \ref{SEreal}
    (a.1)--(d.1)) of a selected group of galaxies in a larger survey
    are studied.

    The discussions in this section show that the only way to get as
    close as possible to the ``true'' total photon count and area of
    an object of interest is to use lower thresholds. Trying to
    correct for the photon count which is lost, when a high threshold
    that is required in the signal-based detection technique is used,
    will only be a source of systematic bias in measurements and the
    resulting scientific results. The problems mentioned in this
    section were the primary motivation in finding a new approach to
    detection and segmentation that would mitigate many of the
    problematic issues described above.

  \section{Finding the Mode}\label{findingmode}

  In Section \ref{localmode}, the existing methods to find the mode of
  an image were reviewed. Here a novel approach to finding the mode of
  a distribution is proposed. It is very accurate as long as the mode
  of the distribution is approximately symmetric around it, like the
  pixel count distributions of Figure \ref{dataandnoise}. This is a
  safe constraint for images of astronomical targets, which generally
  have more fainter pixels than brighter ones (the yellow histograms
  of Figure \ref{dataandnoise}). When the noise is significant, the
  noise will ensure a symmetric mode similar to the examples of Figure
  \ref{dataandnoise}. As the signal becomes more significant, the
  quantile of the mode will shift to lower values.

  Given an \emph{ordered} data set $X$ with $n$ elements, where
  $X_n\geq{X_{n-1}} (n>1)$. A mirrored distribution is defined about
  the \emph{mirror} point located at index $m$ as follows. All the
  data prior to $m$ are placed in opposite order after $m$, as if a
  mirror was placed there. The elements of the mirrored data set are
  thus defined with

  \begin{displaymath}
    M_i = \left\{
    \begin{array}{l l}
      X_i                      & \quad i\leq m \\
      X_m + (X_m-X_{m-(i-m)})   & \quad i>m
    \end{array}\right.
  \end{displaymath}

  The mirrored distribution is therefore identical to the original
  distribution until and including $X_m$. $M$ has $2(m-1)+1$ elements
  and like $X$, it is ordered. Figure \ref{mirrordemo} shows one
  example data set (the pixels of Figure \ref{segtf}(b.2)) with
  mirrored distributions overplotted. The mirror points in Figure
  \ref{mirrordemo} are placed at various quantiles of the original
  distribution in each case.  The histogram is only displayed in this
  demonstration to aid the understanding. It is not used at all during
  the analysis.

  Figure \ref{mirrordemo} shows that when the mirror is very far from
  the mode, the cumulative frequency plot of the mirrored distribution
  will deviate below or above that of the original pixels very
  sharply. As the mirror approaches the mode, the maximum difference
  of the two cumulative frequency plots in the vicinity of the mode
  becomes smaller. Therefore, on the symmetric mode of a distribution,
  the two cumulative frequency plots should remain very similar for a
  large number of pixels (or distance in the ordered array) after the
  mirror.

  Let $\delta(M_i,m)$ represent the difference between the cumulative
  frequency plots of the original and mirrored data sets for $M_i$,
  when the mirror is placed on the index $m$. For any given mirror
  position and positive integer $N$, let $\Delta(m)$ be the maximum
  $|\delta(M_i,m)|$ for all $m<i<m+N$. Then the symmetric mode of a
  distribution can be found by finding $m$ such that $\Delta(m)$ is
  minimized. The Golden section search algorithm \citep{GoldenSection}
  is used to find the mirror point that minimizes $\Delta(m)$.

    \begin{figure*}
    \centering
    \ifdefined\makeeps
    \newcommand{\inputdir}{\figtexdir/mode}

\makeatletter \newcommand{\pgfplotsdrawaxis}{\pgfplots@draw@axis} \makeatother

    \pgfplotsset{axis line on top/.style={
      axis line style=transparent,
      ticklabel style=transparent,
      tick style=transparent,
      axis on top=false,
      after end axis/.append code={
        \pgfplotsset{axis line style=opaque,
          ticklabel style=opaque,
          tick style=opaque,
          grid=none}
        \pgfplotsdrawaxis}
      }
    }

    \begin{tikzpicture}

        \scriptsize
        \begin{groupplot}[group style={group size=4 by 1,
                     horizontal sep=0pt},
                     xmode=normal,ymode=normal,
                     enlargelimits=false,
                     no markers, width=0.25\linewidth,
                     height=0.15\linewidth,
                     scaled x ticks = false,
                     xlabel={Pixel intensity ($e^-$) - $M_m$},
                     scale only axis, axis line on top]

          \nextgroupplot[yticklabels={}, ytickmin=-2, ytickmax=-1]
            \addplot [const plot mark mid, fill=myblue]
                     table [x expr={\thisrowno{0}*10000}, y index=1]
                           {\inputdir/1_seg_crop_modehist.txt}
            \closedcycle;
            \addplot [const plot mark mid, fill=mygreen, fill opacity=0.6]
                     table [x expr={\thisrowno{0}*10000}, y index=2]
                           {\inputdir/1_seg_crop_modehist.txt}
            \closedcycle;

            \addplot [very thick, color=myblack]
                     table [x expr={\thisrowno{0}*10000}, y index=1]
                     {\inputdir/1_seg_crop_modecfp.txt};
            \addplot [thin, color=myblack]
                     table [x expr={\thisrowno{0}*10000}, y index=2]
                     {\inputdir/1_seg_crop_modecfp.txt};

            \draw [dotted]
                  ({axis cs:\moderealonesympoint,0}|-{rel axis cs:0,1})--
                  ({axis cs:\moderealonesympoint,0}|-{rel axis cs:0,0});

          \nextgroupplot[yticklabels={}, ytickmin=-2, ytickmax=-1]
            \addplot [const plot mark mid, fill=myblue]
                     table [x expr={\thisrowno{0}*10000}, y index=1]
                           {\inputdir/6_seg_crop_modehist.txt}
            \closedcycle;
            \addplot [const plot mark mid, fill=mygreen, fill opacity=0.6]
                     table [x expr={\thisrowno{0}*10000}, y index=2]
                           {\inputdir/6_seg_crop_modehist.txt}
            \closedcycle;

            \addplot [very thick, color=myblack]
                     table [x expr={\thisrowno{0}*10000}, y index=1]
                     {\inputdir/6_seg_crop_modecfp.txt};
            \addplot [thin, color=myblack]
                     table [x expr={\thisrowno{0}*10000}, y index=2]
                     {\inputdir/6_seg_crop_modecfp.txt};

            \draw [dotted]
                  ({axis cs:\moderealsixsympoint,0}|-{rel axis cs:0,1})--
                  ({axis cs:\moderealsixsympoint,0}|-{rel axis cs:0,0});

          \nextgroupplot[yticklabels={}, ytickmin=-2, ytickmax=-1]
            \addplot [const plot mark mid, fill=myblue]
                     table [x index=0, y index=1]
                           {\inputdir/mode1_modehist.txt}
            \closedcycle;
            \addplot [const plot mark mid, fill=mygreen, fill opacity=0.6]
                     table [x index=0, y index=2]
                           {\inputdir/mode1_modehist.txt}
            \closedcycle;

            \addplot [very thick, color=myblack]
                     table [x index=0, y index=1]
                     {\inputdir/mode1_modecfp.txt};
            \addplot [thin, color=myblack]
                     table [x index=0, y index=2]
                     {\inputdir/mode1_modecfp.txt};

            \draw [dotted]
                  ({axis cs:\modemodeonesympoint,0}|-{rel axis cs:0,1})--
                  ({axis cs:\modemodeonesympoint,0}|-{rel axis cs:0,0});

          \nextgroupplot[yticklabels={}, ytickmin=-2, ytickmax=-1]
            \addplot [const plot mark mid, fill=myblue]
                     table [x index=0, y index=1]
                           {\inputdir/mode2_modehist.txt}
            \closedcycle;
            \addplot [const plot mark mid, fill=mygreen, fill opacity=0.6]
                     table [x index=0, y index=2]
                           {\inputdir/mode2_modehist.txt}
            \closedcycle;

            \addplot [very thick, color=myblack]
                     table [x index=0, y index=1]
                     {\inputdir/mode2_modecfp.txt};
            \addplot [thin, color=myblack]
                     table [x index=0, y index=2]
                     {\inputdir/mode2_modecfp.txt};

            \draw [dotted]
                  ({axis cs:\modemodetwosympoint,0}|-{rel axis cs:0,1})--
                  ({axis cs:\modemodetwosympoint,0}|-{rel axis cs:0,0});

        \end{groupplot}

        \node[anchor=south west] at (0.16\linewidth,0.1\linewidth)
             {\normalsize a: $\moderealonemodeq\%$};
        \node[anchor=south west] at (0.41\linewidth,0.1\linewidth)
             {\normalsize b: $\moderealsixmodeq\%$};
        \node[anchor=south west] at (0.66\linewidth,0.12\linewidth)
             {\normalsize c: $\modemodeonemodeq\%$};
        \node[anchor=south west] at (0.92\linewidth,0.12\linewidth)
             {\normalsize d: $\modemodetwomodeq\%$};

    \end{tikzpicture}
    \else
    \includegraphics[width=\linewidth]{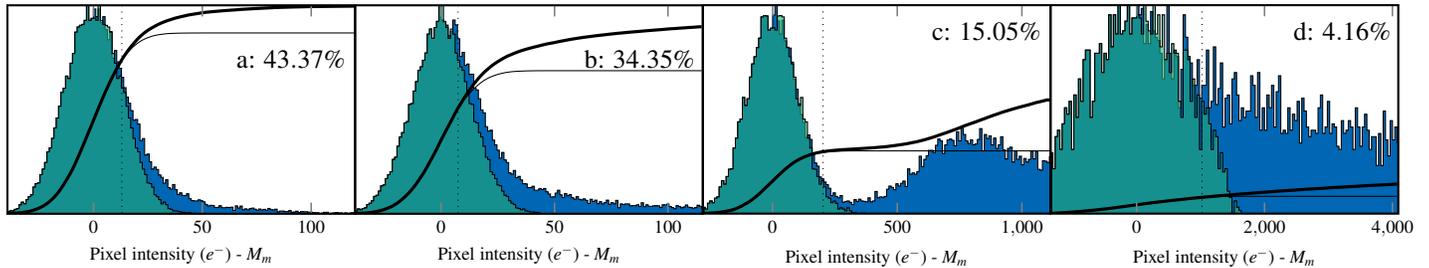}
    \fi
    \caption{\label{mode} Finding the mode by minimizing
      $\Delta(m)$. Similar to Figure \ref{mirrordemo}, with the mirror
      placed on the mode. The dotted line shows the first value where
      $|\delta(M_i,m)|>\imgstatmirrordist\sqrt{m}$, which is used to
      find the ``symmetricity'' of the mode (see text). (a) Same data
      set as Figure \ref{mirrordemo}. (b) The convolved image of
      Figure \ref{objseg}(c.1): a bright and large galaxy skewing the
      distribution, further shifting the mode to lower
      percentiles. (c) The convolved and noisy image of a circular
      $\modeonemag$ magnitude mock S\'{e}rsic profile with
      $n=\modeonen$, $r_e=\modeonere$ pixels, truncated at
      $\modeonet{}r_e$ with center on ($\modeonex$, $\modeoney$) in a
      $200\times200$ pixel image. (d) Similar to (c) but $\modetwomag$
      magnitude with $n=\modetwon$, $r_e=\modetwore$, truncated at
      $\modetwot{}r_e$ placed at ($\modetwox$, $\modetwoy$).}
  \end{figure*}

  If the distribution has several symmetric local maximums, and as
  long as $N$ is large enough, the first (with the lowest quantile)
  will be found, even if the brighter peak has more pixels in its
  proximity (see Figure \ref{mode}(c) for a distribution with two
  local maximums in the histogram). This is exactly what is desired in
  astronomical image processing because if there is a second local
  maximum with a larger count and larger number of pixels, then that
  local maximum is not due to noise. Noise will always produce the
  first (lowest count) local maximum, which can be called the
  \emph{mode}. In most astronomical applications of low {\small S/N}
  objects, the first local maximum will be the only one. If not, it
  will have more pixels in its vicinity than any other local maximum
  in the pixel distribution.

  The mirror cumulative frequency plot found for the mode should be
  approximately equal to the actual data in the vicinity of the
  mode. Beyond that, it should ideally always remain below the actual
  data (see Figure \ref{mode}). However, in minimizing $\Delta(m)$ as
  defined above, the resulting mirror cumulative frequency plot will
  zig-zag around the actual distribution since $|\delta(M_i,m)|$ was
  used. The cumulative frequency plot is fundamentally a counting
  function. Therefore, the error in choosing an index, say $m$,
  follows a Poisson distribution. Thus the standard deviation in
  finding $m$ is $\sqrt{m}$. Hence a limit can be set on how negative
  $\delta(M_i,m)$ can become. Taking $\alpha$ to be a given constant
  specified by the user, if $\delta(M_i,m)<-\alpha\sqrt{m}$,
  $\Delta(m)$ is set to a fixed constant, let's call it
  $C$.\footnote{The value of $C$ is an internal issue and is
    irrelevant to this discussion or the user. It is only necessary
    that it should be larger than the largest possible $\Delta(m)$.}
  The golden section search is modified from its standard algorithm to
  choose the interval with the lowest values when it confronts
  $C$. Recall that such conditions only occur when $m$ is a
  significant overestimation of the mode.

  Figure \ref{mode}(a) shows the result for the same distribution of
  Figure \ref{mirrordemo}. In Figure \ref{mode}(b) this technique is
  applied to the more skewed distribution of the convolved image of
  Figure \ref{objseg}(c.1). A very bright spiral galaxy has caused the
  postage stamp pixel distribution to be more skewed. Figures
  \ref{mode}(c) and (d) show the distributions of two mock images with
  two very bright and large S\'{e}rsic profiles of $n=\modeonen$ and
  $n=\modetwon$ placed in a small postage stamp respectively. These
  are two extreme cases, showing how the quantile of the mode can
  reach extremely low quantiles.

  As the distribution becomes more skewed, the mode shifts to lower
  quantiles; see Figure \ref{mode}. This is the basis of the new
  method to define the threshold of the image (explained fully in
  Section \ref{thresh} and Section \ref{lithresh}). In these examples,
  the boundaries for the golden section search are set to the 0.01 and
  0.51 quantiles of the image. As long as data is present in the image
  (which create a positive skew) the mode will not exceed the median,
  thus the higher limit. In these examples, $N$ is set to be $1/2$ of
  the total number of pixels and in order to be more efficient,
  $\delta(M_i,m)$ is sampled in $1000$ equally spaced points between
  $M_m$ and $M_{m+N}$. $\alpha$ is also taken to be
  $\imgstatmirrordist$.

  Depending on the brightness of the data pixels and the fraction of
  pixels that they occupy, the mode calculated with this technique can
  loose its accuracy or it can cease to exist. Also, in extremely
  skewed distributions, the golden section intervals might miss the
  mode and converge on a lower mirror point. Therefore a measure of
  quality is necessary to accompany the mode.

  Let $a$ be the value of the $0.01$ quantile of the mirrored
  distribution. The minimum is not chosen due to its scatter. Take $b$
  to be the value on the mode and finally, let $c$ be the value of the
  first $M_i$ beyond the mode where $|\delta(M_i,m)|>\alpha\sqrt{m}$,
  same $\alpha$ as above. This point is marked on the plots of Figure
  \ref{mode} with a dotted vertical line. These three values can
  define a measure of ``symmetricity'' ($s$) about the mode with:

  $$s=\frac{c-b}{b-a}$$

  The symmetricity for the examples of Figure \ref{mode} are
  respectively: $\moderealonesym$, $\moderealsixsym$,
  $\modemodeonesym$ and $\modemodetwosym$. Based on these and other
  simulations, $s>0.2$ appears to be a good symmetricity of an
  accurate ``symmetric'' mode, when the mode is above the $\sim0.02$
  quantile of the image. When the output mode approaches such
  quantiles, $b-a$ in the equation above will be too small to give an
  accurate result.

  In addition to the initial cost of sorting of the data set, the rest
  of the process is very simple and not {\small CPU} intensive, since
  only 1000 points are checked in each round of the golden section
  search. Therefore the {\small CPU} cost of running this technique is
  similar with the cost for $\sigma$-clipping which also requires
  sorting, to find the median, but needs multiple passes through the
  whole data.

\section{SExtractor Parameters}\label{SEconfig}
  Below is a list of the \textsf{SExtractor} parameters used in this
  study. We make no claim that these are ``the best'' (or worst)
  possible set of parameters. The values for each parameter were
  selected based on values reported by other research teams which
  extensively used \textsf{SExtractor} in their astrophysical data
  analysis and its manual; see Appendices \ref{existingsky} and
  \ref{existingmethods}.

  The file \texttt{PSF.conv} refers to the {\small PSF} used to create
  mock images when \textsf{SExtractor} was applied to the mock images
  and to that produced by \textsf{TinyTim} when real \emph{\small HST}
  images were used; see Appendix \ref{existingmethods} for more
  details.

  \noindent\makebox[\linewidth]{\rule{\linewidth}{1pt}}

\noindent
\begin{alltt}
\footnotesize
MEMORY_OBJSTACK  \MEMORYOBJSTACK
MEMORY_PIXSTACK  \MEMORYPIXSTACK
MEMORY_BUFSIZE   \MEMORYBUFSIZE

DETECT_TYPE      \DETECTTYPE
GAIN_KEY         \GAINKEY
PIXEL_SCALE      \PIXELSCALE
FITS_UNSIGNED    \FITSUNSIGNED

BACK_TYPE        \BACKTYPE
BACK_SIZE        \BACKSIZE
BACK_FILTERSIZE  \BACKFILTERSIZE
BACKPHOTO_TYPE   \BACKPHOTOTYPE
BACKPHOTO_THICK	 \BACKPHOTOTHICK

WEIGHT_TYPE      \WEIGHTTYPE
WEIGHT_GAIN      \WEIGHTGAIN

FILTER           \FILTER
FILTER_NAME      \FILTERNAME
MASK_TYPE        \MASKTYPE

INTERP_MAXXLAG 	 \INTERPMAXXLAG
INTERP_MAXYLAG 	 \INTERPMAXYLAG
INTERP_TYPE      \INTERPTYPE

DETECT_THRESH    \DETECTTHRESH
ANALYSIS_THRESH  \ANALYSISTHRESH
THRESH_TYPE      \THRESHTYPE
DETECT_MINAREA   \DETECTMINAREA

DEBLEND_NTHRESH  \DEBLENDNTHRESH
DEBLEND_MINCONT  \DEBLENDMINCONT
CLEAN            \CLEAN
CLEAN_PARAM      \CLEANPARAM

FLAG_IMAGE       \FLAGIMAGE
FLAG_TYPE        \FLAGTYPE

SEEING_FWHM      \SEEINGFWHM
STARNNW_NAME     \STARNNWNAME

MAG_ZEROPOINT    \MAGZEROPOINT
PHOT_APERTURES   \PHOTAPERTURES
PHOT_AUTOPARAMS  \PHOTAUTOPARAMS
PHOT_AUTOAPERS   \PHOTAUTOAPERS
PHOT_FLUXFRAC    \PHOTFLUXFRAC
PHOT_PETROPARAMS \PHOTPETROPARAMS
SATUR_LEVEL      \SATURLEVEL
SATUR_KEY        \SATURKEY

CHECKIMAGE_TYPE  \CHECKIMAGETYPE
CHECKIMAGE_NAME  \CHECKIMAGENAME

CATALOG_NAME     \CATALOGNAME
CATALOG_TYPE     \CATALOGTYPE
PARAMETERS_NAME  \PARAMETERSNAME

VERBOSE_TYPE     \VERBOSETYPE
\end{alltt}

  \noindent\makebox[\linewidth]{\rule{\linewidth}{0.7pt}}

\section{NoiseChisel Parameters}\label{NCconfig}
  The full list of parameters that was used for \textsf{NoiseChisel}
  in this paper is printed here. If a different parameter was used, it
  is explained in the text. Note that we do not consider these
  parameters to be ``the best'' for any generic data set. For the
  processing of Figure \ref{largeimage}, \texttt{-{}-nch1=\scccdncha},
  \texttt{-{}-lmesh=\scccdlmesh} and the following two on/off options
  were also activated \texttt{-{}-fullinterpolation} and
  \texttt{-{}-fullsmooth}. For the \emph{\small HST} images,
  \texttt{-{}-skysubtracted} was activated.

  \noindent\makebox[\linewidth]{\rule{\linewidth}{1pt}}

\noindent

\begin{alltt}
    \footnotesize
# Input:
 hdu                 \hdu{}
 khdu                \khdu{}
 skysubtracted       \skysubtracted{}
 minbfrac            \minbfrac{}
 minnumfalse         \minnumfalse{}

# Mesh grid:
 smeshsize           \smeshsize{}
 lmeshsize           \lmeshsize{}
 nch1                1
 nch2                1
 lastmeshfrac        \lastmeshfrac{}
 mirrordist          \mirrordist{}
 minmodeq            \minmodeq{}
 numnearest          \numnearest{}
 smoothwidth         \smoothwidth{}
 fullconvolution     \fullconvolution{}
 fullinterpolation   \fullinterpolation{}
 fullsmooth          \fullsmooth{}

# Detection:
 qthresh             \qthresh{}
 erode               \erode{}
 erodengb            \erodengb{}
 opening             \opening{}
 openingngb          \openingngb{}
 sigclipmultip       \sigclipmultip{}
 sigcliptolerance    \sigcliptolerance{}
 dthresh             \dthresh{}
 detsnminarea        \detsnminarea{}
 detsnhistnbins      \detsnhistnbins{}
 detquant            \detquant{}
 dilate              \dilate{}

# Segmentation:
 segsnminarea        \segsnminarea{}
 segquant            \segquant{}
 segsnhistnbins      \segsnhistnbins{}
 gthresh             \gthresh{}
 minriverlength      \minriverlength{}
 objbordersn         \objbordersn{}
\end{alltt}

\noindent\makebox[\linewidth]{\rule{\linewidth}{0.7pt}}

\ifdefined\usebiblatex
\printbibliography
\else
\bibliography{./tex/Ref}
\fi
\end{document}